\documentclass[aps,prb,twocolumn,superscriptaddress]{revtex4-2}

\usepackage{graphicx}
\usepackage{color}
\usepackage{amsmath, amssymb}
\usepackage{float}
\newcommand{\BETS}{$\kappa$-(BETS)$_2$Mn[N(CN)$_2$]$_3$}
\newcommand{\kBETS}{$\kappa$-BETS-Mn}

\newcommand{\eps}{$\varepsilon^{\prime}$}
\newcommand{\betaICl}{$\beta^\prime$-(ET)$_2$ICl$_2$}

\begin{document}

\title{Ferroelectric and Multiferroic Properties of Quasi-2D Organic Charge-Transfer Salts}

\author{Michael Lang}
\affiliation{Institute of Physics, Goethe University Frankfurt, 60438 Frankfurt (M), Germany}

\author{Peter Lunkenheimer}
\affiliation{Experimental Physics V, Center for Electronic Correlations and Magnetism, University of Augsburg, 86159 Augsburg, Germany}

\author{Owen Ganter}
\affiliation{Department of Physics and Center for Functional Materials, Wake Forest University, Winston-Salem, North Carolina 27109, USA}

\author{Stephen Winter}
\affiliation{Department of Physics and Center for Functional Materials, Wake Forest University, Winston-Salem, North Carolina 27109, USA}

\author{Jens M\"uller}
\affiliation{Institute of Physics, Goethe University Frankfurt, 60438 Frankfurt (M), Germany}

\begin{abstract}
In conventional ferroelectrics the electric dipoles are generated by off-center displacements of ions. In recent years, a new type of so-called electronic ferroelectrics has attracted great attention, where the polarization is driven by electronic degrees of freedom. Of particular interest are materials with strong electronic correlations, featuring a variety of intriguing phenomena and instabilities, which may interact with or even induce electronic ferroelectricity. In this review, we will focus on the class of strongly correlated charge-transfer salts, where electronic ferroelectricity was suggested by theory and has been confirmed by numerous experiments. The paper summarizes some basic physical properties of various relevant quasi-twodimensional salts and gives some background on the experimental tools applied to establish ferroelectricity. We discuss the key experimental observations, including the exciting discovery of multiferroicity, and provide some theoretical considerations on the magnetoelectric couplings that are of relevance here.
\end{abstract}

\maketitle

\section{Introduction}\label{sec1}

Organic charge-transfer salts continue to be the subject of intensive investigations due to their rich phenomenology resulting from reduced dimensions and the combined effects of strong electron-electron and electron-phonon interactions in the presence of geometrical frustration \cite{Seo2004,Toyota2007,Powell2011,Kanoda2011,Clay2019}.

In addition, a particular type of ferroelectricity, and even multiferroicity, have been theoretically predicted \cite{van-den-Brink2008,Ishihara2010} and experimentally observed \cite{Lunkenheimer2012}. Here the ferroelectric polar order is generated by electronic degrees of freedom in contrast to the off-center displacement of ions in conventional ferroelectrics. The much lighter mass of the electrons in these charge-driven, so-called electronic ferroelectrics implies a distinctly different high-frequency dynamics, which opens up new routes for technical applications, see, e.g., \cite{spaldin2019advances} for a recent review. 
The research activities related to the intriguing dielectric properties of charge-transfer salts, building on the discovery of charge-order-driven ferroelectricity in the quasi-1D Fabre salts \cite{Nad2000,Monceau2001} in the year 2000, have gained momentum in recent years as clear signatures for electronic ferroelectricity and even multiferroicity have been reported for various families of the quasi-2D charge-transfer salts. While we will briefly touch on the important early findings in the quasi-1D systems, the focus of our paper will lie on the more recent developments in the field of quasi-2D systems. There are  a few previous review articles available \cite{Lunkenheimer2015a,Tomic2015,Riedl2022} addressing some of the aspects discussed here. 

The paper is organized as follows. In chapter \ref{sec2} we will summarize some basic properties of the relevant families of organic charge-transfer salts including their lattice and electronic structures, and give some theoretical considerations on the materials' symmetry, their magentoelectric coupling and phase diagrams. In chapter \ref{sec-Exp-Techn} we will provide a brief introduction into the experimental techniques used to probe the dielectric properties and to investigate the involvement of other degrees of freedom. Priority is given to the techniques employed by the authors’ own investigations. In chapter \ref{sec-Sign-ferroelectricity} we recall the rich variety of dielectric phenomena observed in the quasi-1D and quasi-2D salts. The experimental results on the $\kappa$-phase salts will be discussed in more detail in chapter \ref{sec-beyond dimer Mott} with regard to the role of intra-dimer degrees of freedom and the nature of relaxor-type ferroelectricity. A short summary is given in chapter \ref{sec-summary}.\\     

\section{Quasi-1D and quasi-2D charge-transfer salts}\label{sec2}

\subsection{Crystallographic and electronic structure}\label{subsec2}
\begin{figure}[!t]
\centering
\includegraphics[width=0.35\textwidth]{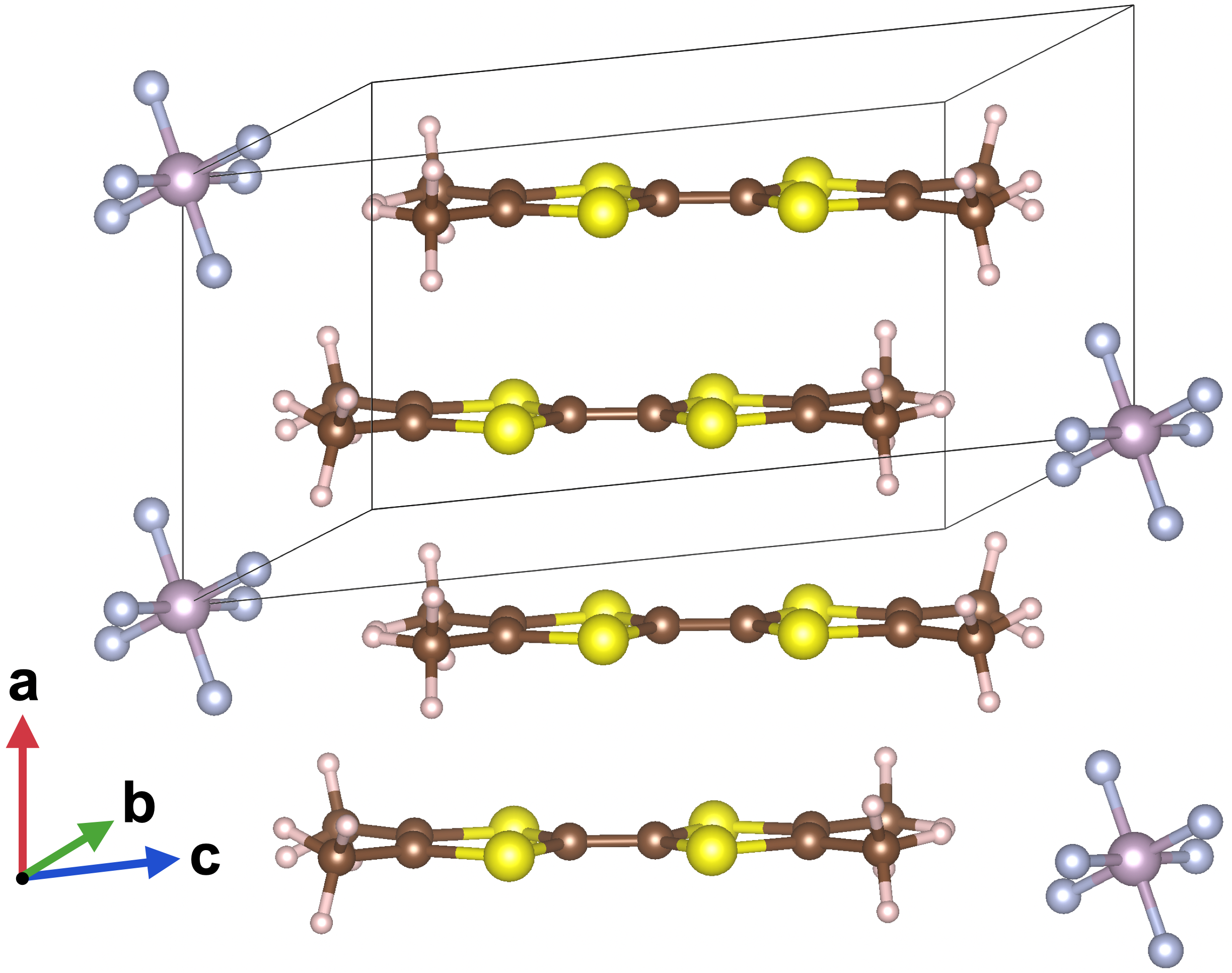} 
\caption{Crystal structure of (TMTTF)$_2$PF$_6$ viewed from a somewhat tilted angle relative to the $b$ direction. The TMTTF molecules pile up along the $a$ direction, the axis of highest conductivity.}\label{TMTTF-structure}
\end{figure}

  Evidence for electronic ferroelectricity in organic charge-transfer (CT) salts was first observed in the family of the quasi-onedimensional (quasi-1D) conductors (TMTTF)$_2X$ \cite{Monceau2001} where TMTTF stands for tetramethyltetrathiafulvalene and $X$ for a monovalent anion. In these salts, planar organic molecules are stacked along the $a$ axis for which the intermolecular overlap of $\pi$ orbitals is strongest (see Fig.\,\ref{TMTTF-structure}), thus forming a quasi-1D band structure. Due to a dimerization, the strength of which depends on the anion $X$, there is one hole per dimer. Upon cooling to below $T_\mathrm{CO} \sim$ 50 - 150\,K, charge ordering sets in where the TMTTF molecules within the dimer become non-equivalent. The charge disproportionation below $T_\mathrm{CO}$ and the accompanying loss of inversion symmetry are key for the formation of polar order in this family of charge-transfer salts. \\

\begin{figure}[!b]
\centering
\includegraphics[width=0.45\textwidth]{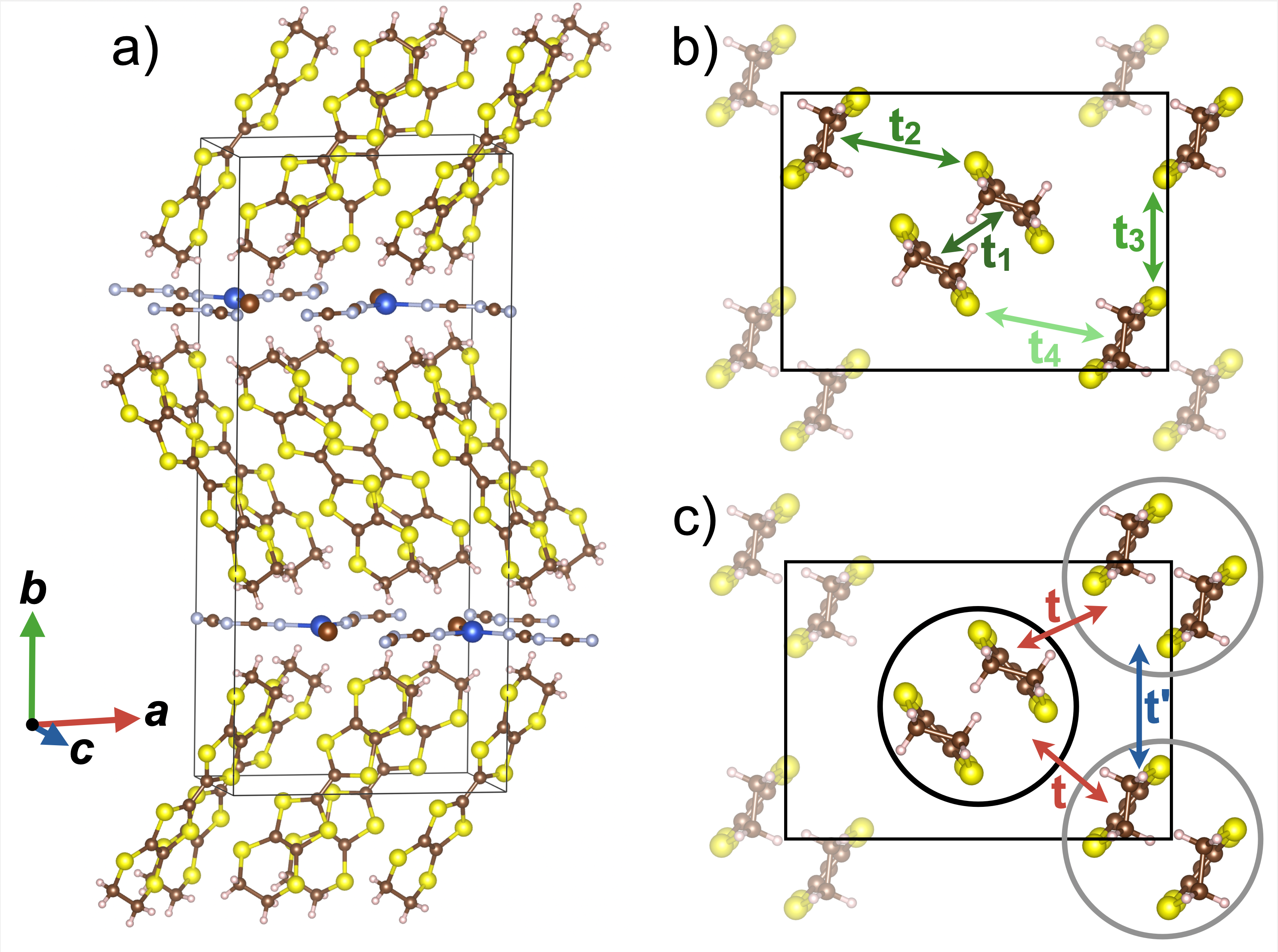} 
\caption{(a) Crystal structure of $\kappa$-(ET)$_2X$ with thick layers of ET molecules separated by thin sheets of inorganic anions $X$. (b) Closer look on the ET layer viewed along the molecules’ long axes with $t_{1,2,3,4}$ referring to the hopping parameters between the highest occupied molecular orbitals. (c) For strong intra-dimer hopping $t_1$, pairs of ET molecules form a dimer (spheres) which are arranged on an anisotropic triangular lattice with hopping terms $t$ and $t'$.}\label{kappa-Cl-structure}
\end{figure}

\begin{figure}[!t]
\centering
\includegraphics[width=0.4\textwidth]{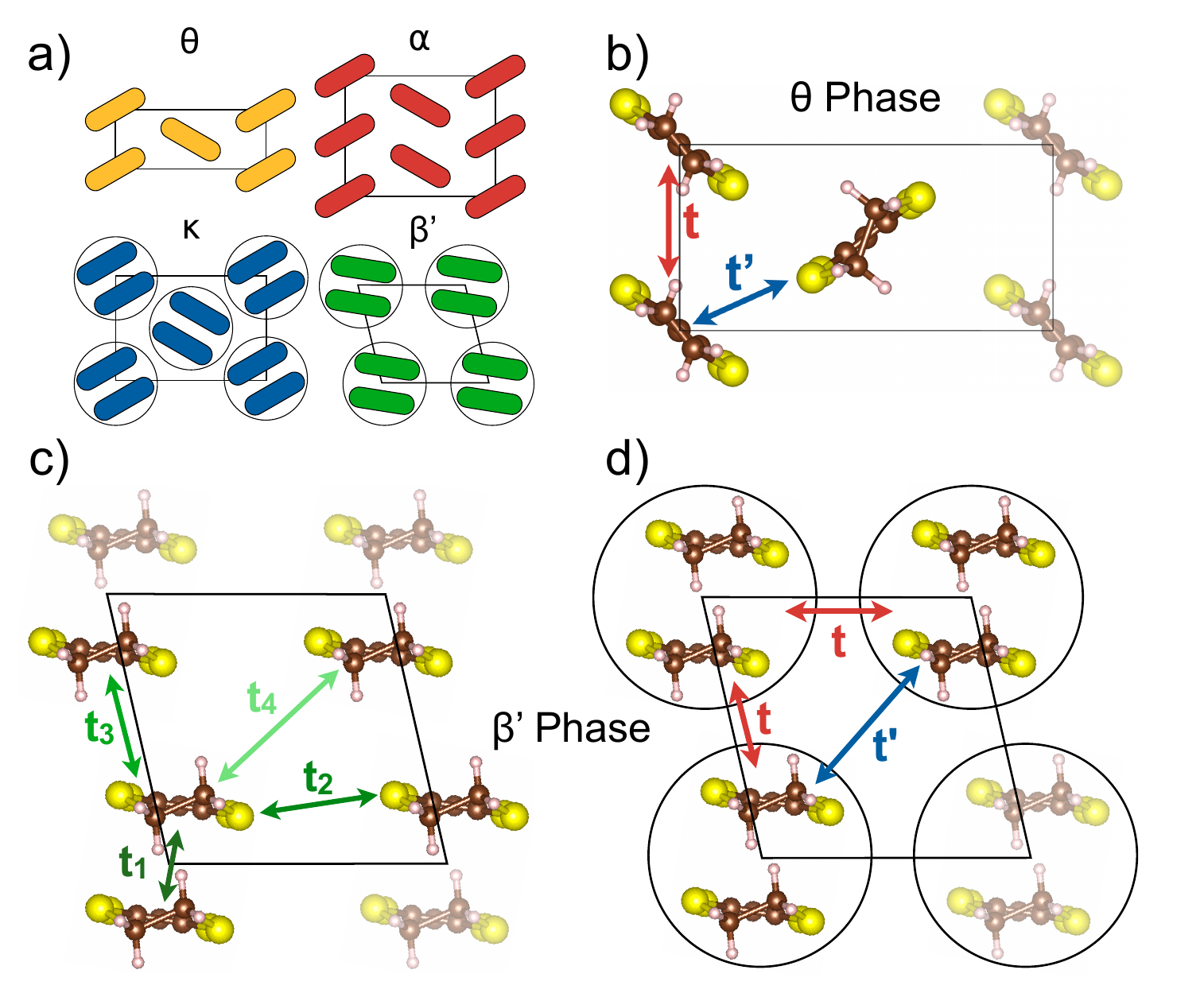} 
\caption{(a) Cartoon depiction of four relevant polymorphs. (b) ET layer for $\theta$-phase packing showing in-stack interaction $t$ and out-of-stack interaction $t'$. (c) ET layer for $\beta'$-phase packing with inter-molecular hopping integrals labelled $t_{1-4}$. (d) inter-dimer hopping integrals labelled as $t$ and $t'$.}\label{cartoon}
\end{figure}

Subsequently, electronic ferroelectricity has also been found in 2D salts of BEDT-TTF and BETS, where BEDT-TTF (or simply ET) stands for bisethylenedithio-tetrathiafulvalene, while BETS stands for bisethylenedithio-tetraselena-fulvalenes. In the BETS molecule the four S atoms in the inner TTF skeleton are replaced by Se. Prominent examples for this type of electron-driven ferroelectricity include some $\kappa$-(ET)$_2X$ salts (Fig.\,\ref{kappa-Cl-structure}) as well as some $\theta$-phase variants, see Fig.\,\ref{cartoon}(a) for a cartoon of the various relevant (ET)$_2X$ polymorphs. \\   

\begin{figure}[!t]
\centering
\includegraphics[width=0.4\textwidth]{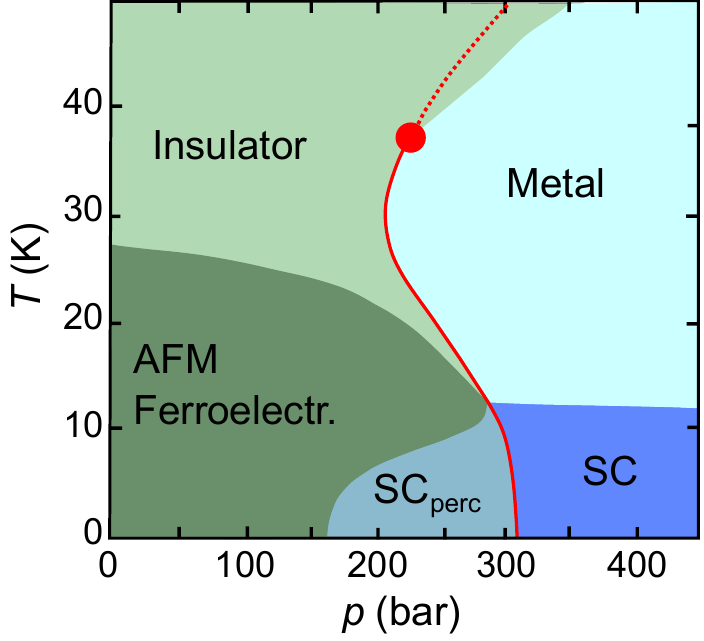}
\caption{Temperature-pressure phase diagram of $\kappa$-(ET)$_2$Cu[N(CN)$_2$]Cl. At ambient and small pressure, the systems shows antiferromagnetic (AFM) order accompanied by ferroelectric order. The evolution of the ferroelectric order under pressure has not yet been investigated. With increasing pressure the system undergoes a first-order Mott metal insulator transition (red solid line) which terminates in a second-order critical endpoint (red sphere). On the high-pressure side of the Mott transition superconductivity (SC) occurs. Taken from \cite{Riedl2022}.}
\label{kappa-Cl-phase-diagram}
\end{figure}

The $\kappa$-phase constitute the most intensively studied class of (ET)$_2X$ salts due to its model character for exploring the behavior of  correlated electrons in reduced dimensions. A plethora of intriguing phenomena has been reported, including the Mott metal-insulator transition \cite{Kanoda1997,Limelette2003,Fournier2003} and its critical properties \cite{Kagawa2005,Kagawa2009,Furukawa2015a,Isono2016,Gati2016}, unconventional superconductivity \cite{Powell2011,Guterding2016,Wosnitza2019,Clay2019,Furukawa2023}, strongly frustrated quantum magnetism \cite{Shimizu2003,Powell2011,Riedl2019,Pustogow2014}, ferroelectricity \cite{Gati2018b} as well as multiferroicity \cite{Lunkenheimer2012}.  Figure \ref{kappa-Cl-phase-diagram} shows exemplarily the temperature-pressure phase diagram for $\kappa$-(ET)$_2$Cu[N(CN)$_2$]Cl ($\kappa$-Cl) highlighting a rich phenomenology by pressure tuning the system across the Mott metal-insulator transition. 
At ambient pressure ($p$ = 0) this system is a Mott insulator which orders antiferromagnetically below $T_\mathrm{N}$ = 27\,K \cite{Miyagawa1995}. Upon cooling, the magnetically-ordered state was found to be accompanied by ferroelectric order \cite{Lunkenheimer2012} making this system multiferroic with $T_\mathrm{FE} \approx T_\mathrm{N}$ suggesting an intimate interrelation between both types of ordering. By the application of moderate pressure of only 300\,bar the system undergoes a first-order Mott transition \cite{Kanoda1997,Lefebvre2000,Limelette2003,Fournier2003}. This first-order line terminates in a second-order critical endpoint at ($T_\mathrm{cr}$, $p_\mathrm{cr}$) $\approx$ (36.5\,K, 234\,bar) \cite{Kagawa2004,Kagawa2005,Kagawa2009,Furukawa2015a,Isono2016,Gati2016}. On the metallic side of the Mott transition, superconductivity (SC) is observed below $T_\mathrm{c} \approx$ 12\,K. \\

\begin{table*}
\caption{\label{tab:parameters}Model parameters for relevant $\kappa$- and $\beta$'-type organic charge-transfer salts. }
\begin{center}
\begin{tabular}{|c|c|c|c|c|c|c|c|c|c|c|}
\hline
Material & $T$ & Ref & $t_1$ & $t_2$ & $t_3$ & $t_4$ & $t$ & $t^\prime$ &  $|t^\prime / t|$ & $|t_1 / t'|$ \\
\hline
$\kappa$-(ET)$_2$Cu[N(CN)$_2$]Cl & 15 & \cite{Koretsune2014} & 207 & 102 & 67 & 43 & 73 & 34 & 0.46 & 6.2 \\
\hline
 $\kappa$-(ET)$_2$Cu$_2$(CN)$_3$ & 293 & \cite{Guterding2015} & 176 & 78.0 & 81.4 & 18.7 & 48.4 & 40.7 &  0.84 & 4.3 \\
\hline
 $\kappa$-(ET)$_2$Hg(SCN)$_2$Cl & 296 & \cite{Gati2018} & 126.6 & 60.0 & 80.8 & 42.0 & 51.0 & 40.4 & 0.79 & 3.1 \\
\hline
 $\kappa$-(BETS)$_2$Mn[N(CN)$_2$]$_3$ & RT & \cite{Riedl2021} & 177 & 8 & 125 & 63 & 36 & 63 & 1.76 & 2.8 \\ 
\hline
 $\beta$'-(ET)$_2$ICl$_2$ & 12 & \cite{Koretsune2014} & 251 & -22 & -36 & 102 & 40 & -18 & 0.45 & 13.9 \\
\hline
\end{tabular}
\end{center}
\label{model parameters}
\end{table*}

In these $\kappa$-phase salts, the most common mode of charge order (CO) is intra-dimer charge disproportionation, pictured in Fig.~\ref{fig:kappa-phase}. The propensity for this mode of CO can be considered as a competition between dimerization strength and inter-dimer Coulomb repulsion. 
By symmetry, there are four distinct hopping parameters $t_{1,2,3,4}$, cf. Fig.~\ref{kappa-Cl-structure}(b). For strong intra-dimer hopping $t_1$, the systems can be modelled by an effective-dimer model [Fig.~\ref{kappa-Cl-structure}(c)] characterized by only two hopping terms $t$ and $t'$ given by $t$ = ($t_2$ + $t_4$)/2 and $t'$ = $t_3$/2 \cite{Komatsu1996,Kandpal2009}. In this limit, the materials can be viewed as effectively half-filled, with one electron per dimer occupying the anti-bonding combination of molecular orbitals, and shared equally between both molecules. The appearance of intra-dimer CO represents a departure from this pure dimer-Mott limit. Table~\ref{tab:parameters} summarizes the hopping parameters for $\kappa$-(ET)$_2$Cu[N(CN)$_2$]Cl and related $\kappa$-phase salts. The strength of dimerization (and therefore the propensity to form intra-dimer CO) can be evaluated from the  $|t_1 / t|$ and $|t_1 / t^\prime |$. In general, $\kappa$-(ET)$_2$Cu[N(CN)$_2$]Cl has a relatively strong dimerization (quantified by a large value of $t_1$), while $\kappa$-(ET)$_2$Hg(SCN)$_2$Cl has the weakest dimerization. The theoretical consequences for CO are reviewed in detail in section \ref{sec_phase_diagrams}. 

The $\theta$-phase [pictured in Fig.~\ref{cartoon}(b)] is the second crystallographic phase where ferroelectricity is prominently observed in 2D CT salts. In this case, the ET molecules form undimerized $\pi$-stacks, which occupy an anisotropic triangular lattice defined by two hopping integrals: $t$ and $t^\prime$. With one hole per two organic molecules, these are viewed as 3/4-filled, and exhibit a variety of CO patterns resulting largely from the balance of inter-molecular Coulomb repulsion terms. The phase diagram is discussed in more detail in section \ref{sec_phase_diagrams}.

Finally, we also address the structure of compounds in the $\alpha$- and $\beta^\prime$-phases, discussed in section \ref{sec-Sign-ferroelectricity}. The orientations of the organic molecules in these phases is shown schematically in Fig.~\ref{cartoon}(a). The $\alpha$-phase can be viewed as a variant of the $\theta$-phase, in which the unit cell is doubled along the $\pi$-stacking direction, and half of the $\pi$-stacks are weakly dimerized. The phenomenology of $\alpha$- and $\theta$-phase materials is similar, and the two may be discussed together. The $\beta^\prime$-phase materials exhibit moderate to strong dimerization analogous to the $\kappa$-phase, but dimers stack in a uniform fashion rather than a herringbone pattern, as shown in Fig.~4(c,d). An example compound, discussed in section \ref{subsec_betaICl}, is $\beta^\prime$-(ET)$_2$ICl$_2$, which exhibits a strong dimerization ($|t_1/t^\prime| \sim 14$). Similar to the $\kappa$-phase materials, the primary mode of CO is intra-dimer charge disproportionation, which competes with dimerization (see section \ref{sec_phase_diagrams}).

\subsection{Symmetry considerations for ferroelectricity and multiferroicity}

\label{sec-magnetoelastic coupling}

Magnetic and polar charge orders can be discussed in terms of their corresponding order parameters, and the free energy governing their evolution. The form of this free energy is restricted by crystal symmetries.

The vast majority of organic CT salts are centrosymmetric in their para-electric states, including the above-mentioned $\kappa$- and $\theta$-phases. Magnetic order parameters $M$ (both antiferromagnetic and ferromagnetic) are odd with respect to time reversal, while ferroelectric polarization $P$ is even with respect to time reversal, and odd with respect to inversion. 
For this reason, the free energy is restricted to the form $\mathcal{F} = a P^2 + bM^2 + cP^2 M^2 + ...$ \cite{toledano1985theory,mostovoy2006ferroelectricity,eerenstein2006multiferroic} for temperature-dependent constants $a,b,c,..$. Since $M$ and $P$ represent symmetry-distinct order parameters, they are forbidden from coupling linearly, such that ordering of electric and magnetic dipoles typically occurs at distinct temperatures. This is true unless (i) the magnetoelectric coupling is weak, and the natural magnetic and charge-order transitions accidentally coincide, or (ii) the magnetoelectric coupling $c$ is sufficiently negative as to merge the two into a single first-order transition.

The character of the magnetoelectric coupling depends on the microscopic details, and the nature of the electronic and magnetic order parameters. For example, as detailed in section \ref{subsec-quasi-2D-kappa-Cl}, it has been suggested that ferroelectric ordering of the intra-dimer dipole moments couples ``attractively'' ($c<0$) to colinear two-sublattice antiferromagnetic order in $\kappa$-(ET)$_2$Cu[N(CN)$_2$]Cl \cite{Lunkenheimer2012,Lang2014}, due to relief of magnetic frustration with weak charge order. In this case, since the antiferromagnetic order preserves inversion and translation symmetries, a weak ferromagnetic moment appears as a consequence of Dzyaloshinskii-Moriya interactions. $\kappa$-(ET)$_2$Cu[N(CN)$_2$]Cl has thus been identified as a rare organic multiferroic, in which electronic and magnetic order parameters are mutually conducive. In contrast, $\kappa$-(ET)$_2$Hg(SCN)$_2$Cl exhibits an apparent re-entrant charge liquid state \cite{hassan2020melting} in which charge order initially sets in at $T_\mathrm{CO} \approx 30$ K, but subsequently melts below 10 K, recovering a paraelectric state. This has been discussed in terms of ``repulsion'' ($c > 0$) 
between charge order and three-sublattice 120$^\circ$ magnetic order, which was previously anticipated theoretically \cite{Naka2010}.  In this view, the appearance of strong magnetic correlations suppresses CO. As discussed in the next sections, electronic charge order tends to reduce dimensionality of the magnetic couplings, resulting in an incompatibility of magnetic order with CO. These findings highlight the subtleties of magnetoelectric coupling, which can give rise to a variety of scenarios in which the magnetic and electric properties are coupled.

\subsection{Theoretical phase diagrams} \label{sec_phase_diagrams}
\paragraph{$\kappa$-phase}

\begin{figure}
\centering
\includegraphics[width=0.9\linewidth]{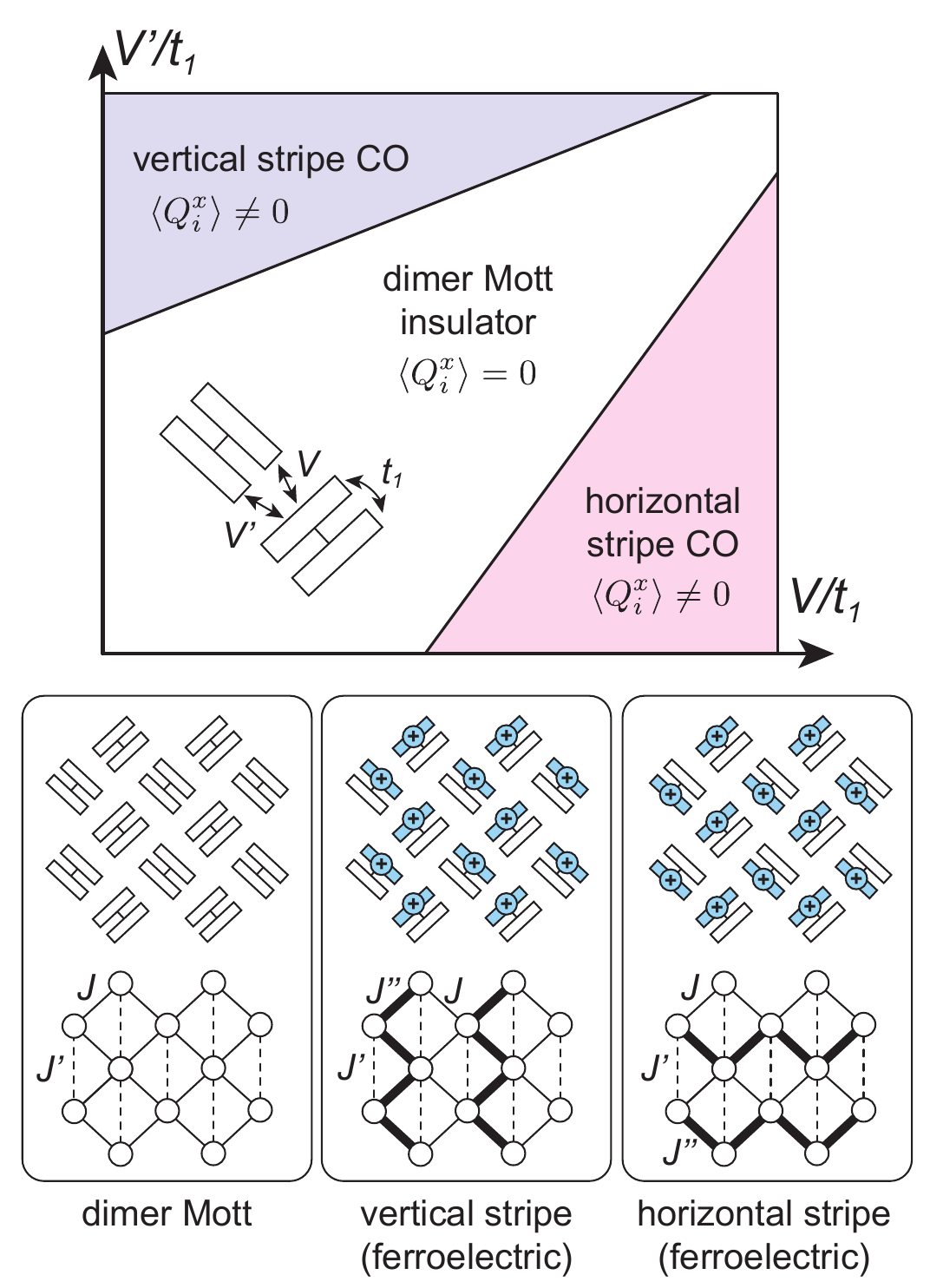}
\caption{Schematic phase diagram for $\kappa$-phase ET materials as a function of inter-dimer Coulomb repulsion $V$ and $V^\prime$. For detailed studies, see, e.g.~\cite{Naka2010,shinaoka2012mott,kaneko2017charge,watanabe2017phase}. Local spin moments are associated with the sites having excess holes, indicated in light blue. Charge order alters the effective magnetic couplings between dimers, breaking the symmetry between different $J$-bonds.}
\label{fig:kappa-phase}
\end{figure}

The phase diagrams of $\kappa$-phase materials in regards to charge order has been explored theoretically by various approaches including mean-field theory \cite{Naka2010} and variational Monte Carlo \cite{shinaoka2012mott,kaneko2017charge,watanabe2017phase}. As discussed in the previous sections, the most common mode of charge order in $\kappa$-phase materials is intra-dimer charge disproportionation, in which one hole occupies each dimer, but the charge is unequally distributed across the two ET molecules. Similar considerations apply to the $\beta^\prime$-phase, although the latter has not been studied in as much detail theoretically. Interdimer CO modes for $\kappa$-phase have been explored theoretically \cite{shinaoka2012mott,kaneko2017charge}, but do not appear in known materials, aside from catachol-based materials where proton dynamics strongly influence the CO patterns \cite{ueda2014hydrogen}. A range of theoretical studies \cite{Hotta2010,Naka2010,Hotta2012,naka2016quantum,jacko2020interplay} therefore adopt a pseudospin parameter $\mathbf{Q}_i$ to represent the charge degree of freedom of dimer $i$. $\langle Q_i^z\rangle = +1/2$ implies an equal charge distribution, in which the hole occupies the antibonding combination of ET molecular orbitals.  $\langle Q_i^x\rangle = \pm 1/2$ implies complete localization of the hole to one molecule or the other (see \cite{Naka2010}). The most important terms in the coupled magnetoelectric Hamiltonian can then be written as \cite{Naka2010,Ishihara2014,naka2016quantum,jacko2020interplay}:
\begin{align}
    \mathcal{H} = & \ \sum_{\langle ij\rangle} J_{ij}\mathbf{S}_i \cdot \mathbf{S}_j - \Gamma \sum_{i} Q_i^z  + \sum_{\langle ij\rangle} W_{ij}Q_i^x Q_j^x
    \nonumber \\ & \ \hspace{15mm}
    + \sum_{\langle ij\rangle} K_{ij}(Q_i^x - Q_j^x)\mathbf{S}_i \cdot \mathbf{S}_j
\end{align}
where $J_{ij}$ is the bare antiferromagnetic coupling between spins in adjacent dimers. $\Gamma$ is proportional to the intra-dimer hopping (denoted by $t_1$ in Fig.~\ref{fig:kappa-phase}), and represents the energetic preference for the hole to delocalize across both molecules. $W_{ij}$ describes the interaction of dipoles on different dimers, mostly due to Coulomb repulsion between electrons occupying molecules in adjacent dimers (denoted by $V, V^\prime$ in Fig.~\ref{fig:kappa-phase}).  Finally, $K_{ij}$ describes the modulation of the intersite magnetic couplings as a function of charge distribution. The sign of the coupling is such that the magnetic interaction between two holes is increased with increasing localization of those holes to near sides of their respective dimers. An extensive discussion of magnetoelectric couplings for $\kappa$-(ET)$_2X$ materials and their $\textit{ab initio}$ estimation can be found in Ref.\,\cite{jacko2020interplay}.

In the context of CO patterns, the fate of the electric dipole moments is largely controlled by the transverse field Ising model defined by $\Gamma$ and $W_{ij}$. The schematic phase diagram, derived from various studies, is depicted in Fig.~\ref{fig:kappa-phase}. This model has three phases. The first is a dimer-Mott phase with no charge disproportionation ($\langle Q_i^x\rangle = 0$), which is favored at large $\Gamma \approx t_1$. The second and third phases are charge-ordered ferroelectric states ($\langle Q_i^x\rangle \neq 0$), favored at large $W_{ij} \sim V$  \cite{Hotta2010,Hotta2012}. The preferred orientation of the net dipole moment and pattern of charge order (vertical vs. horizontal stripes) is determined by the specific details of inter-dimer Coulomb repulsion. The quantum critical point separating the CO and dimer Mott phases is expected to be associated with enhanced relaxor-type dielectric susceptibility \cite{Hotta2023}, discussed in more detail in section \ref{subsubsecDielFerroel}.
The CO breaks the symmetry of the magnetic interactions, weakening the $J^\prime$ coupling and one of the $J$ couplings, while enhancing the second $J$ coupling. As a consequence, the magnetic interactions become increasingly one-dimensional with increasing CO.

For weak CO, the enhancement of magnetic couplings along one axis of the crystal is compatible with the collinear N\'{e}el order observed in $\kappa$-(ET)$_2$Cu[N(CN)$_2$]Cl. For stronger CO, the increasingly one-dimensional magnetic interactions may instead suppress magnetic order. This has been considered as a scenario for $\kappa$-(ET)$_2$Hg(SCN)$_2X$ ($X$ = Cl, Br)  \cite{Gati2018b}, where no sign of magnetic order is detected, yet the majority of spins are strongly antiferromagnetically coupled below $T_{\rm CO}$. In these materials, defects in the charge-ordering pattern (domain walls) likely host quasi-free orphan spins, giving rise to a weak inhomogeneous paramagnetic response below $T_\mathrm{CO}$, which is observed in NMR \cite{Le2020,pustogow2020impurity,Urai2022}, ESR\cite{Hemmida2018} and magnetic susceptibility/torque \cite{yamashita2021ferromagnetism,drichko2022charge}. These effects may be enhanced by intrinsic glassiness of the charge order \cite{Deglint2022}.

\paragraph{$\theta$-phase}

For the undimerized $\theta$-phase materials, the interplay between spin and charge degrees of freedom is comparatively less explored. The phase diagram as a function of inter-molecular hopping and Coulomb repulsion has been addressed via a range of theoretical methods, including mean-field theory \cite{Seo2000,Kaneko2006}, slave-boson approaches \cite{mckenzie2001charge}, random phase approximation \cite{kuroki2006origin}, variational Monte Carlo \cite{watanabe2006novel}, exact diagonalization \cite{merino2005quantum}, and density matrix renormalization group \cite{nishimoto2008coexistence}. The $\alpha$-phase materials (such as $\alpha$-(ET)$_2$I$_3$) can be viewed as slightly modulated versions of $\theta$-phase materials, with analogous modes of CO.

\begin{figure}
\centering
\includegraphics[width=0.8\linewidth]{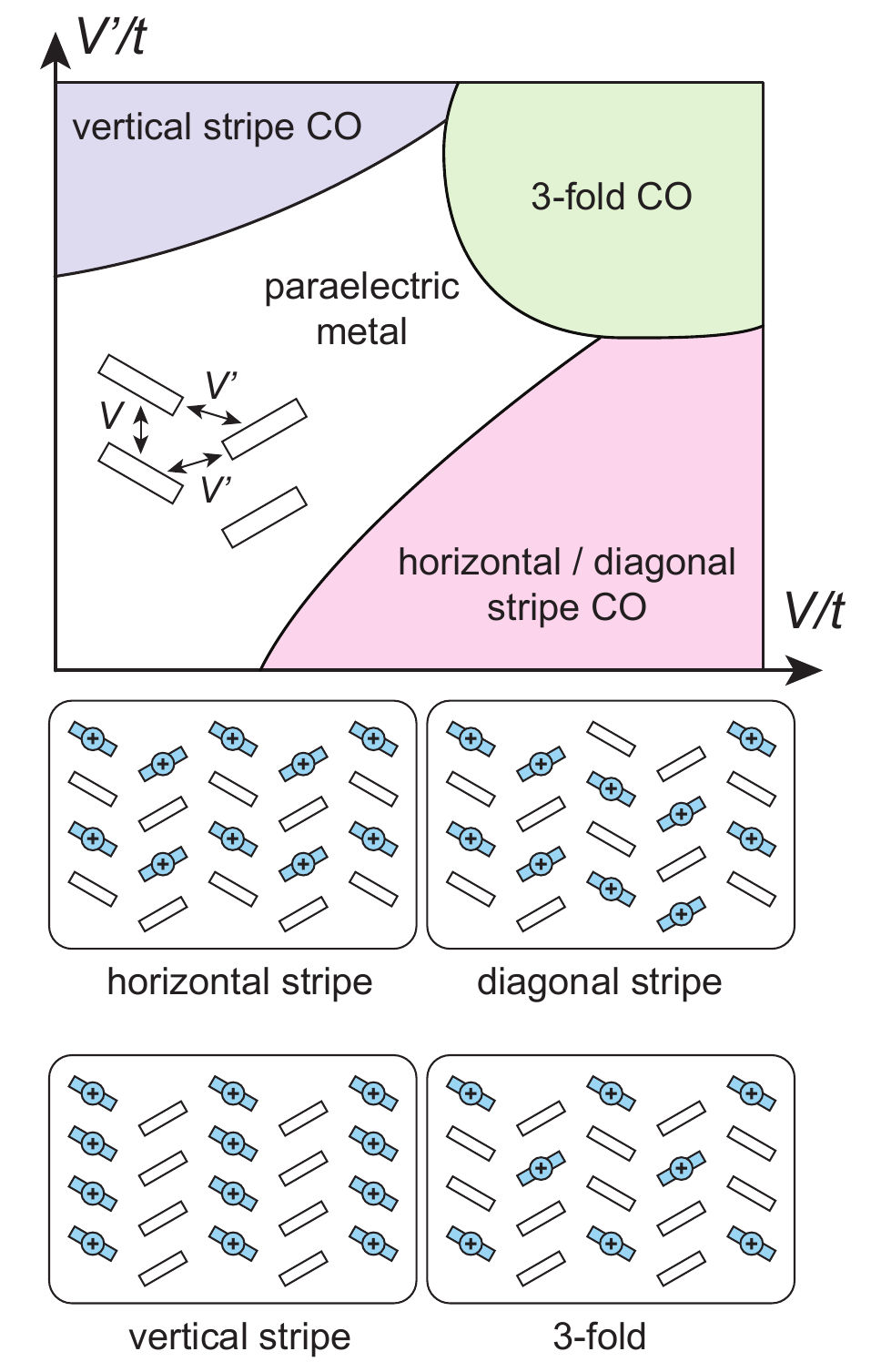}
\caption{Schematic phase diagram for $\theta$-phase ET materials as a function of nearest neighbor Coulomb repulsion $V$ (within a $\pi$-stack), and $V^\prime$ (between $\pi$-stacks). For detailed studies, see, e.g.~\cite{Seo2000,Kaneko2006,watanabe2006novel,nishimoto2008coexistence}. Local spin moments are associated with the sites having excess holes, indicated in light blue.}
\label{fig:theta-phase}
\end{figure}

With the inclusion of short-ranged (nearest neighbor) repulsion, the phase diagram features three prominent CO patterns, with (i) vertical charge stripes, (ii) diagonal/horizontal charge stripes, and (iii) a 3-fold pattern in which excess holes are localized to 1/3 of the sites in a triangular lattice pattern (depicted schematically in Fig.~\ref{fig:theta-phase}). The latter region hosts a ``pinball liquid'' state, which remains metallic in conjunction with CO \cite{hotta2006strong,hotta2006phase,seo2006charge} due to itineracy of the electrons associated with the electron-rich sites. The relative stabilities of the various phases can also be influenced by the details of electron-lattice coupling \cite{udagawa2007charge,tanaka2007effects,miyashita2007charge}, as the susceptibility of the crystal to local structural distortions coinciding with CO can play a role.  The phase diagram becomes somewhat more complicated with consideration of longer-range Coulomb repulsion, which can stabilize CO patterns with longer period \cite{Naka2014,yoshimi2020}, composed of mixed domains of diagonal, horizontal, and vertical stripes. This complex energy landscape is ultimately conducive to glassy charge dynamics \cite{yoshimi2012coulomb,Mahmoudian2015,Sasaki2017,Sato2017}, which provides theoretical intuition for the properties of $\theta$-(ET)$_2$MM$^\prime$(SCN)$_4$ compounds, reviewed in section \ref{subsec-quasi-2D-theta}.

Regarding the magnetic properties of $\theta$-phase compounds: in the striped CO phases, mean-field studies find various magnetic orders coexisting with CO \cite{Seo2000,Kaneko2006}. However, similar to the $\kappa$-phase materials, strong CO tends to result in magnetic connectivities that are low-dimensional: either 1\,D, in the case of perfect vertical or diagonal/horizontal stripes, or 0\,D when complicated (glassy) domain structures leads to many breaks in the spin chains. As a consequence, magnetic order tends to be suppressed by CO. For $\theta$-(ET)$_2$RbZn(SCN)$_4$ ($\theta$-RbZn) and $\theta$-(ET)$_2$CsZn(SCN)$_4$ ($\theta$-CsZn),  magnetic susceptibility, ESR and NMR have been employed to probe the magnetic response \cite{Mori1998,mori2000discontinuous,nakamura2000possible,Miyagawa2000,chiba2008charge}. For $T<T_\mathrm{CO}$, $\theta$-RbZn forms a robust horizontal/diagonal CO pattern, in which the spins at the electron-poor sites form 1D chains. The susceptibility is well modelled by an antiferromagnetic Heisenberg chain \cite{Mori1998} down to $T\sim 20$ K, at which point a spin gap opens, and the compound forms a singlet ground state. The precise nature of this singlet state is not completely settled, but it is thought to correspond to a spin-Peierls transition (with further distortion of the lattice). For $\theta$-CsZn, the glassy charge state at low temperatures gives rise to an enhanced Curie tail in the magnetic suceptibility, and strongly inhomogeneous NMR relaxation. These effects can be understood analogously with the orphan-spin response in $\kappa$-(ET)$_2$Hg(SCN)$_2X$; due to the effective coupling of spins and charges, the homogeneity of the magnetic response is a direct probe of the homogeneity of the charge order.

\section{Experimental techniques}\label{sec-Exp-Techn}

In this chapter we briefly introduce some experimental techniques used to characterize the properties of the various charge-transfer salts. Preference is given to the methods applied by the authors’ own investigations including, in particular, dielectric spectroscopy, fluctuation (noise) spectroscopy and polarization measurements. 

\subsection{Dielectric spectroscopy}\label{subsecDielSpec}

\subsubsection{Quantities and experimental methods}\label{subsubsecQuantities}

Dielectric spectroscopy measures the response of a material to ac electric fields by recording the real and imaginary part of its dielectric permittivity, $\varepsilon^{*} = \varepsilon' - i\varepsilon''$ (the minus sign ensures that $\varepsilon''$ is positive), for varying frequencies of the applied field. The measurements are usually done for different temperatures, but also other external parameters can be varied, e.g., pressure or magnetic field. Dielectric spectroscopy simultaneously also provides information on the complex conductivity, $\sigma^{*} = \sigma' + i\sigma''$, which is related to the permittivity via $\sigma^{*} = i\varepsilon^{*}\varepsilon_0\omega$ (where $\varepsilon_0$ is the permittivity of free space and $\omega$ the circular frequency), resulting in $\sigma' = \varepsilon''\varepsilon_0\omega$ and $\sigma'' = \varepsilon'\varepsilon_0\omega$. In general, ferroelectric transitions lead to an anomaly in the temperature dependence of the dielectric constant, $\varepsilon'$($T$). Depending on the type of the ferroelectric state of a material, characteristic differences in $\varepsilon'$($T$), but also in the frequency dependence of this quantity are found. The dielectric loss, $\varepsilon''$($T,\nu$), mainly provides information on relaxational processes arising from dipolar dynamics, a characteristic property of some classes of ferroelectric materials as will be explained below. The real part of the conductivity $\sigma'$ in principle provides the same information as $\varepsilon''$. However, it is better suited to analyze charge-transport contributions in the material as found in many organic charge-transfer salts. For example, plain dc conductivity leads to a frequency-independent contribution in $\sigma'$($\nu$) and hopping conductivity to a $\nu^{s}$ power law with $s <$ 1 \cite{Emmert2011,Jonscher1983,Lunkenheimer2015a,Lunkenheimer2010}. 

In standard dielectric-spectroscopy experiments, usually covering frequencies in the Hz - MHz range, commercially available autobalance bridges or frequency-response analyzers are employed, the latter providing superior experimental resolution, especially at very low values of the dielectric loss or conductivity \cite{Kremer2003}. They require the application of two metallic contacts to the samples, from which two or four wires (i.e., two attached to each contact) establish the electrical connection to the device. Using four wires (sometimes termed pseudo-fourpoint contact geometry) reduces parasitical cable contributions to the measured quantities. For dielectric spectroscopy at higher frequencies, from MHz up to several tens of GHz, various experimental methods exist \cite{Boehmer1989,Klein1993,Schneider2001,Kremer2003,Scheffler2005,Scheffler2012}. Among them, coaxial reflectometry is most commonly applied: There, an electromagnetic wave travelling along a coaxial line is reflected by the sample at the end of the line. The latter is mounted in a sample holder ensuring the bridging of the inner and outer conductors by the sample material \cite{Boehmer1989,Kremer2003}. For this method, two metallic contacts have to be applied to the sample, too. A network or impedance analyzer at the other end of the line either measures the complex reflection coefficient, which depends on the sample impedance (the complex resistance), or uses an $I$-$V$ method that more directly determines the impedance via the voltage-current ratio. Via textbook formulae, from the impedance the complex capacitance and the admittance (the complex conductance) can be calculated. Using the sample dimensions (length and cross-section area), from these quantities then the permittivity and conductivity can be derived \cite{Kremer2003,Jonscher1983}. To enable a correction for parasitic contributions arising from the coaxial line and the sample holder, several calibrations of the setup with different known impedances are mandatory \cite{Boehmer1989}. In one of these calibration steps, the sample should be replaced by a short with the same shape. This helps to correct for the inductance of the sample holder which can hamper the measurement results, especially at the highest frequencies.

At frequencies beyond about 50\,GHz, usually methods analogous to optical experiments are used, i.e., the electromagnetic wave is not restricted to a waveguide (e.g., a coaxial line) but propagates through free space. Then the reflection of the wave from and/or its transmission through the sample is measured, from which the dielectric properties can be deduced. In the present work, such experiments are not treated.

\subsubsection{Contact preparation and electrode effects}\label{subsubsecContacts}

The metallic contacts at the samples, which are required for dielectric measurements from the lowest frequencies up to several tens of GHz, can be produced by sputtering, evaporation, or by applying conductive suspensions as graphite, silver, or gold paints and pastes. When suspensions are used, one has to assure that their solvent does not affect the sample material. This is especially relevant for the organic systems treated in the present work. In experiments performed in the Augsburg laboratory on various organic charge-transfer salts \cite{Lunkenheimer2012,Lunkenheimer2015a,Canossa2021,Fischer2021,Avino2017,Fischer2018}, graphite paint with diethyl succinate solvent or gold paint with propylene glycol methyl ether was found to be suitable for contact preparation, avoiding any sample deterioration. Quasi-2D materials, as predominantly treated in the present work, often form single crystals of platelike shape. Covering the opposite sides of such crystals with the contact material leads to a parallel-plate capacitor with well-defined geometry and field distribution. The electric field then is oriented perpendicular to the lattice planes. To achieve a field direction parallel to the planes, coplanar contacts can be applied, however, leading to ill-defined field distribution, which partly is also oriented perpendicular to the planes, resulting in some uncertainty concerning the absolute values of the permittivity and conductivity. As an alternative, caplike contacts can be applied, "wrapping" opposite ends of the sample, thereby ensuring that the electrical current is flowing parallel to the planes, across the whole cross section of the sample. 

It should be noted that ferroelectrics either are highly insulating or at least have rather low conductivity (otherwise, the polarization would be shielded by the conduction electrons). Thus, the conductivity of the employed metallic contact material can be neglected in light of the much higher sample resistance, and it is irrelevant whether graphite, silver, gold, or any other metal is used. However, especially for semiconducting samples, including various organic charge-transfer salts, at the contact-sample interface insulating depletion zones can develop, due to the formation of Schottky diodes \cite{Lunkenheimer2002,Lunkenheimer2010}. This can considerably affect the measured dielectric sample response, leading to non-intrinsic frequency dependencies, and care has to be taken to avoid any misinterpretation of the detected effects. Measurements of samples of different thickness and/or using different contact materials can help to resolve such problems \cite{Lunkenheimer2010}. Moreover, an analysis of the experimental data in terms of equivalent circuits has proven very useful for the deconvolution of the intrinsic and non-intrinsic contributions to the dielectric spectra \cite{Lunkenheimer2002,Emmert2011,Lunkenheimer2010}. An elaborate example of such an analysis, performed for an organic charge-transfer salt revealing relaxor ferroelectricity, can be found in the Supplementary Information of Ref. \cite{Thomas2024}.

\subsubsection{Dielectric response of ferroelectrics}\label{subsubsecDielFerroel}

\begin{figure*}[t]
\centering
\includegraphics[width=0.85\textwidth]{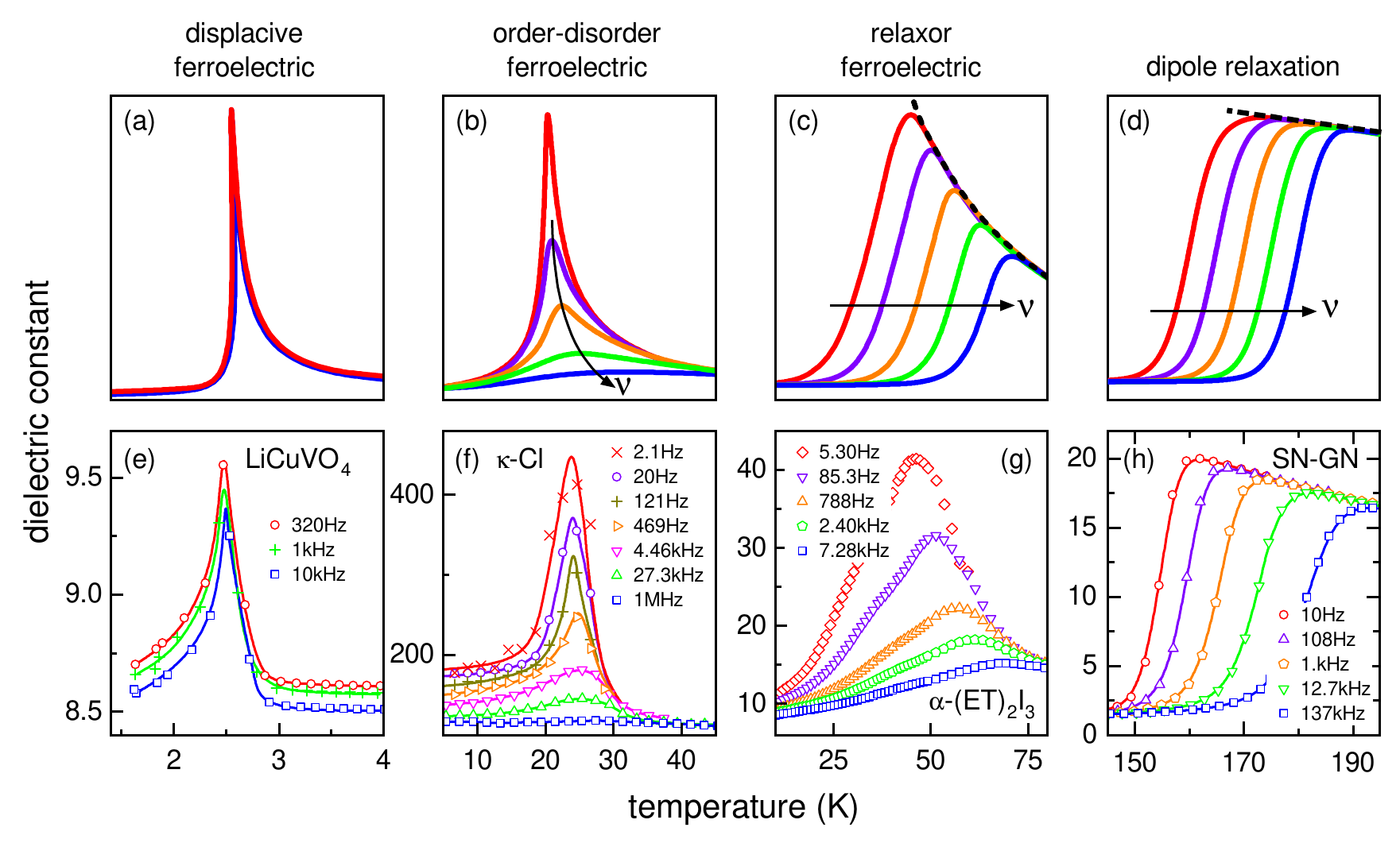}
\caption{(a) - (c) Schematic temperature dependence of the dielectric constant $\varepsilon'$ for the three most common classes of ferroelectrics as measured at different frequencies \cite{Krohns2019}. Frame (d) schematically shows $\varepsilon'$($T$) for a non-ferroelectric material with internal dipolar reorientational degrees of freedom, leading to a relaxation process. The main difference to $\varepsilon'$($T$) of a relaxor ferroelectric is revealed in the static dielectric constant $\varepsilon_\mathrm{s}$($T$), indicated by the dashed lines in (c) and (d). Frames (e) - (h) show  $\varepsilon'$($T,\nu$) of four materials representing experimental examples for the schematic plots in (a) - (d): (e) the inorganic multiferroic LiCuVO$_4$ \cite{Schrettle2008}, (f) the quasi-2D organic multiferroic $\kappa$-(ET)$_2$Cu[N(CN)$_2$]Cl \cite{Lunkenheimer2012}, (g) the quasi-2D organic relaxor-ferroelectric $\alpha$-(ET)$_2$I$_3$ \cite{Lunkenheimer2015b}, and (h) the plastic crystal SN-GN (the corresponding $\varepsilon'(\nu)$ spectra of this material were published in \cite{Bauer2010}). The lines in (e), (f), and (h) are guides for the eye.}\label{paper-ferroele-eps}
\end{figure*}

There are three main classes of ferroelectric order: displacive, order-disorder, and relaxor ferroelectricity. In the following, we give a brief overview of their main properties, by concentrating on their dielectric response. For more details, we refer the reader to Refs.\,\cite{Lunkenheimer2015a,Cross1987,Samara2003,Blinc1974,Lines2001,Waser2005}. In displacive systems, above their ferroelectric phase transition no permanent dipole moments exist. Below the transition, the materials cross over into a structure with reduced symmetry, where (in the most common cases) local shifts of ions to an off-symmetry position give rise to permanent dipoles, revealing long-range polar order. Displacive ferroelectrics are expected to exhibit a well-defined peak in the temperature dependence of the dielectric constant at their phase transition temperature $T_\mathrm{FE}$. At the right flank of this peak, their $\varepsilon'$($T$) should follow a Curie-Weiss law, $\varepsilon' \propto 1/(T-T_\mathrm{CW})$. Often the Curie-Weiss temperature $T_\mathrm{CW}$ is of similar size as $T_\mathrm{FE}$. In the typical Hz - MHz frequency range of standard dielectric experiments, the dielectric properties of displacive ferroelectrics usually show no or only weak frequency dependence. This $\varepsilon'$($T,\nu$) behavior is schematically indicated in Fig.\,\ref{paper-ferroele-eps}(a) \cite{Krohns2019}. One should be aware that the exact peak shape depends on the character of the transition, i.e., whether it is of first order or second order, which implies considerable variations; the curve shown applies to a second-order transition. It should be noted that displacive ferroelectricity was only rarely reported for organic charge-transfer salts \cite{Wiscons2018}. An experimental example of an inorganic displacive ferroelectric is provided in Fig.\,\ref{paper-ferroele-eps}(e). It shows $\varepsilon'$($T$) of LiCuVO$_4$, a multiferroic whose polar order is driven by magnetic ordering, leading to ion displacement due to the inverse Dzyaloshinskii-Moriya interaction \cite{Schrettle2008}. This is an improper ferroelectric where the order parameter driving the phase transition is not the polarization. As revealed in Fig.\,\ref{paper-ferroele-eps}(e), $\varepsilon'$ of this material exhibits a sharp peak at $T_\mathrm{FE} \approx$ 2.5\,K and only a weak frequency dependence as expected for displacive ferroelectrics. 

While in displacive ferroelectrics the dipoles involved in the polar order are only formed below $T_\mathrm{FE}$, in order-disorder ferroelectrics they already exist above the transition, however, with dynamically-disordered orientations \cite{Blinc1974,Lines2001}. Their alignment below $T_\mathrm{FE}$ then gives rise to ferroelectricity. Depending on the material, the dipoles can be due to charged particles (ions, electrons, holes), oscillating between two (or more) polar sites, or can be caused by dipolar molecules that rotate at $T > T_\mathrm{FE}$ but order in a parallel fashion at $T < T_\mathrm{FE}$. Fig.\,\ref{paper-ferroele-eps}(b) schematically depicts the temperature dependence of the dielectric constant found in this class of ferroelectrics for different frequencies \cite{Krohns2019}. As in displacive ferroelectrics, here a peak at the transition shows up, too, but it exhibits significant frequency dependence, leading to a gradual suppression of the peak with increasing frequency. The peak temperature usually weakly shifts to higher temperatures with increasing frequency. An analysis of the frequency dependence of $\varepsilon'$ and $\varepsilon''$ indicates that the observed frequency dispersion is due to reorientational fluctuations of the dipoles within double- or multi-well potentials, leading to a so-called dielectric relaxation process \cite{Kremer2003b}. From the dielectric spectra the relaxation time $\tau$ can be deduced, characterizing the dynamics of the dipoles. In order-disorder ferroelectrics, its temperature dependence $\tau$($T$) often follows a critical slowing down upon approaching the ferroelectric transition from high or low temperatures, instead of simple thermally-activated Arrhenius behavior, expected for dipoles not involved in polar ordering. Further analysis of the spectra reveals a nearly monodispersive, Debye-type nature of this dynamics, i.e., there is only a narrow distribution of relaxation times in these systems \cite{Ediger2000}. This is expected for well-ordered crystalline materials, in contrast to dipolar systems undergoing glassy freezing, where these distributions can be very broad \cite{Ediger2000,Kremer2003b,Lunkenheimer2018}. An experimental example of $\varepsilon'$($T,\nu$) for order-disorder ferroelectricity in an organic charge-transfer salt is given in Fig.\,\ref{paper-ferroele-eps}(f) \cite{Lunkenheimer2012}. In $\kappa$-(ET)$_2$Cu[N(CN)$_2$]Cl (abbreviated $\kappa$-Cl in the figure), holes fluctuating within molecular dimers were proposed to give rise to dipolar moments, which order below the transition. A detailed discussion of these data will be provided in section \ref{subsec-quasi-2D-kappa-Cl}.

Finally, Fig.\,\ref{paper-ferroele-eps}(c) schematically shows the $\varepsilon'$($T$) behavior at different frequencies as typically measured for relaxor ferroelectrics \cite{Cross1987,Krohns2019,Samara2003}. The peculiar dielectric behavior of this class of ferroelectrics is usually assumed to arise from the glasslike freezing of clusterlike short-range ferroelectric order \cite{Cross1987,Samara2003}. Again, a peak in $\varepsilon'(T)$ is observed, but compared to the other classes of ferroelectrics, it is smeared-out and the peak temperature is strongly frequency dependent. The behavior in Fig.\,\ref{paper-ferroele-eps}(c) in some respects resembles that of materials with dipolar reorientations that do not undergo dipole order upon cooling but instead exhibit glassy freezing, finally resulting in a low-temperature state with static orientational disorder. The $\varepsilon'$($T$) of such materials is schematically shown in Fig.\,\ref{paper-ferroele-eps}(d). It is revealed by many dipolar liquids undergoing a glass transition into an amorphous state \cite{Lunkenheimer2018} but also by certain crystalline systems, e.g., by plastic crystals, which are formed by dipolar molecules that can rotate even in the crystalline state \cite{Lunkenheimer1996b,Brand2002}. An example of such a system is provided in Fig.\,\ref{paper-ferroele-eps}(h) showing $\varepsilon'$($T,\nu$) of a mixture of 60\% succinonitrile and 40\% glutaronitrile (SN-GN) in its plastic-crystalline phase \cite{Bauer2010}. The main difference of relaxor ferroelectrics compared to such systems is the much stronger temperature dependence of their static dielectric constant $\varepsilon_s$ as indicated by the dashed lines in Fig.\,\ref{paper-ferroele-eps}(c) and (d). For dipolar systems without order, a Curie-law ($\varepsilon_s \propto 1/T$) is expected \cite{Boettcher1974}. In relaxor ferroelectrics, this temperature dependence often (but not always) deviates from the typical Curie-Weiss behavior detected in canonical ferroelectrics. The strong increase of $\varepsilon_s$($T$) with decreasing temperature in relaxor ferroelectrics indicates increasing interdipolar correlations that lead to local polar order.

Analyzing the dielectric spectra of relaxor ferroelectrics often reveals deviations from monodispersive Debye behavior \cite{Viehland1991}. This points to disorder, leading to a distribution of relaxation times \cite{Ediger2000}. Indeed, relaxor ferroelectricity is often found in systems with substitutional disorder, a prominent example being the perovskite PbMg$_{1/3}$Nb$_{2/3}$O$_3$ \cite{Smolenskii1961}. However, there are also nominally well-ordered materials for which relaxor ferroelectricity was reported, including several organic charge-transfer salts \cite{Lunkenheimer2015b,Iguchi2013,Abdel-Jawad2010,Canossa2021,Fischer2021,Matsui2003}. The temperature dependence of the relaxation time $\tau$($T$) of relaxor ferroelectrics can be usually described \cite{Viehland1990,Levstik1998,Viehland1991,Glazounov1998} by the empirical Vogel-Fulcher-Tammann (VFT) law \cite{Vogel1921,Fulcher1925,Tammann1926}:

\begin{equation}
\tau = \tau_0 \cdot \exp{\left( \frac{B}{T-T_\mathrm{VF}} \right)}. \label{Vogel-Fulcher}
\end{equation}

\noindent Here, $\tau_0$ is a preexponential factor, $B$ is an empirical constant, and $T_\mathrm{VF}$ is the Vogel-Fulcher temperature where $\tau$ diverges. Instead of deriving the relaxation times from the spectra, for relaxor ferroelectrics, $\tau$($T$) is often approximated by evaluating the peak temperatures $T_\mathrm{p}$ in the $\varepsilon'$($T$) plots measured at different frequencies [Fig.\,\ref{paper-ferroele-eps}(c)]. Then $T$ in Eq.\,\eqref{Vogel-Fulcher} is $T_\mathrm{p}$ and $\tau$ is calculated from the measurement frequency $\nu$ of the evaluated $\varepsilon'$($T$) curve via $\tau$ = 1/(2$\pi\nu$). Notably, Eq.\,\eqref{Vogel-Fulcher} is commonly applied to characterize the molecular dynamics in glass-forming liquids \cite{Lunkenheimer2000,Ediger1996,Lunkenheimer2018}, indicating that glassy freezing in some respects must play a role in relaxor ferroelectrics. Indeed, the latter are usually assumed to reveal the glasslike freezing-in of short-range, nano-scale ferroelectric order. As an experimental example of a relaxor ferroelectric, Fig.\,\ref{paper-ferroele-eps}(g) shows $\varepsilon'$($T,\nu$) of the organic charge-transfer salt $\alpha$-(ET)$_2$I$_3$ which reveals the typical signature of relaxor ferroelectricity \cite{Lunkenheimer2015b}. These data will be discussed in detail in section \ref{subsec-quasi-2D-alpha}.

\subsubsection{Polarization measurements}\label{subsecPolarization}
\begin{figure*}[t]
\centering
\includegraphics[width=0.8\textwidth]{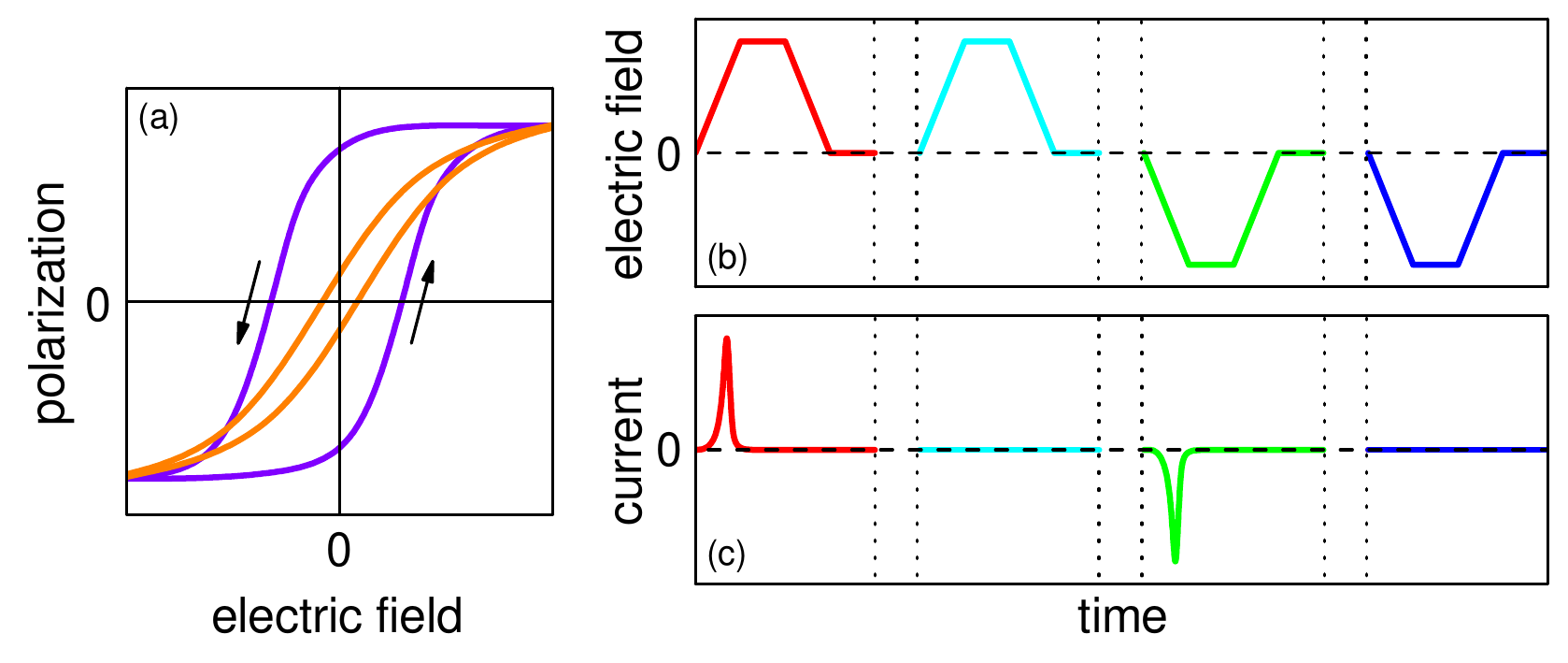}
\caption{(a) Examples of a wide and slim $P$($E$) hysteresis loop found in different ferroelectrics. Relaxor ferroelectrics exhibit slim loops. (b) Schematic plot of the time-dependent electric field as applied in typical PUND measurements. (c) The current response to the field shown in (b) as expected for a ferroelectric \cite{Krohns2019}.}\label{pap-PE_PUND}
\end{figure*}
A ferroelectric phase transition leads to the development of spontaneous electrical polarization $P$ at $T < T_\mathrm{FE}$. This can be investigated by measurements of $P$ as a function of the electric field $E$, by so-called positive-up-negative-down (PUND) measurements, and by pyrocurrent measurements. Strictly speaking, the definition of ferroelectricity not only involves the occurrence of spontaneous polarization below $T_\mathrm{FE}$ (signifying a pyroelectric material) but in addition also requires that the polarization can be reversed by an external electric field \cite{Blinc1974}. This can be best checked by $P$($E$) or PUND measurements. Nowadays, various devices for such measurements are commercially available. In the Augsburg group, an aixACCT TF2000 ferroelectric analyzer is used, partly in combination with a Trek 609C-6 amplifier. It allows for measurements of the time-dependent polarization in response to applied ac voltages up to 1.1\,kV, with frequencies up to 1\,kHz. As the applied voltages in such experiments are high, some care should be taken to avoid electrical breakthrough to metallic parts of the cryostat and sample holder and the experiments should be performed under low pressure or in vacuum to prevent electric gas discharge.

For ferroelectrics, $P$($E$) measurements reveal typical hysteresis curves as schematically shown in Fig.\,\ref{pap-PE_PUND}(a). Basically, they reflect the shift of polar domain walls and the observed saturation at high fields indicates the approach of a mono-domain state. In principle, $P$($E$) measurements work best at temperatures somewhat below $T_\mathrm{FE}$. At lower temperatures, the domains in ferroelectrics can be large and the fields needed to induce domain-wall motions can be too high to avoid electrical breakdown \cite{Blinc1974,Lines2001,Scott2000}. However, many organic charge-transfer salts are semiconducting and field-induced currents due to their enhanced dc conductivity at elevated temperatures can render the detection of hysteresis loops due to polarization-switching impossible. Therefore, in such systems the optimum temperature for such measurements has to be identified (usually by trial and error), but in some cases $P$($E$) measurements are simply not feasible. One should also be aware that non-intrinsic effects arising, e.g., from the above-mentioned depletion zones of Schottky diodes may lead to the erroneous detection of $P$($E$) hysteresis loops. This can be avoided by performing additional experimental checks \cite{Loidl2008}. Depending on the material, the width and saturation value of the loops can vary considerably. Relaxor ferroelectrics usually exhibit very slim hysteresis loops. Their small remanent polarization [detected at $E$ = 0; cf. orange curve in  Fig.\,\ref{pap-PE_PUND}(a)], implies that most polar nanodomains again become randomly oriented after switching off the field \cite{Samara2003}. 

For PUND measurements, a succession of time-dependent electric field pulses as schematically indicated in Fig.\,\ref{pap-PE_PUND}(b) is applied to the sample. Normally, such measurements can be done using the same state-of-the-art devices as for the $P$($E$) loops. If the field at the increasing flank of the first pulse becomes sufficiently high, polarization switching sets in and the concomitant reorientation of the electrical dipoles, involving the motion of charges, leads to a current-pulse which is detected by the device  [Fig.\,\ref{pap-PE_PUND}(c)]. If the time between the first and second pulse with same field direction is sufficiently short to avoid a decay of the induced polarization, the current response to the second pulse should be featureless because all dipoles are already oriented. The third pulse with inverted field switches the polarization to the opposite direction, again leading to a current peak, while the fourth pulse does not trigger any current response. PUND measurements are an excellent way to distinguish between intrinsic and non-intrinsic polarization effects as the latter should lead to identical current response for the first and second pulses as well as for the third and fourth ones \cite{Scott2000}.

Pyrocurrent measurements also make use of the motion of charges associated with the reorientation of dipoles. In the most common form of these experiments, the sample is first brought close to a monodomain polar state by cooling it to a temperature below $T_\mathrm{FE}$ while applying a high electrical dc field. Afterwards, the sample is heated again through the transition without field. This leads to a pyrocurrent around $T_\mathrm{FE}$ when the dipoles reorient into disordered positions while crossing the transition. The polarization change at $T_\mathrm{FE}$ can be calculated by integrating the time-dependent current response. For the application of this method, the sample should be highly insulating. Otherwise, currents from ohmic charge transport superimpose the very small pyrocurrent.

\subsection{Fluctuation (noise) spectroscopy}
\label{subsec-fluc-spec}

As will be demonstrated in this review, conductance/resistance fluctuation (noise) spectroscopy often proves highly insightful as a complementary technique to study charge and/or polarization dynamics of dielectric/ferroelectric compounds at temperatures where the samples are too conductive in order to perform dielectric spectroscopy. For example, at elevated temperatures where the conductivity of otherwise semiconducting/insulating materials may still be relatively high, or in systems that undergo a metal-insulator transition upon cooling which is then accompanied by some type of dipolar order, it is often practical to perform conductance/resistance noise spectroscopy often providing complementary information due to the same or related relaxation processes picked up by dielectric spectroscopy. Naturally, it is desirable to have an overlap of the temperature regimes where both methods can be performed. A concrete example is the identification of structural, glass-like ordering of the ethylene endgroup degrees of freedom in various organic CT salts by noise spectroscopy \cite{JMueller2015,Thomas2022} at temperatures $T \sim 100$\,K where the samples are too conductive to investigate the relaxation by dielectric spectroscopy.

The noise power spectral density (PSD) of the resistance ($R$) fluctuations (the discussion is fully analogue for fluctuations of the conductance $G$) $\delta R(t) = R(t) - \langle R(t) \rangle$, where $\langle R(t) \rangle$ is the time-averaged mean value that may be considered equal to zero, is defined as the time-averaged squared modulus of the Fourier transformed fluctuating signal:
\begin{equation}
S_R(f) = 2 \lim\limits_{T \rightarrow \infty} \frac{1}{T} \left| \int\limits_{-T/2}^{T/2}\delta R(t) e^{-{\rm i}2\pi f t}{\rm d}t \right|^2,
\end{equation}
where the variance of the signal $\langle \delta R(t)^2 \rangle$ is normalized by $\langle \delta R(t)^2 \rangle = \int_0^\infty S_R(f){\rm d}f$. The quantity $S_R$ describes how much power is contributed by different parts of the frequency spectrum or simply answers the question 'How strongly does a certain frequency contribute to the system's noise?' \cite{JMueller2011,JMueller2018}.

\begin{figure}
\centering
\includegraphics[width=0.475\textwidth]{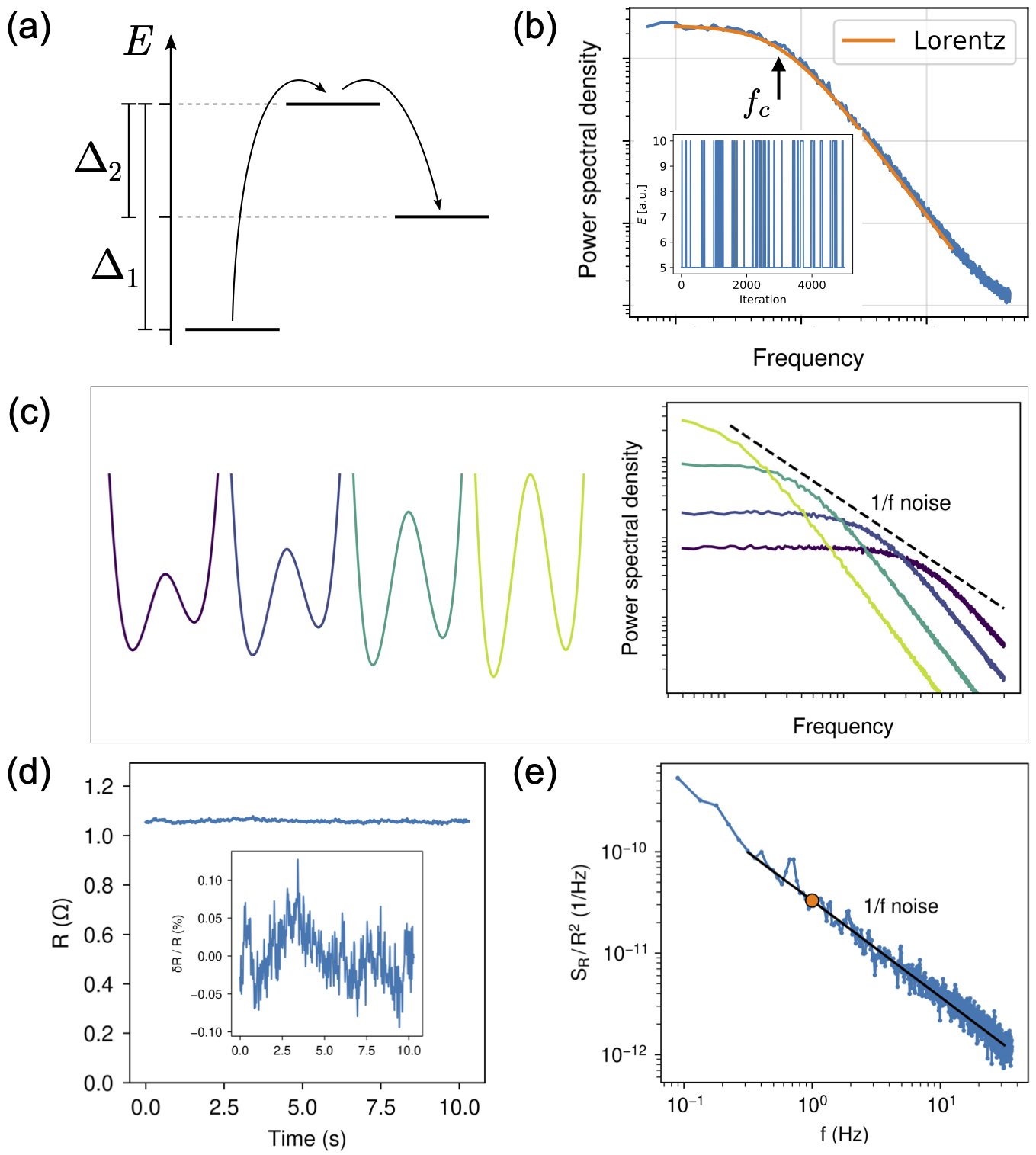}
\caption{(a) Schematics of a fluctuating two-level system (TLS) and (b) the simulated time series of the energy (inset) and the corresponding Lorentzian noise spectrum, where the corner frequency $f_\mathrm{c}$ is indicated. (c) A superposition of many Lorentzians with a distribution of activation energies gives rise to an overall $1/f$-type spectral behavior. (d) A measured time series of a fluctuating resistance with the relative resistance change shown in the inset. Note that $\delta R/R$ is rather small. (e) The corresponding normalized resistance noise PSD with a generic behavior $S_R/R^2 \propto 1/f^\alpha$ characterized by $S_R/R^2(f = 1\,{\rm Hz})$ (orange circle) and the frequency exponent $\alpha \approx 1$ (black line). After \cite{Thyzel2024}.}
\label{noise-methods_1}
\end{figure}
A simple description of the $1/f$-type fluctuations, which are ubiquitously observed in condensed-matter physics \cite{Kogan1996,Weissman1988,Raquet2001,JMueller2011,JMueller2018}, is achieved by the superposition of two-level fluctuations each contributing a Lorentzian spectrum with a certain weight given by the energetics of the two-level system (TLS), schematically shown in Fig.\ \ref{noise-methods_1}(a), fluctuating between the ground state and an excited state. The simulated time train shown in (b) displays so-called random telegraph noise (RTN) and the corresponding power spectrum is a Lorentzian, where the corner frequency $f_\mathrm{c}$ is directly related to the energy diagram of the TLS, e.g.\ $f_\mathrm{c} = f_\mathrm{0} \exp{(-E_\mathrm{a}/k_\mathrm{B}T)}$ for a thermally activated system, where $f_\mathrm{0}$ is an attempt frequency and $E_\mathrm{a}$ the effective activation energy of the system, which can be extracted from measuring noise spectra at different temperatures. In Fig.\ \ref{noise-methods_1}(c) it is shown that an ensemble of different switching processes with a certain energy distribution may lead to $1/f$ noise, indicated by the superposition of four TLS each contributing a Lorentzian. Figure \ref{noise-methods_1}(d) displays a representative time-dependent measurement of the resistance of an organic charge-transfer salt in the metallic regime fluctuating about a mean value $\langle R(t) \rangle$. As shown in the inset, where $\langle R(t) \rangle$ (the long-term average that usually is of interest in a standard resistance measurement) is subtracted, the relative fluctuations $\delta R/R$ are usually small, i.e.\ of order 0.01\,\%\ or less. However, the time trace clearly shows that both fast and rather random (uncorrelated) events and long-term correlations are present. The latter are of interest since they reflect the microscopic kinetics of the charge fluctuations. The calculated normalized PSD shown in Fig.\ \ref{noise-methods_1}(e) can be fitted by a spectrum $S_R/R^2 \propto 1/f^\alpha$ with $\alpha \approx 1$, where both the magnitude of the spectrum --- either conveniently taken at 1\,Hz or by considering the variance of the signal, i.e.\ the spectral weight in a given frequency interval over which the PSD is integrated --- and the frequency exponent $\alpha(T) = - \partial\ln S_R(T)/\partial\ln f$ are evaluated. $\alpha = 1$ corresponds to a homogeneous distribution of the energies of fluctuators contributing to the $1/f$-type noise, and $\alpha > 1$ and $\alpha < 1$ correspond to slower and faster fluctuations in comparison, respectively \cite{JMueller2018}.

An important theorem in time-dependent statistical physics is the Wiener-Khintchine theorem connecting the power spectrum $S_\mathrm{x}(f)$ of a fluctuating quantity $x$ to a time-like property of the statistically varying function $\delta x(t)$ \cite{MacDonald1962}. This time-like property is given by the system's autocorrelation function $\Psi_{\mathrm{xx}}(\tau) = \langle x(t) \cdot x(t + \tau) \rangle$ describing the microscopic kinetics of the fluctuation process: 
\begin{equation}
S_x(f) = 4 \int\limits_{0}^{\infty}\Psi_{xx}(\tau) \cos{(2 \pi f \tau)}{\rm d}\tau.
\end{equation}  
The Fourier transform can be inverted, i.e.\ measuring the noise PSD in principle provides access to the correlation function $\Psi_{xx}(\tau)$ which is a non-random characteristic of the microscopic kinetics of the system's random fluctuations, describing how the fluctuations evolve in time on average \cite{Kogan1996,JMueller2018}.

For an advanced analysis, in addition it is possible for any random process $\delta x(t)$ to define the so-called second spectrum $S^{(2)}_x(f)$ as the power spectrum of the fluctuations of $S_x(f)$ with time, i.e., the Fourier transform of the autocorrelation function of the time series of $S_x(f)$, allowing to access higher-order correlation functions. In practice, an additional frequency $f_2$ related to the time over which $S_x(f)$ fluctuates is thereby introduced. The second spectrum $S^{(2)}_x(f_2,f)$ probes deviations from Gaussian behavior since it is independent of the frequency $f_2$ if the fluctuations are uncorrelated, i.e.\ caused by independent TLS. In contrast, a distinct frequency dependence $S^{(2)} \propto 1/f_2^\beta$ with $\beta \approx 1$, is observed for correlated (interacting) fluctuators \cite{Restle1983,Seidler1996}. A frequency-dependent second noise spectrum indicates ergodicity breaking observed in spin glass physics \cite{Weissman1993} and --- if observed accompanying the slowing down of charge carrier dynamics at a metal-insulator transition --- often is interpreted as glassy freezing of the fluctuations \cite{Jaroszynski2002,Jaroszynski2004,Kar2003,Hartmann2015}.

\begin{figure}
\centering
\includegraphics[width=0.45\textwidth]{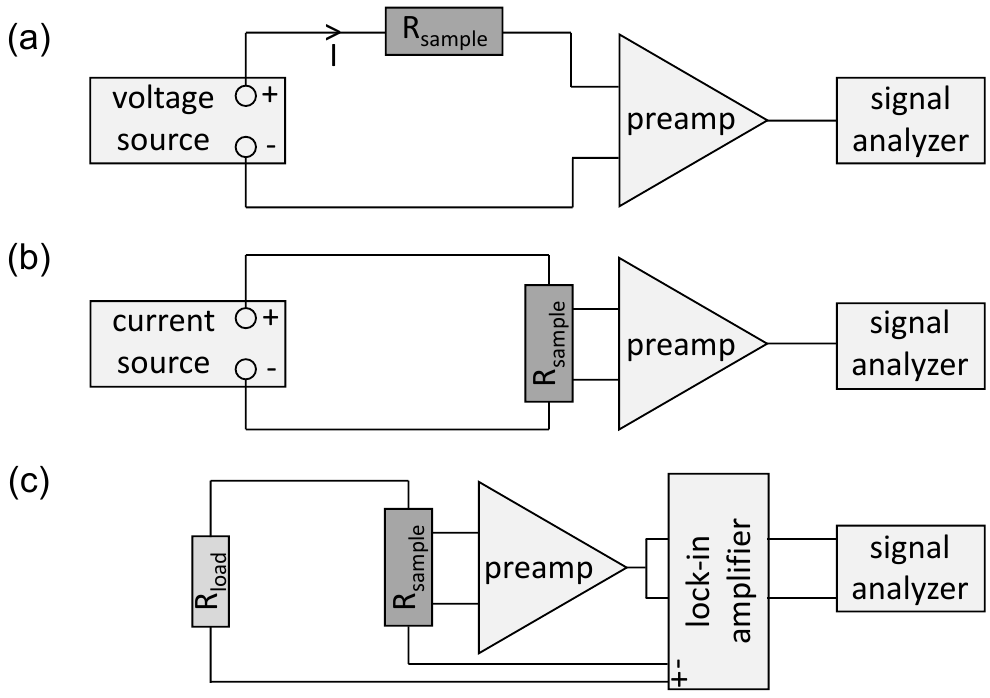}
\caption{Different setups for fluctuation spectroscopy. (a) Two-point dc setup for measurements current (conductance) fluctuations in high-impedance samples. (b) Four-point dc setup and (c) four-point ac setup for measurements of voltage (resistance) fluctuations in more conductive samples. The fluctuating signal in each case is fed into a signal analyzer (e.g. SR785) to obtain the noise power spectral density (PSD).}
\label{noise-methods_2}
\end{figure}
As shown in Fig.\ \ref{noise-methods_2}, fluctuation spectroscopy measurements discussed in this review article were performed with a four-point ac ($T>T_{\rm{MI}}$) and four-point dc method ($T<T_{\rm{MI}}$) by using a preamplifier (Stanford Research 560) and a signal analyzer (Stanford Research 785) (see \cite{JMueller2011,JMueller2018} for more detailed information), which provides the PSD of the voltage fluctuations by computing the Fast Fourier Transform. For measurements of the second spectrum, we employed a fast data acquisition card (DAQ, National Instruments PCI-6281) instead of the signal analyzer in order to determine the power spectral densities from the recorded time signal by a software.
In many conducting systems (like the ones discussed here) the measured PSD of the voltage noise above the equilibrium one increases with the current $I$ as $S_V \propto I^2$ (or the current noise with the voltage $V$ as $S_I \propto V^2$). This is usually interpreted as the modulation noise caused by fluctuations (random modulation) of the sample’s resistance $\delta R(t)$ (or conductance $\delta G(t)$). According to Kirchhoff’s law, the spectral densities of current noise at fixed voltage and voltage noise at fixed current are simply given by  $(S_I/I^2)_V = {\rm const.} = S_G/G^2 = (S_V/V^2)_I = {\rm const.} = S_R/R^2$ where $R$ is the differential resistance and $G = R^{-1}$ the conductance \cite{Kogan1996}. Thus, it is common to consider the resistance or conductance noise power spectral density normalized to the resistance or conductance squared in order to address the relative change in the fluctuation properties.

\begin{figure}
\centering
\includegraphics[width=0.45\textwidth]{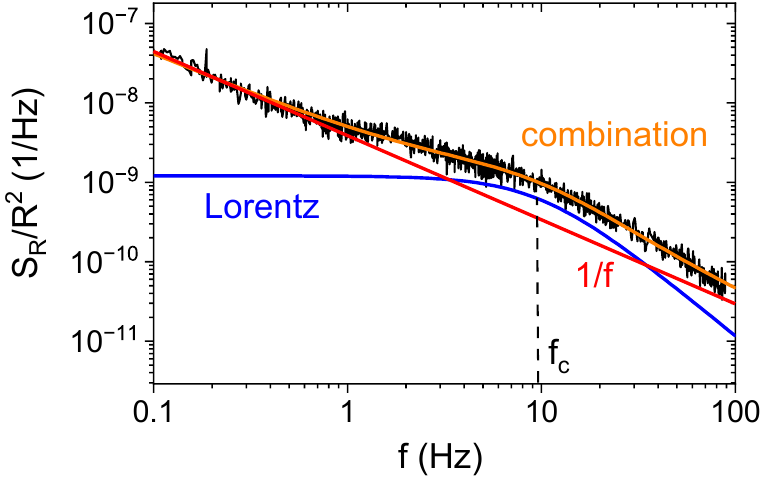}
\caption{Typical noise PSD $S_R/R^2(f)$ where a Lorentzian contribution (blue line) is superimposed on the underlying $1/f^\alpha$ noise (red line). Orange line is a fit to Eq.\ \eqref{noise_spectra}.}
\label{noise-superposition}
\end{figure}
In general, albeit a macroscopic probe, noise measurements are suitable to infer microscopic information, since (i) the noise scales inversely with the system size and (ii) the signature of a single fluctuating TLS may be enhanced over the $1/f$-type contribution in the corresponding 'noise window' (given by the present temperature, electric and magnetic field, and frequency). Figure \ref{noise-superposition} shows a typical example of such a behavior where the measured spectrum can be described by 
\begin{equation}
\frac{S_R(f,T)}{R^2} = \frac{a}{f^\alpha} + \frac{b}{f^2 + f_\mathrm{c}^2},
\label{noise_spectra}
\end{equation}
where $a(T)$ and $b(T)$ are the amplitudes of the $1/f$-type and Lorentzian noise contributions, respectively, and $\alpha(T)$ and $f_c(T)$ their frequency exponent and corner frequency. When analyzing the shift of fluctuating TLS with temperature, electric or magnetic field superimposed on underlying $1/f^\alpha$ noise with $\alpha \approx 1$, it is often convenient to plot $f \times S_R/R^2$ vs.\ $f$, cf.\ Figs.\ \ref{beta-prime-conductance-spectra} and \ref{beta-prime-conductance-Lorentzians} below. This often allows estimating the size of switching entities (single particles or clusters), thus gaining information not only on the ground state, but also on the fluctuations to the excited state, as well as the time and length scales at which the fluctuations occur. This is of fundamental importance for non-uniform electronic or magnetic states in condensed-matter systems \cite{Raquet1999,JMueller2009a}.

\subsection{Thermal expansion}\label{subsec-therm-expan}

\begin{figure}[t]
\centering
\includegraphics[width=0.325\textwidth]{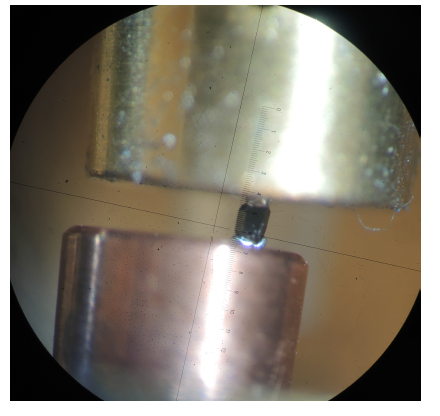}
\caption{A single crystal of $\kappa$-(BEDT-TTF)$_2$Mn(CN)$_4$ (black color) with a length along the $c$ axis of $L_\mathrm{c}$ = 0.708\,mm mounted in the dilatometer cell between a (gold-plated) Cu piston above the sample and a Cu block below the sample \cite{Hartmann2022}. The crystal is embedded from above and below in gallium (silver color) serving as a mechanical fit: in mounting the crystal, gallium spheres heated to above their melting temperature, were placed above and below the crystal. The piston is then screwed down, thereby deforming the spheres, and the gallium is cooled below its solidification temperature.}\label{thermal expansion-setup-2}
\end{figure}

Measurements of the relative length changes $\Delta L_\mathrm{i}$($T$)/$L_\mathrm{i}$, along the different crystal axes $i = a, b, c,$ can provide important information on the involvement of lattice degrees of freedom for the various types of orderings. These measurements were performed by using an ultrahigh-resolution capacitive dilatometer (built after \cite{Pott1983}) with a resolution  $\Delta L_\mathrm{i}$/$L_\mathrm{i} \geq$ 10$^{-10}$, where $L_\mathrm{i}$ is the sample length along the $i$ axis. The obtained $\Delta L$($T$)/$L$ data were differentiated numerically to determine the thermal expansion coefficient $\alpha$($T$) = $1/L$ d$L$/d$T$ using the following procedure: the data for $\Delta L$($T$)/$L$ were divided into equidistant intervals of typically $\Delta T$ = 0.15\,K for low temperatures (typically $T \leq $ 12\,K) and  $\Delta T$ = 0.35\,K for high temperatures (typically $T \geq$ 12\,K). In each of the intervals the mean slope was determined from a linear regression. In a standard experiment, the length changes were measured upon heating with a slow sweep rate $q \leq$ 1.5\,K/h to ensure thermal equilibrium. Figure \ref{thermal expansion-setup-2} shows an example of a single crystal of $\kappa$-(ET)$_2$Mn(CN)$_4$ with typical dimensions in the sub-millimeter range and how this crystal is mounted in the dilatometer cell.

\subsection{Inelastic neutron scattering }\label{subsec-inelas-neutr}

\begin{figure}[t]
\centering
\includegraphics[width=0.4\textwidth]{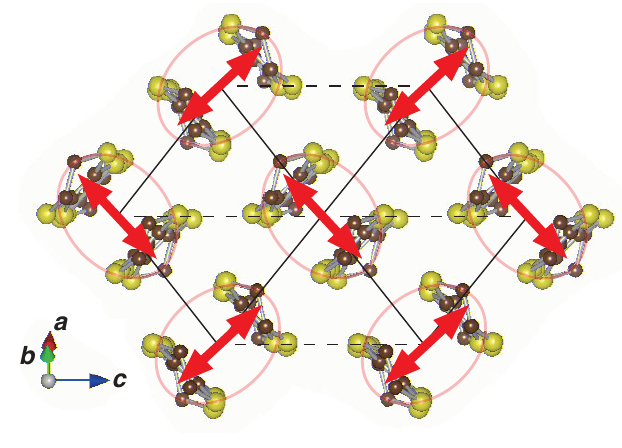}
\caption{Top view of the ET layer of $\kappa$-(ET)$_2$Cu[N(CN)$_2$]Cl (a) with the ellipses indicating the ET dimers. The thick red arrows show schematically the breathing mode with the polarization vector \textbf{$\xi$}. A wave vector \textbf{Q} = $(603)$ was chosen for phonon measurements in the ${(h0l)}$ scattering plane to maximize the scattering intensity. Taken from Ref.\,\cite{Matsuura2019}.}\label{neutron-dimer}
\end{figure}

Fluctuations in the electronic channel accompanying the formation of ferroelectricity may also give rise to phonon renormalization effects. The latter can be sensitively probed by inelastic neutron scattering. For these experiments deuterated single crystals of $\kappa$-(D$_8$-ET)$_2$Cu[N(CN)$_2$]Cl were used \cite{Matsuura2019}. To improve the signal-to-noise ratio a composite sample consisting of several co-aligned single crystals with a total mass of 7 - 9\,mg were used. In order to minimize disorder related to the freezing of orientational degrees of freedom of the ET molecules’ terminal ethylene end groups \cite{JMueller2002,JMueller2015}, the crystals were cooled slowly with $q = -1$\,K/min around the freezing temperature $T_\mathrm{glass}$ = 75\,K. In order to study renormalization effects of the intra-dimer breathing mode (red arrow in Fig.\,\ref{neutron-dimer}) in response to ordering phenomena in the spin and charge channel, a momentum transfer between the initial and final state of the neutron of $\textbf{Q} = (603)$ was selected. Since this wave vector is practically parallel to the polarization vector $\xi$ of the breathing mode, it ensures a high scattering intensity which is proportional to $(Q\cdot \xi) ^{2}$. Constant-$Q$ scans, performed at various temperatures, reveal clear phonon modes at energies $E$ = 2.6, 6, 8 and 11\,meV which were fitted by using damped harmonic oscillator (DHO) functions

\begin{equation}
\mathrm{DHO}_\mathrm{i}(Q,\omega) = \frac{\Gamma_\mathrm{i}\hbar\omega}{[\hbar^{2}(\omega^{2}-{\omega_\mathrm{i}}^{2})]^{2}+({\Gamma_\mathrm{i}\hbar\omega})^{2}}, \label{DHO}
\end{equation}

\noindent where $\Gamma_\mathrm{i}$ and $\hbar\omega_\mathrm{i}$ denote the damping factor and phonon energy of the $i$th mode, respectively. The fits were convolved with the experimental resolution of 0.5\,meV \cite{Matsuura2019}. From these fits, the phonon damping $\Gamma_\mathrm{q}$, i.e., the inverse of the lifetime, and its variation with temperature could be determined. Results are discussed in section \ref{subsec_kCL-neutron}.

\section{Signatures of ferroelectricity revealed in quasi-1D and quasi-2D systems}\label{sec-Sign-ferroelectricity}

\subsection{Quasi-1D (TMTTF)$_2X$}\label{subsec-quasi-1D}

\begin{figure}[t]
\centering
\includegraphics[width=0.45\textwidth]{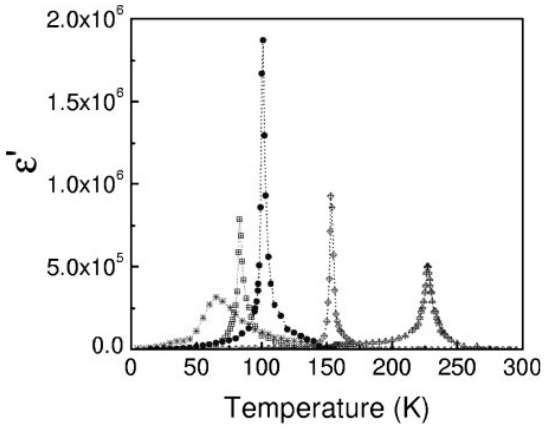}
\caption{Temperature dependence of the dielectric constant taken at a frequency of 100\,kHz for (TMTTF)$_2X$ with $X$ = PF$_6$ (stars), AsF$_6$ (circles), SbF$_6$ (diamonds), BF$_4$ (squares), and ReO$_4$ (triangles). Figure taken from Ref.\,\cite{Nad2006} }\label{dielectrics-TMTTF2}
\end{figure}

Clear signatures for ferroelectricity as a consequence of charge ordering were observed for the quasi-1D (TMTTF)$_2X$ salts for a variety of anions $X$ such as PF$_6$, AsF$_6$, SbF$_6$, BF$_4$, and ReO$_4$ \cite{Monceau2001,Nad2006}. These salts undergo a charge-order transition at $T_\mathrm{CO}$ = 60 - 160\,K, accompanied by a charge disproportionation 2$\delta$ between neighboring TMTTF molecules, as evidenced by nuclear magnetic resonance (NMR) \cite{Takahashi2006,Chow2000,Matsunaga2013} and infrared vibrational spectroscopy \cite{Dumm2004,Dressel2012}. At $T_\mathrm{CO}$ the dielectric constant $\varepsilon'$ shows a pronounced peak, cf. Fig.\,\ref{dielectrics-TMTTF2}, reflecting the ferroelectric character of the charge-ordered state. Upon approaching $T_\mathrm{CO}$ from both sides, the $\varepsilon'$($T$) data follow to a good approximation a Curie-Weiss-like temperature dependence $\varepsilon'$($T$) = $A$/$|T- T_\mathrm{CO} |$ \cite{Nad2006} with a Curie-Weiss constant $A$ for $T < T_\mathrm{CO}$ which is twice that at $T > T_\mathrm{CO}$. This behavior is in fact expected for a second-order ferroelectric transition \cite{Lines2001}.
Detailed investigations of the frequency dependence of the dielectric response for (TMTTF)$_2$AsF$_6$ \cite{Nad2006,Staresinic2006} showed the typical behavior for order-disorder-type ferroelectrics [cf. Fig.\,\ref{paper-ferroele-eps}(b)], where the electric dipoles that are disordered at high temperatures order with a net overall polarization below a phase-transition temperature $T_\mathrm{FE}$ \cite{Lines2001}: the data revealed a sharp peak in $\varepsilon'$($T$) at low frequencies of 30 and 100\,kHz, the amplitude of which becomes reduced with increasing frequency into the MHz regime, while the peak position remains nearly unchanged.

\begin{figure}[t]
\centering
\includegraphics[width=0.4\textwidth]{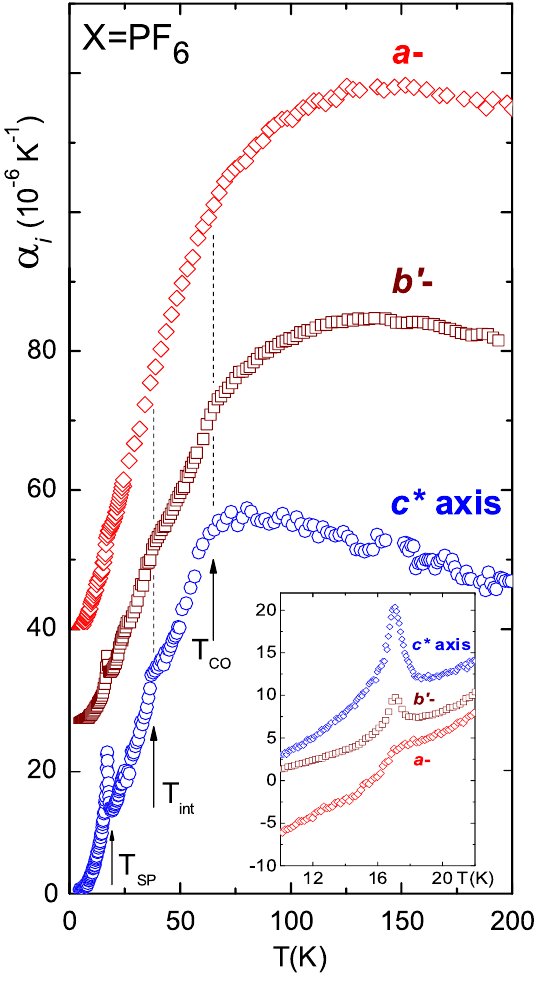}
\caption{Uniaxial thermal expansion coefficients $\alpha_i$ measured along three orthogonal axes of (TMTTF)$_2$PF$_6$. The curves have been offset for clarity. Arrows at $T_\mathrm{SP}$ and $T_\mathrm{CO}$ mark the spin-Peierls and charge-ordering transition temperatures, respectively. The arrow at $T_\mathrm{int}$ marks an anomaly of unknown origin. The inset shows details of the thermal expansion anomalies at the Peierls transition. Taken from Ref.\ \cite{deSouza2008}.}
\label{thermal expansion TMTTF2X}
\end{figure}

The charge-order transition in these salts, which for a long time was labeled the mysterious \textit{``structureless transition''} due to a lack of clear structural signatures, has been attributed to the combined effect of strong on-site, $U$, and inter-site, $V$, Coulomb interactions \cite{Seo1997}. Hence, these systems have been considered prime examples for electronic ferroelectricity. However, as demonstrated by high-resolution thermal expansion measurements, there are distinct lattice effects accompanying the transition at $T_\mathrm{CO}$ \cite{deSouza2008}, cf. Fig.\,\ref{thermal expansion TMTTF2X}. Surprisingly, the strongest effects were observed along the $c^{*}$ axis, which is perpendicular to the stacking axis of the TMTTF molecules. In fact, finite lattice effects are expected given that atomic displacements, breaking the system’s inversion symmetry, are necessary prerequisites for ferroelectricity to occur. Based on the observed thermal expansion anomalies, a scheme was proposed where the charge order within the TMTTF stacks gives rise to a uniform ($q$ = 0) displacement of the anions $X^{-}$. The proposed shift of the anions lifts the inversion symmetry thereby creating the ferroelectric character of the transition. For more details on the rich phenomenology revealed for these quasi-1D molecular conductors, see e.g., the reviews in Refs.\,\cite{Nad2006,Tomic2015,Lunkenheimer2015a}. 

\subsection{Quasi-2D $\alpha$-(BEDT-TTF)$_2$I$_3$}\label{subsec-quasi-2D-alpha}

\begin{figure}[t]
\centering
\includegraphics[width=0.45\textwidth]{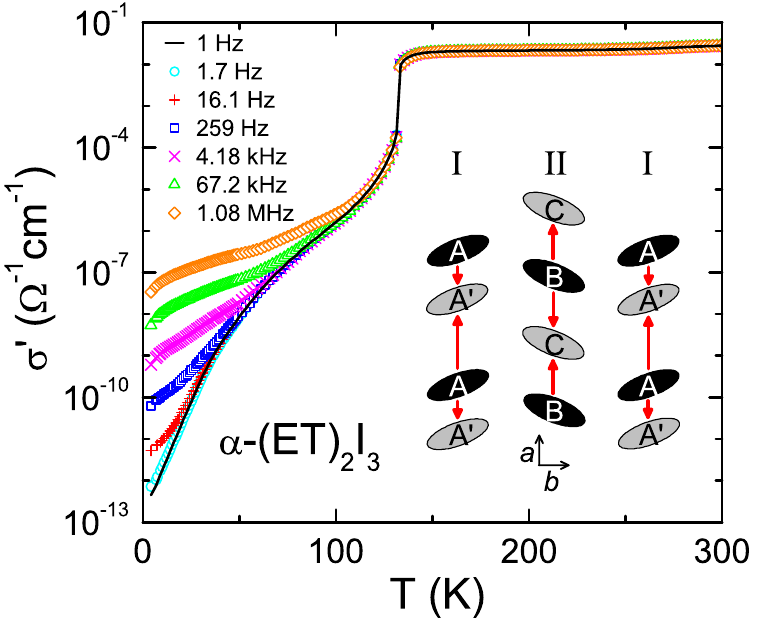}
\caption{Temperature dependence of the ac conductivity of $\alpha$-(ET)$_2$I$_3$ measured along the out-of-plane $c$ axis at varying frequencies. Inset: Sketch of the ET plane with view along the long axis of the molecules. Black molecules indicate a higher charge value. The dimerization in stack I is strongly exaggerated. The thick red arrows indicate the dipolar moments resulting from the charge disproportionation, summing up to a non-zero polarization. Taken from Ref.\,\cite{Lunkenheimer2015b}. 
 }\label{alpha-I3-conductivity}
\end{figure}

In $\alpha$-(ET)$_2$I$_3$ the ET molecules are arranged in a herringbone fashion, with two crystallographically different stacks I and II along the $a$ axis, see inset of Fig.\,\ref{alpha-I3-conductivity} \cite{Lunkenheimer2015b}. In stack I the ET molecules denoted A and A' are weakly dimerized whereas the molecules B and C in stack II are not dimerized. Charge ordering, giving rise to a metal-insulator transition \cite{Bender1984}, was observed at $T_\mathrm{CO}$ = 135\,K \cite{Takano2001,Kakiuchi2007}, cf. the conductivity data in Fig.\,\ref{alpha-I3-conductivity} revealing a rapid drop below $T_\mathrm{CO}$. In fact, a charge-ordered ground state was predicted theoretically for this salt with a quarter-filled hole band based on the combined effect of on-site and inter-site Coulomb interactions \cite{Seo2000,Seo2004}. The pattern of CO can be viewed as analogous to the horizontal stripe patterns of the $\theta$-phase compounds (Fig.~\ref{fig:theta-phase}). While a non-uniform charge distribution was observed already in the metallic state above $T_\mathrm{CO}$ \cite{Yue2010}, the charge disproportionation becomes more pronounced below $T_\mathrm{CO}$. 
Based on optical second-harmonic generation (SHG) measurements \cite{Yamamoto2008,Yamamoto2010}, providing evidence for a non-centrosymmetric crystal structure --- a prerequisite for ferroelectricity ---, this salt was indeed shown to be a candidate for electronic ferroelectricity.  

\begin{figure}[t]
\centering
\includegraphics[width=0.45\textwidth]{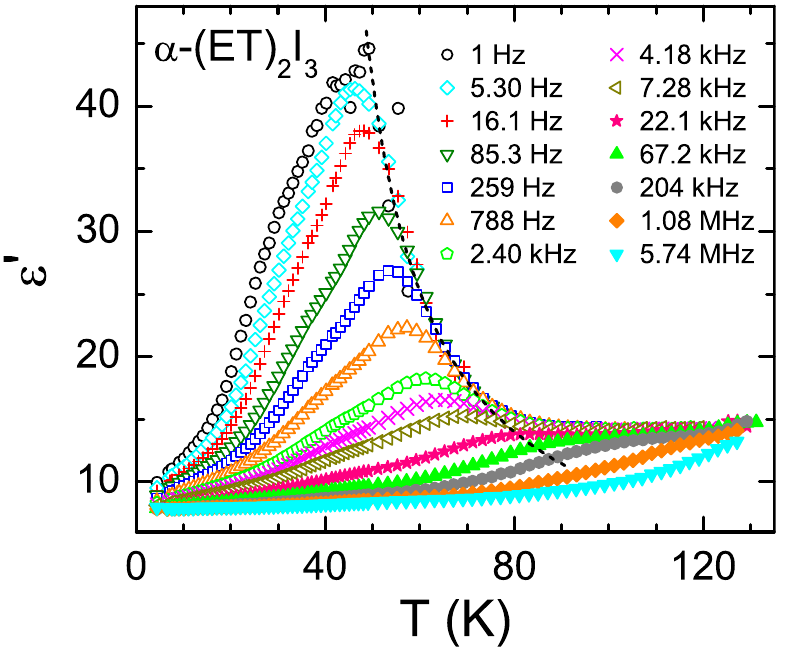}
\caption{Temperature dependence of the dielectric constant $\varepsilon'$($T$) of $\alpha$-(ET)$_2$I$_3$, measured for electric field perpendicular to the ET planes at varying frequencies. The dashed line indicates Curie-Weiss behavior with a Curie-Weiss temperature  $\Theta_\mathrm{CW}$ = 35\,K. Taken from Ref.\,\cite{Lunkenheimer2015b}.
 }\label{alpha-I3-epsilon}
\end{figure}

Measurements of the dielectric constant $\varepsilon'$($T$) for electrical field $E$ aligned perpendicular to the ET planes (Fig.\,\ref{alpha-I3-epsilon}) revealed clear signatures for relaxor-type ferroelectricity \cite{Lunkenheimer2015b}, cf. Fig.\,\ref{paper-ferroele-eps}(c). The data exhibit a pronounced peak in $\varepsilon'$($T$) around 40 - 50\,K at low frequencies which shrinks in size and becomes shifted to higher temperatures with increasing frequency. The variation of the peak position with frequency was found be in accord with a VFT law, Eq.\ \eqref{Vogel-Fulcher} \cite{Vogel1921,Fulcher1925,Tammann1926}, which is well established in glass physics \cite{Lunkenheimer2000} and often seen in relaxor ferroelectrics \cite{Viehland1990,Levstik1998}. The high-temperature flank of the $\varepsilon'$($T$) peaks can be described by a Curie-Weiss law (broken line in Fig.\,\ref{alpha-I3-epsilon}) with a Curie-Weiss temperature $\Theta_\mathrm{CW}$ = 35\,K, providing at least a rough estimate of the freezing temperature. Notably, dielectric measurements for fields $E$ parallel to the ET planes failed to reveal a peak in $\varepsilon'$($T$) \cite{Ivek2010,Ivek2011}. The reason for that is likely related to the distinctly higher in-plane conductivity even in the charge-ordered state, making dielectric measurements difficult. It should be noted that the dipolar polarization in $\alpha$-(ET)$_2$I$_3$ should be predominantly oriented parallel to the ET planes. However, in Ref. \cite{Lunkenheimer2015b} it was argued that the dipolar moment in this and related materials should also have a component oriented perpendicular to the planes, detected by the dielectric measurements performed in that work.

\begin{figure}[t]
\centering
\includegraphics[width=0.45\textwidth]{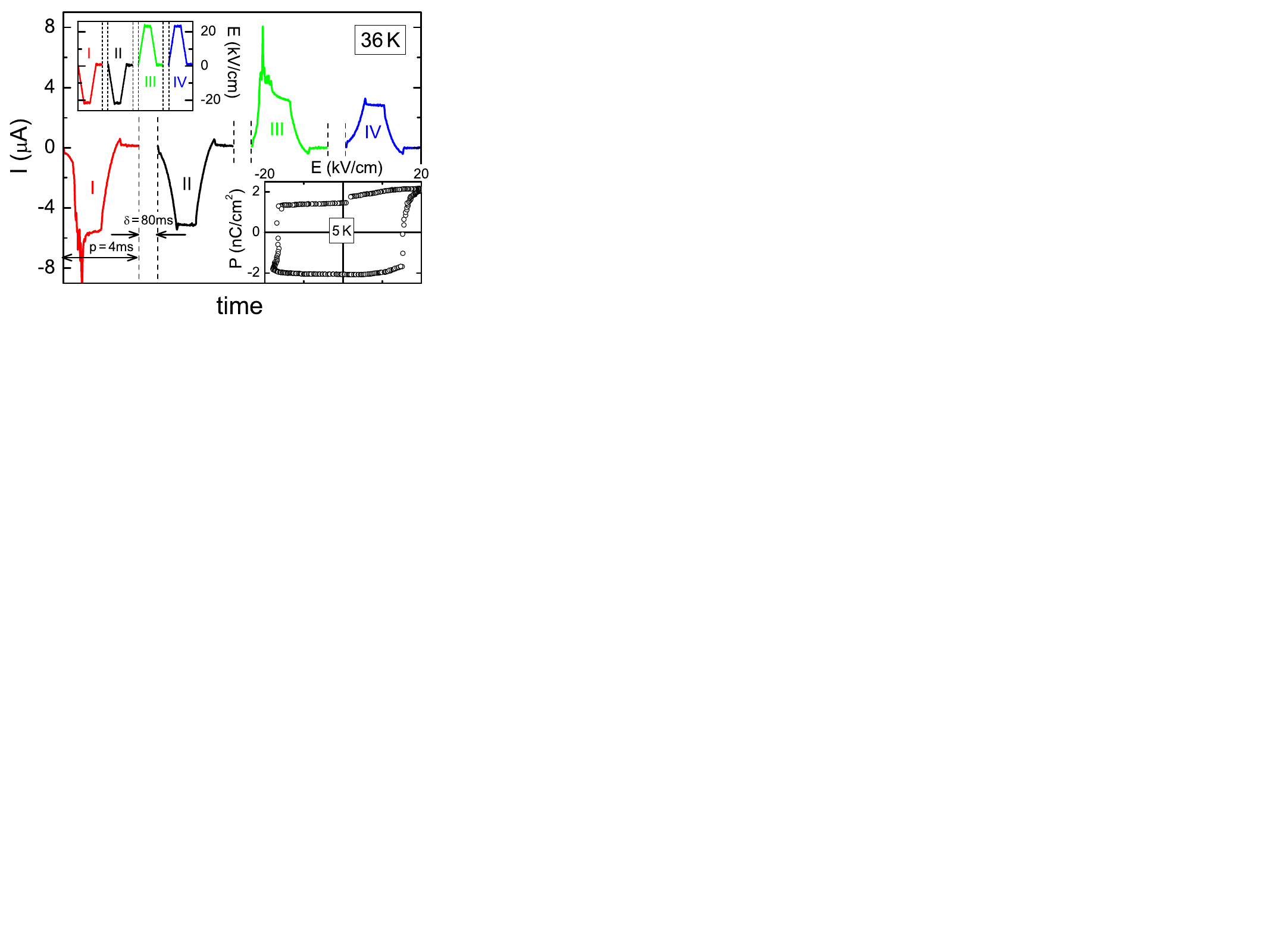}
\caption{Time-dependent current as detected in $\alpha$-(ET)$_2$I$_3$ by PUND measurements at  36\,K. $\delta$ denotes waiting time and $p$ the pulse width. Left inset shows the excitation signal whereas the right inset shows the polarization-field hysteresis curve at 5\,K. Taken from Ref.\ \cite{Lunkenheimer2015b}.}
\label{aplpha-I3-PUND}
\end{figure}

Further evidence for ferroelectric order was provided by probing the macroscopic polarization using PUND measurements (see chapter \ref{subsecPolarization} for details on the technique). Upon the application of a sequence of trapezoid electrical field pulses (left inset of Fig.\,\ref{aplpha-I3-PUND}) a current was observed with characteristic peaks in response to the first and third pulse when the electric field $|E|$ exceeds a threshold level of the order of 20\,kV/cm, indicating the switching of the macroscopic polarization [cf. Figs.\,\ref{pap-PE_PUND}(b) and (c)]. In contrast, no such peaks were observed for the second and fourth pulse as the polarization was already switched by the preceding pulse. 

The results on the dielectric response and polarization provided clear evidence for ferroelectricity in $\alpha$-(ET)$_2$I$_3$. The dielectric response was found to be typical for relaxor-type ferroelectricity. In contrast to the findings for the various (TMTTF)$_2X$ salts discussed above, where ferroelectricity sets in at $T_\mathrm{CO}$, here it appears to occur deep in the charge-ordered state. 
In Ref. \cite{Lunkenheimer2015b} it was argued that this decoupling of polar and charge order and the occurrence of relaxor ferroelectricity in $\alpha$-(ET)$_2$I$_3$ arises from the alteration of undimerized and dimerized stacks in the $\alpha$-type structure which could counteract the formation of canonical long-range ferroelectricity directly below $T_\mathrm{CO}$.
Instead, between $T_\mathrm{CO}$ and about 80\,K, in Fig.\,\ref{alpha-I3-epsilon} conventional dipolar relaxation behavior is found [cf. Fig.\,\ref{paper-ferroele-eps}(d)]. In Ref. \cite{Lunkenheimer2015b} it was proposed to arise from the relaxation of single dipoles or several ferroelectrically-correlated dipoles within one chain. 

\subsection{Quasi-2D $\theta$-(BEDT-TTF)$_2$RbZn(SCN)$_4$}\label{subsec-quasi-2D-theta}

In the $\theta$-polymorph the ET molecules form a quarter-filled electron system on a triangular lattice implying some degree of (charge) frustration. This salt shows charge order, as predicted theoretically \cite{Kino1996,Senthil2008}, accompanied by a structural transition at $T_\mathrm{CO}$ = 200\,K \cite{Mori1998,Miyagawa2000,Chiba2001,Alemany2015,Hashimoto2022,Saito2024}. Interestingly, through rapid cooling the charge-order transition can be avoided and instead glasslike charge dynamics were observed which were assigned to a novel charge-glass state \cite{Kagawa2013}. 

\begin{figure}[t]
\centering
\includegraphics[width=0.45\textwidth]{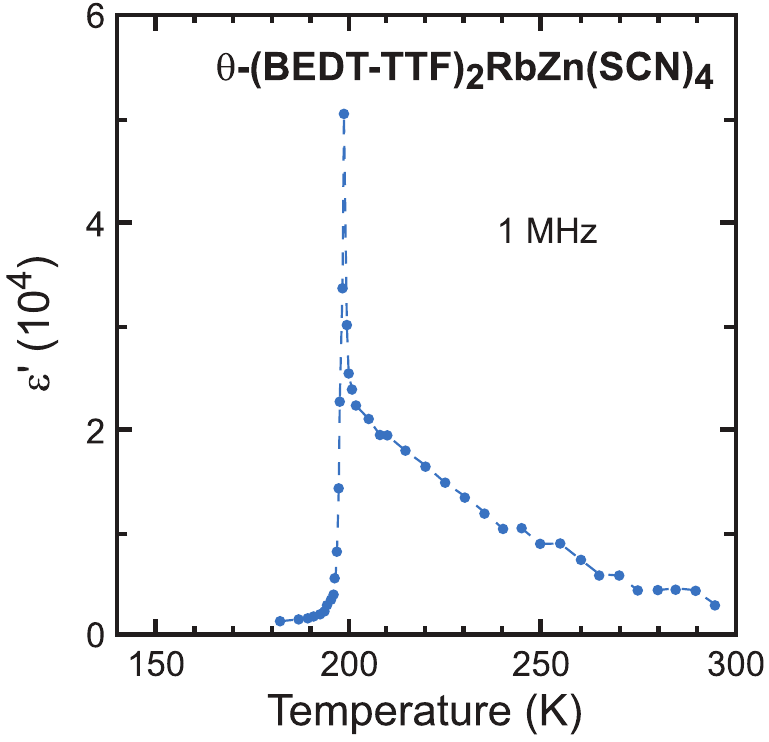}
\caption{Temperature dependence of the dielectric constant $\varepsilon'$($T$) of $\theta$-(BEDT-TTF)$_2$RbZn(SCN)$_4$ taken at a frequency of 1\,MHz upon cooling
at a rate of 0.1\,K/min. Taken from \cite{Tomic2015} after Ref.\ \cite{Nad2006a}.
 }\label{theta-ET2X-epsilon}
\end{figure}

\begin{figure}[t]
\centering
\includegraphics[width=0.45\textwidth]{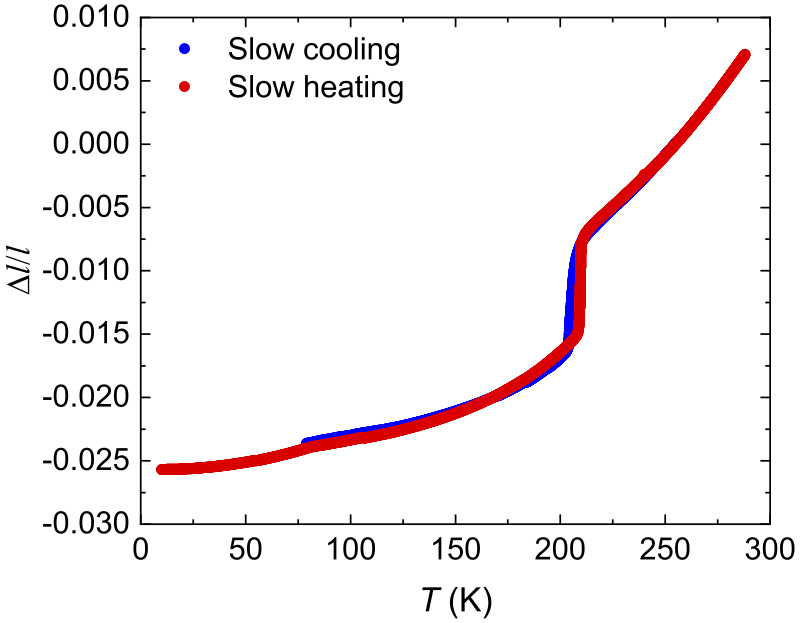}
\caption{Relative length change for $\theta$-(ET)$_2$RbZn(SCN)$_4$ measured along the in-plane $c$ axis upon cooling and heating. Figure taken from Ref.\ \cite{Saito2024}.
 }\label{theta-ET2X-delta l}
\end{figure}

Measurements of the dielectric constant revealed a gradual growth upon cooling from room temperature, followed by a sharp peak at $T_\mathrm{CO}$, with an almost divergent behavior in the immediate vicinity of $T_\mathrm{CO}$, cf. Fig.\,\ref{theta-ET2X-epsilon}. The smooth increase of $\varepsilon'$($T$) in the metallic regime is likely related to the observation of slowly fluctuating charge disproportionation in this temperature regime \cite{Takahashi2004}. On the low-temperature side, a jump-like drop shows up within a narrow temperature window below which $\varepsilon'$($T$) levels off at low values. Interestingly, in Ref. \cite{Nad2006a} the amplitude of the $\varepsilon'(T)$ peak was found to become reduced with increasing frequency. This points to order-disorder type of ferroelectricity where fluctuating dipoles already exist above the transition (see chapter \ref{subsubsecDielFerroel}), in accord with the suggested formation of short-range charge disproportionation already above $T_\mathrm{CO}$ \cite{Takahashi2004}. Although the $\varepsilon'$($T$) data in Fig.\,\ref{theta-ET2X-epsilon} reveal clear signatures of ferroelectricity associated with the charge-order transition, there are distinct differences to the phenomenology observed in the (TMTTF)$_2X$ and $\alpha$-(ET)$_2$I$_3$ salts. This can be attributed to the strong first-order character of the structural transition in $\theta$-(BEDT-TTF)$_2$RbZn(SCN)$_4$ \cite{Nad2006a} accompanied by a lattice modulation, which is reflected also in pronounced discontinuous changes of the $c$-axis lattice parameter (Fig.\,\ref{theta-ET2X-delta l}) and the hysteresis upon cooling and warming in $\varepsilon'$ \cite{Nad2006a}
and $\Delta l$/$l$ \cite{Saito2024}.

\subsection{Quasi-2D $\beta^\prime$-(BEDT-TTF)$_2$ICl$_2$}
\label{subsec_betaICl}

\subsubsection{Dielectric spectroscopy}

\begin{figure}[t]
\centering
\includegraphics[width=0.475\textwidth]{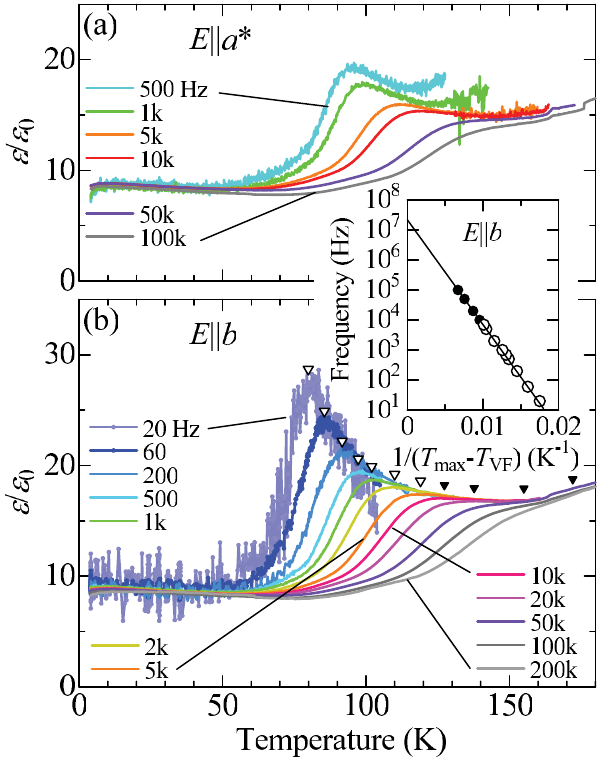}
\caption{Temperature dependence of the dielectric constant $\varepsilon'$($T$) of $\beta$'-(ET)$_2$ICl$_2$ for electric field perpendicular (a) and within (b) ($E \| b$) the ET layers. Data were taken for frequencies between 500\,Hz and 100\,kHz. Reproduced with permission from \cite{Iguchi2013}.}\label{beta-prime-epsilon}
\end{figure}
\begin{figure}[t]
\centering
\includegraphics[width=0.45\textwidth]{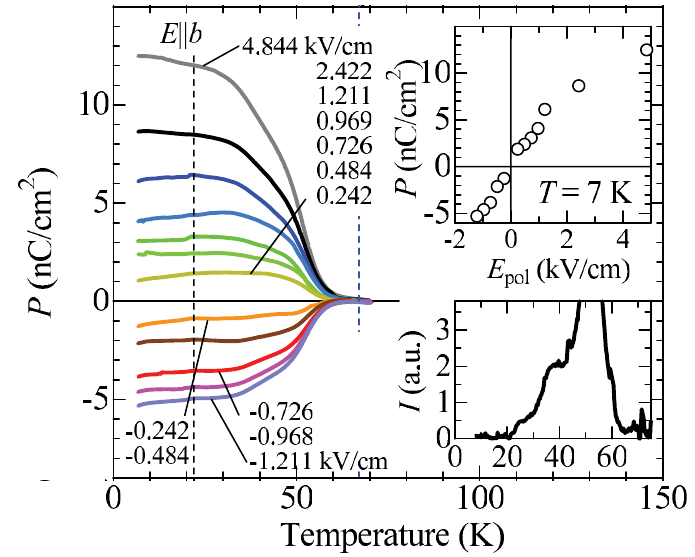}
\caption{Temperature dependence of the electric polarization of $\beta$’-(ET)$_2$ICl$_2$ for $E \| b$ and a poling electric field from $-$1.211 to +4.844\,kV/cm obtained by pyrocurrent measurements. Reproduced with permission after Ref.\ \cite{Iguchi2013}.}
\label{beta-prime-polarization}
\end{figure}

Measurements of the dielectric constant $\varepsilon'$($T$) of this material for electric fields perpendicular, Fig.\,\ref{beta-prime-epsilon}(a), and parallel, Fig.\,\ref{beta-prime-epsilon}(b), to the ET layers revealed clear anomalies in the temperature range 80 - 150\,K with a strong frequency dependence, characteristic of relaxor-type ferroelectricity (see section \ref{subsubsecDielFerroel}) \cite{Cross1987,Samara2003,Bokov2006}. By analyzing the high-temperature flank of the anomaly, yielding a Curie-Weiss-like temperature dependence, a Curie-Weiss temperature of 67\,K was obtained \cite{Iguchi2013}. Further evidence for ferroelectricity in this salt came from studies of the electric polarization obtained by measurements of the pyrocurrent for poling fields parallel to the $b$ axis, cf. Fig.\,\ref{beta-prime-polarization}. A small but clearly resolvable polarization was obtained below a critical temperature $T_\mathrm{FE}$ = 62\,K which is close to the above-mentioned Curie-Weiss temperature of 67\,K. Measurements with poling fields up to 3\,kV/cm aligned along the out-of plane $a^{*}$ axis failed to detect a pyrocurrent, consistent with the notion that the electric dipoles originate from the ET dimers. The observed phenomenology, i.e., the strong frequency-dependent dielectric response alongside with the small polarization indicates the formation of glassy polar domains in this salt, typical for relaxor ferroelectrics. As it exhibits antiferromagnetic ordering below $T_\mathrm{N}\approx22$\,K \cite{Yoneyama1999} but only short-range polar order, the spin and dipolar degrees of freedom in $\beta$’-(ET)$_2$ICl$_2$ obviously are not closely coupled.

\subsubsection{Conductance fluctuation (noise) spectroscopy}
\label{subsec_betaICl_noise}
The system \betaICl\ is a strong dimer-Mott insulator, where electric-field-induced charge disproportionation has been demonstrated by charge-sensitive Raman scattering \cite{Hattori2017}. The material is well-suited for complementary studies of dielectric and conductance noise spectroscopy, where the latter method aims to probe the effects of collective fluctuations also at temperatures far above the Curie-Weiss temperature and the onset of electric polarization at 67\,K and 62\,K, respectively, a regime where samples are too conductive for performing the former technique. As will be discussed below, the conductance fluctuations provide evidence for an important consequence of large dielectric fluctuations, namely the formation of polar nano-regions (PNR) and electronic (e.g., paraelectric-ferroelectric) phase separation \cite{Ishihara2014}, which may help to explain the up-to-now puzzling occurrence of either long-range ferroelectric order or relaxor-type ferroelectricity in the class of quasi-2D organic charge-transfer salts \cite{Abdel-Jawad2010,Lunkenheimer2012,Lang2014}. 

\begin{figure}
\centering
\includegraphics[width=0.475\textwidth]{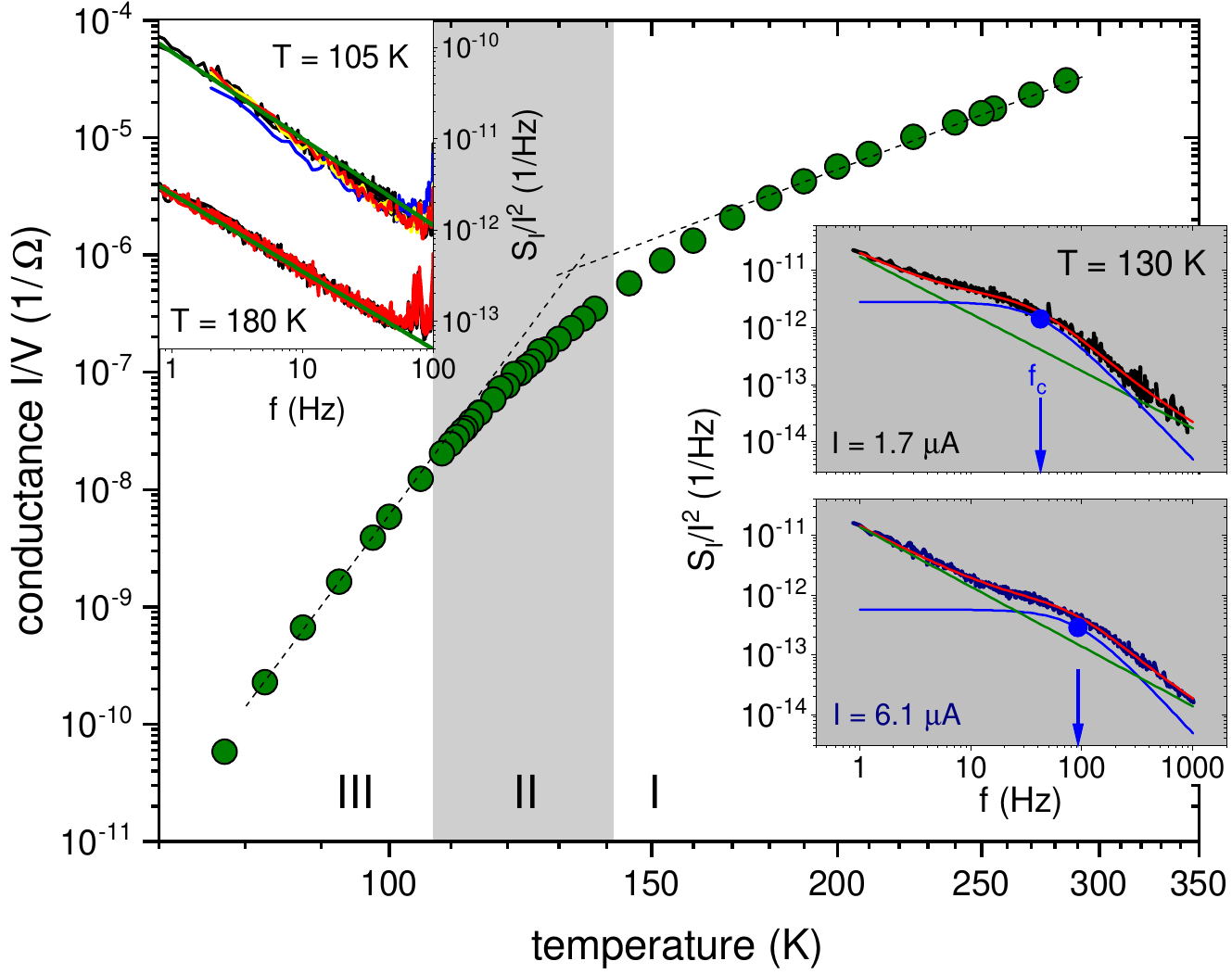}
\caption{Conductance of \betaICl\ vs.\ temperature. (Dashed lines indicate a crossover to more insulating behavior upon cooling.) In the areas labeled I and III the normalized current noise PSD is $1/f$-like and independent of the applied electric field (left inset, green lines are fits to $S_I/I^2 \propto 1/f^\alpha$; for $T = 180$\,K and 105\,K, two and four different currents, respectively, are shown). In the grey shaded temperature regime II ($110 - 140\,{\rm K}$), a single Lorentzian is superimposed on the underlying $1/f$-type noise and $S_I/I^2(f)$ depends on the external electric field. Right insets show representative spectra for regime II. Lines are fits to Eq.\,(\ref{noise_spectra}) with $\alpha = 1$. Taken from Ref.\ \cite{JMueller2020}.}  
\label{beta-prime-conductance-spectra}
\end{figure}
Figure\,\ref{beta-prime-conductance-spectra} shows the dc conductance $G = V/I$ vs.\ $T$ of \betaICl, where $V$ is the voltage applied across the sample and $I$ the measured electric current flow. 
The $\log{G}$ vs.\ $\log{T}$ representation indicates a crossover to more insulating behavior (stronger decrease of the conductance) below about $130 - 140$\,K, see dashed lines.

At discrete temperatures, the current noise PSD $S_I(f,T)$ measured for various excitation voltages $V$ allows to identify three distinctly different temperature regimes (see also Fig.\,\ref{beta-prime-noise-scaling}(a) below): For temperatures $T \gtrsim 140\,{\rm K}$ (regime I) and $T \lesssim 110\,{\rm K}$ (regime III), the normalized noise spectra are of generic $1/f^\alpha$-type and independent of the applied electric field $\mathcal{E} = V/l$ (with $l$ being the distance between the electric contacts along the sample's $c$-axis), i.e.\ the expected scaling $S_I \propto I^2$ is obeyed, see representative noise spectra in these regimes in the left inset of Fig.\,\ref{beta-prime-conductance-spectra}. In between, for $110\,{\rm K} \lesssim T \lesssim 140\,{\rm K}$ (regime II), we find a single Lorentzian spectrum --- characteristic for a fluctuating TLS --- superimposed on the underlying $1/f^\alpha$ noise, cf.\ Fig.\ \ref{noise-superposition} and Eq.\ \eqref{noise_spectra} in section \ref{subsec-fluc-spec} above. 

Strikingly, in regime II the observed normalized noise spectra depend on $\mathcal{E}$, i.e.\ the $S_I \propto I^2$ scaling is not valid. 
The corner frequency $f_\mathrm{c}$ is related to the characteristic energy of the two-level fluctuator and for thermally activated states of a double-well potential one has 
\begin{equation}
f_\mathrm{c} = f_\mathrm{0} \exp{\left( -\frac{E_\mathrm{a}}{k_\mathrm{B} T}\right)}
\label{Arrhenius}
\end{equation}
with an activation energy $E_\mathrm{a}$, an attempt frequency $f_\mathrm{0}$ and Boltzmann's constant $k_\mathrm{B}$ \cite{Raquet2001}.

The anomalous current dependence of the noise spectra in regime II is shown exemplarily in the right inset of Fig.\,\ref{beta-prime-conductance-spectra}.
Clearly, $f_\mathrm{c}$ and hence the characteristic energy of the fluctuation process, shift with $I$, i.e.\ depend on the electric field $\mathcal{E}$ with $f_\mathrm{c}$ increasing from $42$\,Hz at $T = 130$\,K for a sample current of $I = 1.7$\,$\mu$A to $f_\mathrm{c} = 93$\,Hz for $I = 6.1$\,$\mu$A (blue arrows).

\begin{figure*}[t]
\includegraphics[width=\textwidth]{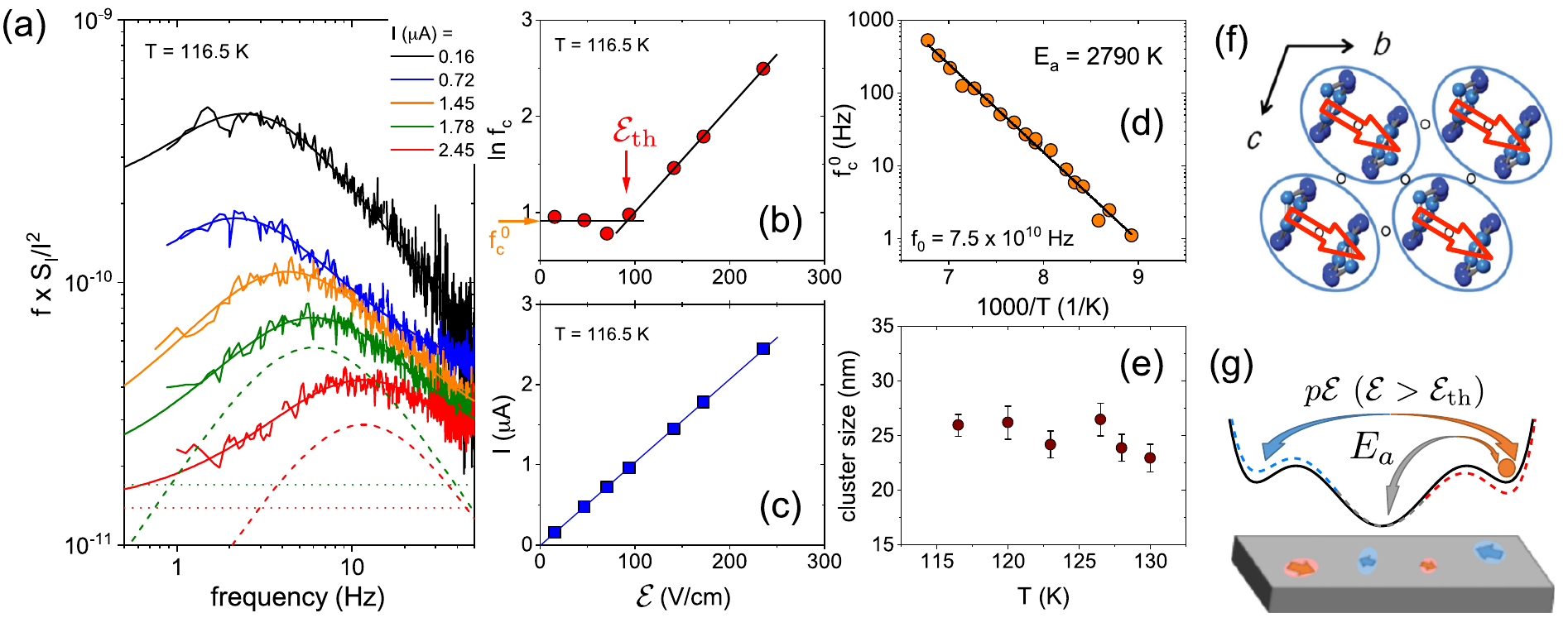} 
\caption{(a) Representative noise spectra  of \betaICl\ as $f \times S_\mathrm{I}/I^2$ vs.\ $f$ in temperature regime II at fixed $T = 116.5$\,K and different sample currents as indicated. Smooth solid lines are fits to Eq.\,(\ref{noise_spectra}). For two selected currents, the $1/f$ and Lorentzian contributions are shown (dotted and dashed lines, respectively). (b) Electric field dependence of the corner frequency $f_\mathrm{c}$. Lines are guides for the eyes indicating the threshold field $\mathcal{E}_\mathrm{th}$ and the zero-field value $f_\mathrm{c}^0 \equiv f_\mathrm{c}(\mathcal{E} \rightarrow 0)$. (c) demonstrates the linear $I$-$V$ regime. (d) Arrhenius plot of $f_\mathrm{c}^0$ vs.\ $1/T$ revealing an activation energy $E_\mathrm{a}/k_\mathrm{B} = 2790$\,K. (e) Cluster size calculated from $p$ estimated from Eq.\ \eqref{mod_Arrhenius}, see text. (f) Arrangement of ET dimers in the conducting planes and orientation of the electric dipoles. (g) Schematics of the suggested zero-field/electric field-induced dielectric/ferroelectric fluctuations, see text for details. Taken from \cite{JMueller2020}.}
\label{beta-prime-conductance-Lorentzians}
\end{figure*}
This unusual current/electric field dependence of the noise spectra in the temperature regime II is illustrated in more detail in Fig.\,\ref{beta-prime-conductance-Lorentzians}(a) for a representative temperature of $T = 116.5$\,K in regime II. Solid lines are fits to Eq.\ \eqref{noise_spectra}. Note that in the representation $f \times S_I/I^2$ vs.\ $f$, the $1/f$-term is a constant (for $\alpha \approx 1$) and the Lorentzian term exhibits a peak centered at $f_\mathrm{c}$, see the respective dotted and dashed curves representing these contributions for two selected currents. An important finding is a threshold behavior of the corner frequency $f_\mathrm{c}$, shown Fig.\,\ref{beta-prime-conductance-Lorentzians}(b), which stays roughly constant for small values of the electric field $\mathcal{E}$ (orange arrow) until it increases exponentially above a threshold field $\mathcal{E}_\mathrm{th}$ (red arrow). Thus, a distinct low-$\mathcal{E}$ and high-$\mathcal{E}$ behavior is observed. Phenomenologically, we describe the characteristic frequency of the two-level fluctuations by Eq.\,(\ref{Arrhenius}) for $\mathcal{E} \leq \mathcal{E}_\mathrm{th}$ and 
by an additional term, see also \cite{Raquet2000}, resulting in
\begin{equation}
f_\mathrm{c} = f_\mathrm{0} \exp{\left( \frac{p(\mathcal{E} - \mathcal{E}_\mathrm{th}) -E_\mathrm{a}}{k_\mathrm{B} T}\right)} \quad {\rm for} \quad \mathcal{E} \geq \mathcal{E}_\mathrm{th}.
\label{mod_Arrhenius}
\end{equation}
Importantly, for all applied electric fields shown here, the $I$-$V$ characteristics, see Fig.\,\ref{beta-prime-conductance-Lorentzians}(c), remains linear, which excludes a trivial heating effect to account for the shift of $f_\mathrm{c}$.  
For the subthreshold value $f_\mathrm{c}^0 \equiv f_\mathrm{c}(\mathcal{E} \rightarrow 0)$, we find an excellent fit to an Arrhenius behavior, Eq.\,(\ref{Arrhenius}),  yielding $E_\mathrm{a}/k_\mathrm{B} = (2790 \pm 70)$\,K,
see Fig.\,\ref{beta-prime-conductance-Lorentzians}(d),
an energy very close to the intra-dimer transfer integral and optical charge gap of $\sim 2900$\,K \cite{Koretsune2014,Hashimoto2015,Tajima2008}, which indicates that the observed switching processes in temperature regime II are of electronic origin. We suggest that the competing inter- and intra-dimer Coulomb interactions \cite{Naka2010,Hotta2012} cause coherent fluctuations of electrons within a cluster of dimers (i.e.\ the gravity center of the hole), switching between the unpolarized dimer-Mott state and a charge disproportionated state, see scheme [$E_a$] in Fig.\,\ref{beta-prime-conductance-Lorentzians}(g). 

The above-threshold behavior 
can be described by
Eq.\ \eqref{mod_Arrhenius} yielding a constant slope $p$, that can be determined from a linear fit to the data in Fig.\,\ref{beta-prime-conductance-Lorentzians}(b). This indicates a {\it discrete} value of a fluctuating dipole moment that corresponds to a cluster of a certain size. At $T = 116.5$\,K we find $p = 1.73 \times 10^{-23}\,{\rm C \cdot cm}$ and a threshold field of $\mathcal{E}_\mathrm{th} \sim 92\,{\rm V/cm}$. 
With the dipole moment $p_\mathrm{d} = 0.13\,ed$ parallel to the $b$-axis 
estimated in \cite{Iguchi2013} [$e$ and $d$ = 3.6\,{\rm \AA} are
the electron charge and distance between the ET molecules in a dimer, respectively, see Fig.\,\ref{beta-prime-conductance-Lorentzians}(f)],
we find that the total number $N$ of elementary dipoles within the cluster --- taking into account  the projection onto the measured $c$-axis --- amounts to $N = p/p_\mathrm{d}^{c\text{-axis}} \approx 9 \times 10^4\,{\rm dimers}$.
Assuming a spherical object, this corresponds to a fluctuating cluster volume with radius $r_{\rm cluster} \approx 26\,{\rm nm}$. Thus, the observed low-frequency dynamics below $\sim 140$\,K is caused by the formation of fluctuating PNR, which above a threshold field $\mathcal{E}_\mathrm{th}$ undergo switching between two states of polarization $+p$ and $-p$, see scheme [$p\mathcal{E}$] in Fig.\,\ref{beta-prime-conductance-Lorentzians}(g).

This dipole-dimer picture has the two polar states ($+p$ and $-p$) and one non-polar state ($n$) whose energies and corresponding lifetimes become modified in a finite electric field $\mathcal{E} > \mathcal{E}_\mathrm{th}$. A simple estimate \cite{JMueller2020} shows that $f_\mathrm{c}$ does not change due to the lifetime of the non-polar dimer Mott state but rather {\it always increases} with increasing electric field due to a second-order effect related to the lifetimes $\tau_\mathrm{+p}$ and $\tau_\mathrm{-p}$. This can be viewed as a transition between the $+p$ and $-p$ states and determines the origin of the threshold field $\mathcal{E}_\mathrm{th}$.
Figure\,\ref{beta-prime-conductance-Lorentzians}(e) shows the estimates of the cluster size for other temperatures revealing a roughly constant or slightly increasing size of the PNR with decreasing temperature.\\
Here, it is important to note that a very similar coupling of two-level excitations to the electric field indicating the presence of fluctuating PNR has been observed for the relaxor ferroelectrics $\kappa$-(ET)$_2$Cu[N(CN)$_2$]Cl and $\kappa$-(BETS)$_2$Mn[N(CN)$_2$]$_3$, see sections \ref{subsec_kCL-noise} and \ref{subsec-quasi-2D-kappa-X} below.

A more complete picture of the charge carrier dynamics coupled to the dielectric behavior in \betaICl\ probed by current noise arises when considering also the underlying $1/f$-type fluctuations. 
At high temperatures (regime I) the frequency exponent $\alpha(T \gtrsim 140\,{\rm K})\approx 0.8$ \cite{JMueller2020}, i.e.\ there is considerably more spectral weight at higher than at lower frequencies. As discussed above, at about 140\,K, upon entering regime II, fluctuating PNR are formed with one such fluctuator being enhanced in our 'noise window', i.e.\ it strongly couples to the conductance fluctuations, which become a sensitive probe to changes in the {\it dielectric} properties. The fluctuating electrical polarization potentials of the PNR act on the transport channel as small capacitor gates that locally modulate the conductance of the underlying resistor network 
\cite{Shklovskii2003,Burin2006}. The superposition of many fluctuating PNR then constitutes part of the underlying $1/f$-type.
This picture of an enhanced localization of charge carriers and increasingly slow dynamics due to the formation of dimer units that fluctuate coherently as extended clusters is corroborated by the rather abrupt increase of the frequency exponent $\alpha(T)$ from almost constant values of $\alpha(T) \approx 0.8$ in temperature regime I to values of $\alpha \gtrsim 1$ below about 140\,K \cite{JMueller2020}
implying a substantial shift of the current fluctuations' spectral weight to low frequencies below this temperature, which may indicate some clusters merging and possibly switching cooperatively. 

\begin{figure}[ht]
\centering
\includegraphics[width=0.45\textwidth]{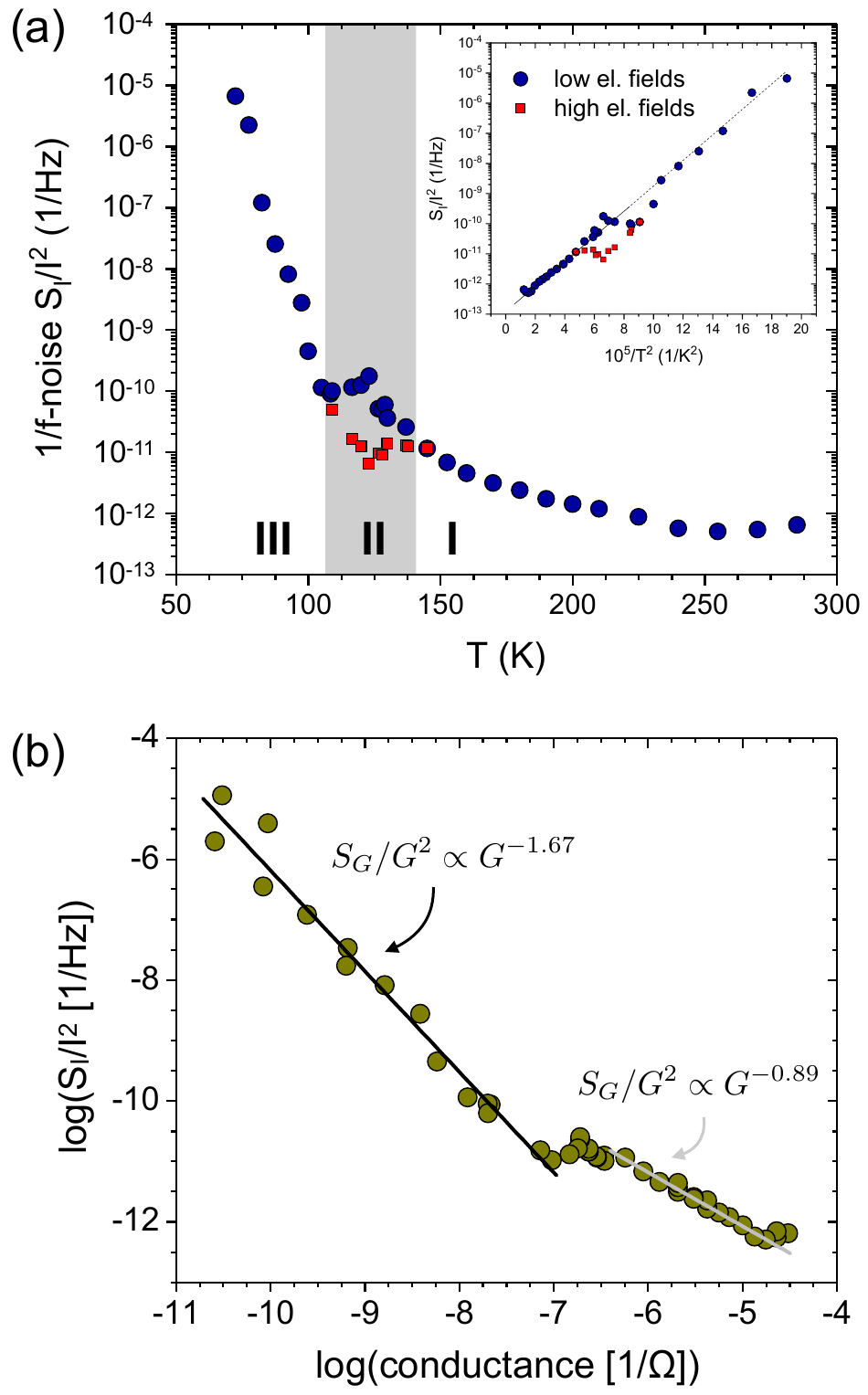}
\caption{(a) Normalized current noise PSD $S_I/I^2$ of $\beta^\prime$-(ET)$_2$ICl$_2$
at 1\,Hz vs $T$. Shown is only the term $a(T)$ in Eq.\,(\ref{noise_spectra}). Blue and red symbols denote the asymptotic low- and high-electric-field behavior, respectively, in regime II. Inset: Scaling plot of the same data suggesting VRH transport \cite{Shklovskii2003}. Blue and red symbols denote the low- and high-electric field limit.
(b) Scaling of the normalized conductance noise $S_G/G^2$ with the conductance $G$ (temperature here is an implicit parameter). The scaling exponents $w$ differ in the different temperature regimes I and III reflecting the change in transport mechanism. Reproduced after Supplementary Information of Ref.\ \cite{JMueller2020}.}
\label{beta-prime-noise-scaling} 
\end{figure}
As shown in Fig.\,\ref{beta-prime-conductance-Lorentzians}(a), the $1/f$-noise magnitude decreases with increasing electric field, with a tendency for saturation at high fields (not shown), with $\alpha$ approaching values of 1 indicating the stabilization of an increasing number and/or size of PNR that increases the total volume of slow fluctuators, thereby reducing the $1/f$-type noise in regime II which may be viewed as the superposition of many independently-fluctuating clusters with a broad energy distribution \cite{Raquet2001}.

Upon further decreasing the temperature, in regime III the slow dynamics of the now stabilized PNR
dominates the $1/f$-noise, which below about 110\,K shows a drastic increase, see Fig.\ \ref{beta-prime-noise-scaling}, which is consistent with fluctuations remaining localized due to the strong Coulomb interaction and the fluctuating electric dipole potential modulating the conductance of neighboring elements in a conducting resistor network with variable range hopping (VRH). 
This, in turn corresponds to a change of the electronic transport mechanism,
which can be inferred from the scaling behavior (same data) $S_I/I^2$ {\it vs.}\ $G$, see Fig.\ \ref{beta-prime-noise-scaling}(b). A linear behavior in such a plot is obeyed at high temperatures (regime I) and low temperatures (regime II) and implies a power-law scaling $S_\mathrm{G}/G^2 \propto G^{-w}$ as expected for a percolation scenario in a random resistor network. The scaling exponent $w$ depends on the details of the network/percolation scenario and usually is determined numerically \cite{Stauffer1994,Kogan1996}, from which in some cases the mechanism of percolative transport can be deduced, see, e.g., \cite{JMueller2009a}. A roughly linear scaling is observed both for the temperature regimes I and II, however with a significantly different slope indicating a drastic change in the electronic transport mechanism in accordance with the analysis of the low-frequency charge carrier dynamics discussed in \cite{JMueller2020}.

Regime II ($110\,{\rm K} \lesssim T \lesssim 140\,{\rm K}$) marks the transformation to the ergodic relaxor state, in which polar regions on nanometer scale with randomly distributed directions of dipole moments, the PNR, appear. (In relaxor ferroelectrics, this temperature can be twice as high as their freezing temperature giving rise to the peak in permittivity \cite{Bokov2006}.) 
This corresponds to the transformation from conductive to dielectric behavior, which is in agreement with the experimental observation that the dielectric response begins to appear in regime II \cite{Iguchi2013}, being accompanied by a slowing down of the charge fluctuations seen in the conductance noise. In \betaICl, the low-frequency noise due to switching of PNR (in regime II) and the low-frequency dispersion of the dielectric response (in regime I)
may have the same origin.

\subsection{Quasi-2D $\beta$’-(cation)[Pd(dmit)$_2$]$_2$}\label{subsec-quasi-2D-beta}
\begin{figure*}[ht]
\centering
\includegraphics[width=0.8\textwidth]{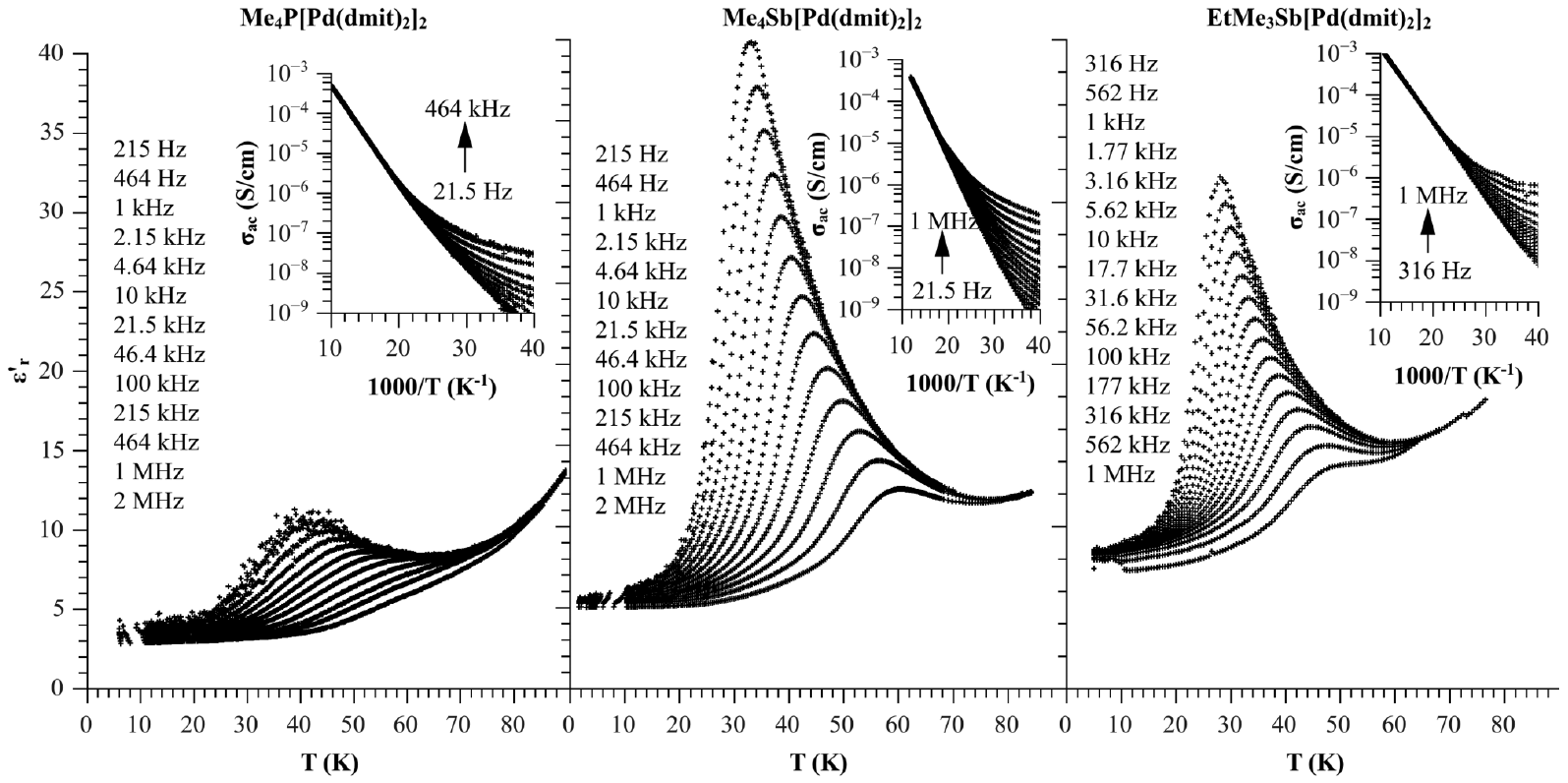}
\caption{Temperature dependence of the out-of-plane dielectric constant of $\beta$’-(cation)[Pd(dmit)$_2$]$_2$ salts measured at varying frequencies. The insets show the corresponding data for the ac conductivity. Taken from Ref.\ \cite{Abdel-Jawad2013}.}\label{beta-prime-dmit}
\end{figure*}
Figure \ref{beta-prime-dmit} shows results of the dielectric constant $\varepsilon'$($T$) measured along the out-of-plane direction of $\beta'$-(cation)[Pd(dmit)$_2$]$_2$ salts for different cations (dmit denotes 1,3-dithiol–2-thione–4,5-dithiolate) \cite{Abdel-Jawad2013}. The dielectric response is similar for all compounds investigated, yielding an anomaly in $\varepsilon'$ which is strongly frequency dependent and shows a Curie-Weiss-like increase upon cooling, similar to the response found in relaxor-type ferroelectrics (chapter \ref{subsubsecDielFerroel}) \cite{Cross1987,Samara2003,Bokov2006}. The Me$_4$P and Me$_4$Sb salts (Me = CH$_3$; left and middle frame of Fig.\,\ref{beta-prime-dmit}) are known to exhibit antiferromagnetism below $T_\mathrm{N}=40$~K and 18~K, respectively \cite{Tamura2009}, and thus can be formally regarded as multiferroic, albeit with short-range polar order only. In contrast, $\beta'$-EtMe$_3$Sb[Pd(dmit)$_2$]$_2$ (Et = C$_2$H$_5$; right frame of Fig.\,\ref{beta-prime-dmit}) was reported to be a quantum spin liquid, avoiding any magnetic order due to quantum fluctuations \cite{Watanabe2014}. In Ref. \cite{Watanabe2014}, this spin liquid state was suggested to be stabilized by the random freezing of dipolar degrees of freedom within the [Pd(dmit)$_2$]$^-$ dimers in this compound, as indicated by its dielectric behavior. Indeed, clusterlike, short-range relaxor ferroelectricity as evidenced by $\varepsilon'$($T,\nu$) of this material (Fig.\,\ref{beta-prime-dmit}) is usually assumed to involve glasslike freezing of dipole motions \cite{Cross1987,Samara2003,Bokov2006}.
However, one should be aware that the dielectric behavior of all three materials in Fig.\,\ref{beta-prime-dmit} is qualitatively similar, but only one of them was considered to exhibit a quantum spin liquid ground state. In this respect, $\beta'$-Me$_4$P[Pd(dmit)$_2$]$_2$ and $\beta'$-Me$_4$Sb[Pd(dmit)$_2$]$_2$ resemble 
$\beta'$-(ET)$_2$ICl$_2$ revealing long-range spin, but only short-range polar order (chapter \ref{subsec_betaICl}).

\section{Beyond the dimer-Mott limit -- the case of $\kappa$-phase (BEDT-TTF)$_2X$}\label{sec-beyond dimer Mott}

The observation of ferroelectric signatures in the $\kappa$ polymorph, which began in 2010 by Abdel-Jawad and collaborators \cite{Abdel-Jawad2013} reporting a relaxor-type dielectric response in the spin liquid candidate system $\kappa$-(ET)$_2$Cu$_2$(CN)$_3$, came as a surprise. Until this observation, the key aspects of these $\kappa$-phase salts, such as their magnetic properties and proximity to the Mott metal-insulator transition, could be well understood by treating the systems in the dimer-Mott limit \cite{Kino1995,Kanoda1997,Toyota2007,Powell2011}. As discussed in section \ref{sec_phase_diagrams}, this limit corresponds to the case of strong dimerization, i.e., a dominant intra-dimer hopping term $t_1$ being much larger than inter-dimer hoppings $t$ and $t^\prime$, see Fig.~\ref{fig:kappa-phase}. In such a scenario, no ferroelectric response would be expected as the intra-dimer degrees of freedom are completely frozen.

\subsection{$\kappa$-(BEDT-TTF)$_2$Hg(SCN)$_2$Cl   }\label{subsec-quasi-2D-kappa-HgCl}

We start the discussion with $\kappa$-(ET)$_2$Hg(SCN)$_2$Cl where strong evidence was provided for electronic ferroelectricity. Based on density functional theory (DFT) calculations \cite{Gati2018b}, this salt has a moderate strength of dimerization $t_1$/$t$’ $\approx $ 3, placing it in between the quarter-filled charge-ordered systems such as $\theta$-(ET)$_2$RbZn(SCN)$_4$ (undimerized) and half-filled dimer-Mott systems such as $\kappa$-(ET)$_2$Cu[N(CN)$_2$]Cl ($t_1$/$t$’ $\approx $ 6).

\begin{figure}[t]
\centering
\includegraphics[width=0.45\textwidth]{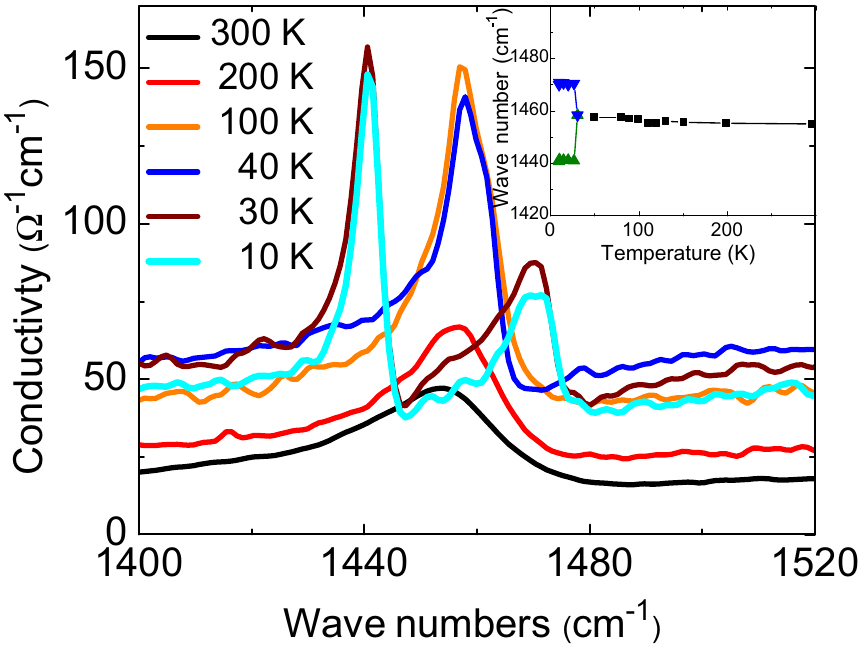}
\caption{Temperature dependence of the infrared conductivity spectra of $\kappa$-(ET)$_2$Hg(SCN)$_2$Cl measured perpendicular to the ET planes in the region of the B$_{1u}$($\nu_{27}$)  mode. The inset shows the temperature dependence highlighting a single mode above 30\,K which splits into two components below 30\,K. Taken from Ref.\,\cite{Drichko2014}.  
 }\label{kappa-HgCl-infrared}
\end{figure}

The $\kappa$-(ET)$_2$Hg(SCN)$_2$Cl salt is known to undergo a metal-insulator transition around 30\,K \cite{Aldoshina1993,Yasin2012,Drichko2014}. By measurements of the infrared reflectance, probing the charge-sensitive B$_{1u}$($\nu_{27}$) mode \cite{Yamamoto2005}, this transition was shown to be accompanied by charge order \cite{Drichko2014}. Figure \ref{kappa-HgCl-infrared} shows conductivity spectra in the frequency range around the B$_{1u}$($\nu_{27}$) mode, which is associated with the out-of-phase vibrations of the C=C bonds of the inner rings of the ET molecule.  At room temperature a single B$_{1u}$($\nu_{27}$) mode is found at 1454\,cm$^{-1}$  which splits into two components at 1441 and 1470\,cm$^{-1}$ upon cooling through the metal-insulator transition (cf. inset of Fig.\,\ref{kappa-HgCl-infrared}), indicating the appearance of two differently charged ET molecules. Based on an empirical relation between the $\nu_{27}$ frequency and the molecule’s charging state \cite{Yamamoto2005}, a charge disproportionation of 0.2\,$e$ was obtained \cite{Drichko2014}.  

\begin{figure}[t]
\centering
\includegraphics[width=0.45\textwidth]{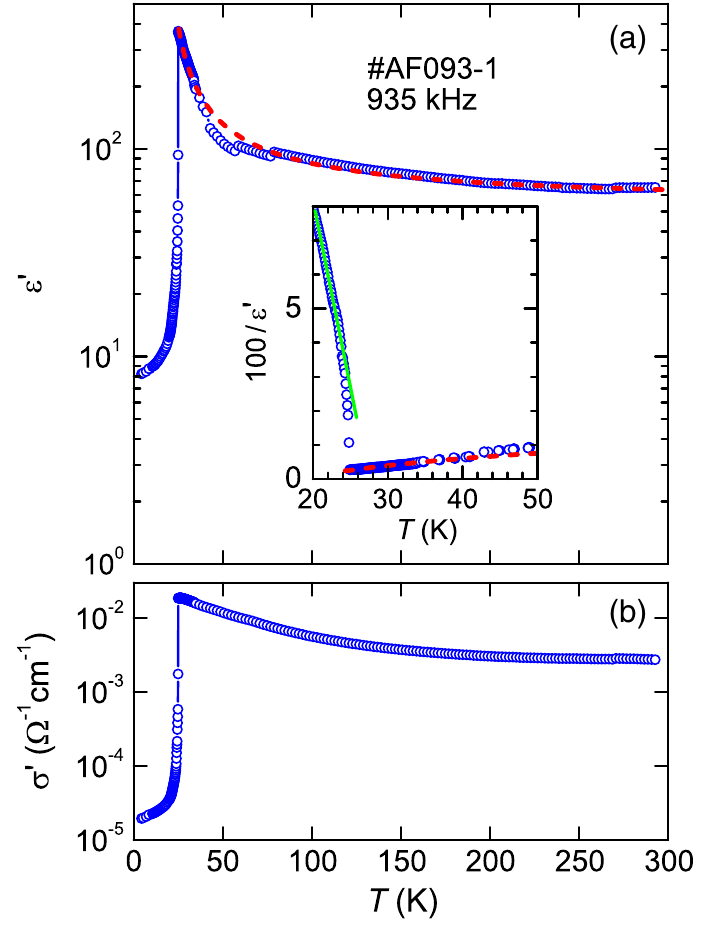}
\caption{Temperature dependence of the dielectric constant $\varepsilon'(T)$ (a) and ac conductivity $\sigma'(T)$ (b) of $\kappa$-(ET)$_2$Hg(SCN)$_2$Cl measured at 935\,kHz. The red dashed line in (a) is a Curie-Weiss fit with some offset. The inset shows the inverse dielectric constant with the lines corresponding to Curie-Weiss behavior above and below the phase transition. Taken from Ref.\,\cite{Gati2018b}.  
 }\label{kappa-HgCl-epsilon}
\end{figure}

Measurements of the dielectric constant $\varepsilon'$($T$) of this salt, performed with the electric field applied along the out-of-plane $a$-axis, show an increase upon cooling and a sharp peak at $T_\mathrm{FE} \approx$ 25\,K, indicative of a ferroelectric transition [Fig.\,\ref{kappa-HgCl-epsilon}(a)] \cite{Gati2018b}. Corresponding data for ac conductivity on the same crystal [Fig.\,\ref{kappa-HgCl-epsilon}(b)] revealed metallic behavior at higher temperatures, followed by a rapid drop at the metal-insulator transition $T_\mathrm{MI}$ around 25\,K by about three orders of magnitude. These data demonstrate the coincidence of the metal-insulator and ferroelectric transitions, i.e., $T_\mathrm{MI}$ = $T_\mathrm{FE}$, for this salt. The data for the dielectric constant, after correcting for a background contribution, were found to follow a Curie-Weiss behavior  $\varepsilon'$- $\varepsilon'$$_\mathrm{off}$ =  $C$⁄($T - T_\mathrm{CW}$), with a Curie-Weiss temperature of $T_\mathrm{CW}\approx17$\,K and a Curie constant $C\approx2500$\,K [dashed line in Fig.\,\ref{kappa-HgCl-epsilon}(a)]. Measurements of $\varepsilon'$($T$) at high frequencies up to about 1\,GHz (Supplementary Information of Ref. \cite{Gati2018b}) revealed a successive reduction of the $\varepsilon'$ peak amplitude with
increasing frequency. This behavior, along with the rather small Curie constant, is consistent with order-disorder ferroelectricity [cf. Fig.\,\ref{paper-ferroele-eps}(b)]. 

Within a simple model \cite{Mason1948} the Curie constant is related to the size of the dipoles by 

\begin{equation}
\begin{split}
\varepsilon' = \frac{C}{(T-T_\mathrm{CW})} = & \frac{1}{3 \varepsilon_0}n \left[\frac{p^{2}}{k_\mathrm{B} (T-T_\mathrm{CW})}\right]\\
&\times \left[1 + \left(\frac{T_\mathrm{CW}}{C}\right) \times \left(\varepsilon_\mathrm{L} -1\right) \right], \label{eq2}
\end{split}
\end{equation}

\noindent with $C$ the Curie constant, $T_\mathrm{CW}$ the Curie-Weiss temperature, $\varepsilon_0$ the dielectric permittivity of vacuum, $n$ the dipole density, $p$ the dipole moment, $k_\mathrm{B}$ the Boltzmann constant, and $\varepsilon_\mathrm{L}$ the low-temperature dielectric constant. This yields $p\approx$ 0.4\,$ed$, with $e$ the electronic charge and $d \approx 4.0$\,\r{A}  the distance between two ET molecules within the dimer \cite{Gati2018b}. In light of the simplifications leading to Eq.\ \eqref{eq2} and the uncertainties of the absolute value of $\varepsilon'$, this value reasonably compares with an expected out-of-plane dipole moment of 0.13\,$ed$, corresponding to the observed charge disproportionation of $\pm$0.1\,$e$ \cite{Drichko2014} and the inclination of the ET molecules away from the $a$-axis, resulting in a tilt of the dipole moment by about $50^{\circ}$ \cite{Gati2018}. It is worth mentioning that for the related $\kappa$-(ET)$_2$Hg(SCN)$_2$Br salt which shows a metal-insulator transition at $T_\mathrm{MI} = 90$\,K the situation is less clear. Here relaxor-type ferroelectricity \cite{Ivek2017} together with indications for a dipole liquid \cite{Hassan2018}, effects of disorder \cite{Le2020}, a spin-glass type ground state \cite{Hemmida2018} as well as weak ferromagnetic signatures \cite{Hemmida2018,Yamashita2021} were reported.

\begin{figure}[t]
\centering
\includegraphics[width=0.475\textwidth]{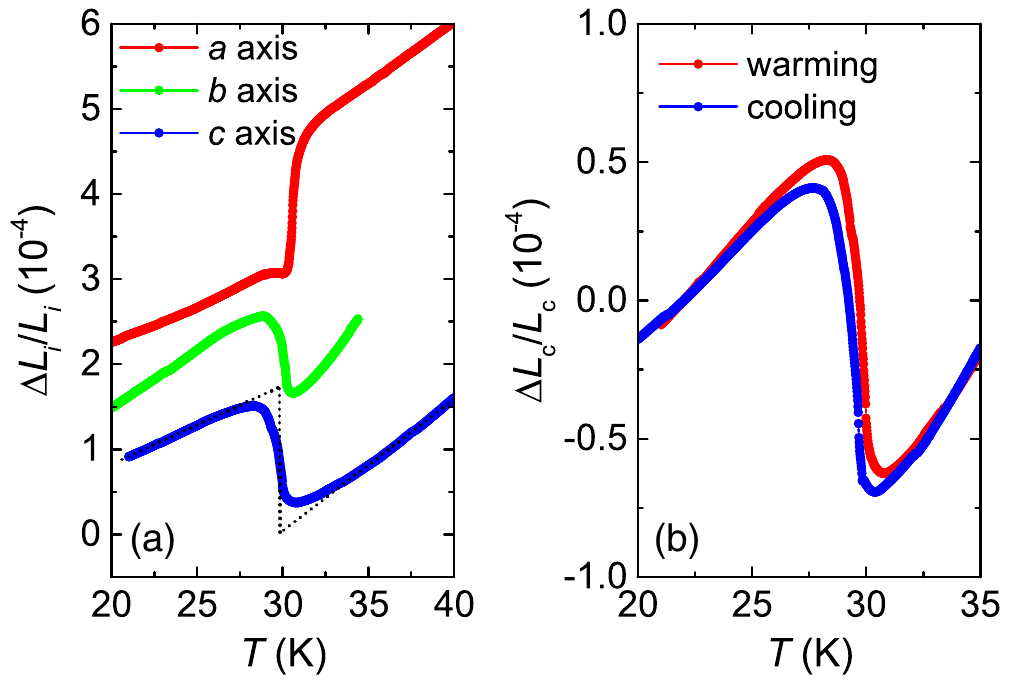}
\caption{(a) Relative length change $\Delta L_i$/$L_i$ as a function of temperature along the $i = a, b, c$ axis of $\kappa$-(ET)$_2$Hg(SCN)$_2$Cl around the charge-order metal-insulator transition. The individual data sets were shifted for clarity. Dotted line indicates an idealized sharp jump for the $c$-axis data. (b) Relative length change along the $c$ axis measured upon warming and cooling. Taken from Ref.\,\cite{Gati2018}.  
 }\label{kappa-HgCl-delta L}
\end{figure}

Measurements of the relative length change $\Delta L_\mathrm{i}$/$L_\mathrm{i}$ along the out-of-plane $a$-axis and the in-plane $b$- and $c$-axes revealed slightly broadened jumps at $T_\mathrm{MI}$ = $T_\mathrm{FE}$ [Fig.\,\ref{kappa-HgCl-delta L}(a)] with a small hysteresis between warming and cooling [Fig.\,\ref{kappa-HgCl-delta L}(b)], demonstrating the first-order character of the phase transition. This is consistent with the dielectric response yielding a Curie-Weiss behavior both above and below $T_\mathrm{FE}$ with strongly different slopes $|$d(1⁄$\varepsilon'$)⁄(d$T$)$|$ [cf. inset of Fig.\,\ref{kappa-HgCl-epsilon}(a)] together with a Curie-Weiss temperature $T_\mathrm{CW} < T_\mathrm{FE}$ \cite{Lines2001}.

\begin{figure}[t]
\centering
\includegraphics[width=0.4\textwidth]{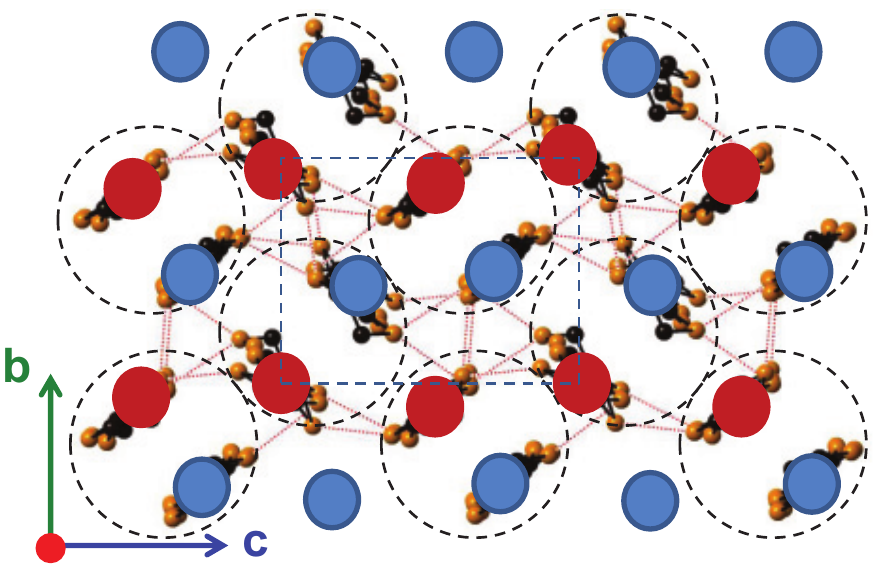}
\caption{Favored  charge-order pattern for  $\kappa$-(ET)$_2$Hg(SCN)$_2$Cl below $T_\mathrm{MI}$ = $T_\mathrm{FE}$ based on the anisotropy of infrared optical conductivity data \cite{Drichko2014} and consistent with thermal expansion data \cite{Gati2018}. Red (blue) spheres indicate charge rich (poor) ET molecules. Taken from Ref.\,\cite{Drichko2014}. 
 }\label{kappa-HgCl-charge order pattern}
\end{figure}

The observed lattice effects at the phase transition highlight a particularly strong effect along the out-of-plane $a$-axis, indicating an involvement of the anion layer in the formation of the charge-ordered state. In fact, given the ionic character of these salts, a change in the charge distribution within the ET layers will inevitably be accompanied by shifts of the counter ions in the anion layers. Based on the observed anisotropic lattice effects, a charge-order pattern was suggested where the charge-rich molecules are arranged in stripes along the $c$-axis and alternate with charge-poor stripes along the $b$-axis, cf. Fig.\,\ref{kappa-HgCl-charge order pattern} \cite{Drichko2014}. This is the horizontal stripe pattern depicted in Fig.\ \ref{fig:kappa-phase}. This pattern, which is consistent with the infrared-conductivity spectra in Ref.\ \cite{Drichko2014}, breaks inversion symmetry both within and between the layers, and thereby enables three-dimensional ferroelectricity to form.

The presence of spin degrees of freedom in the related dimerized salts $\beta$’-(ET)$_2$ICl$_2$ and $\kappa$-(ET)$_2$Cu[N(CN)$_2$]Cl, giving rise to antiferromagnetic order, raises the question about the interaction of charge order with the magnetic degrees of freedom in $\kappa$-(ET)$_2$Hg(SCN)$_2$Cl. In fact, based on previous electron spin resonance (ESR) measurements \cite{Yasin2012}, it was suggested that antiferromagnetic order accompanies the charge order below $T_\mathrm{FE}$ in $\kappa$-(ET)$_2$Hg(SCN)$_2$Cl. However, detailed ESR studies together with specific heat measurements performed in Ref.\ \cite{Gati2018} failed to detect any signatures for a magnetic transition around $T_\mathrm{MI}$. It has been suggested in Ref.\ \cite{Gati2018} that as a consequence of the charge order the magnetic interactions are modified, thereby promoting a quasi-1D spin liquid state. See section \ref{sec_phase_diagrams} for a further discussion on this issue.     

Fluctuation spectroscopy measurements of $\kappa$-(ET)$_2$Hg(SCN)$_2$Cl \cite{Thomas2019} revealed a strong change of the magnitude of resistance fluctuations of at least three orders of magnitude coinciding with the charge ordering transition, in agreement with intrinsic electronic inhomogeneities inherent to the first-order transition. Entering the ferroelectric phase upon cooling is accompanied by a shift of spectral weight to lower frequencies. Only below the metal-insulator transition the normalied noise spectra exhibit an anomalous current dependence, where with increasing electric field a discrete shift of the $1/f$-type spectra or a change of the amplitude and corner frequency of Lorentzian spectra were observed,  similar to the findings on other $\kappa$-phase compounds discussed in this review. Also, the second spectrum shows a frequency dependence only for temperatures below $T_{\rm MI}$, implying non-Gaussian, spatially correlated fluctuations.

\subsection{$\kappa$-(BEDT-TTF)$_2$Cu$_2$(CN)$_3$  }\label{subsec-quasi-2D-kappa-CuCN}
\begin{figure}[t]
\centering
\includegraphics[width=0.475\textwidth]{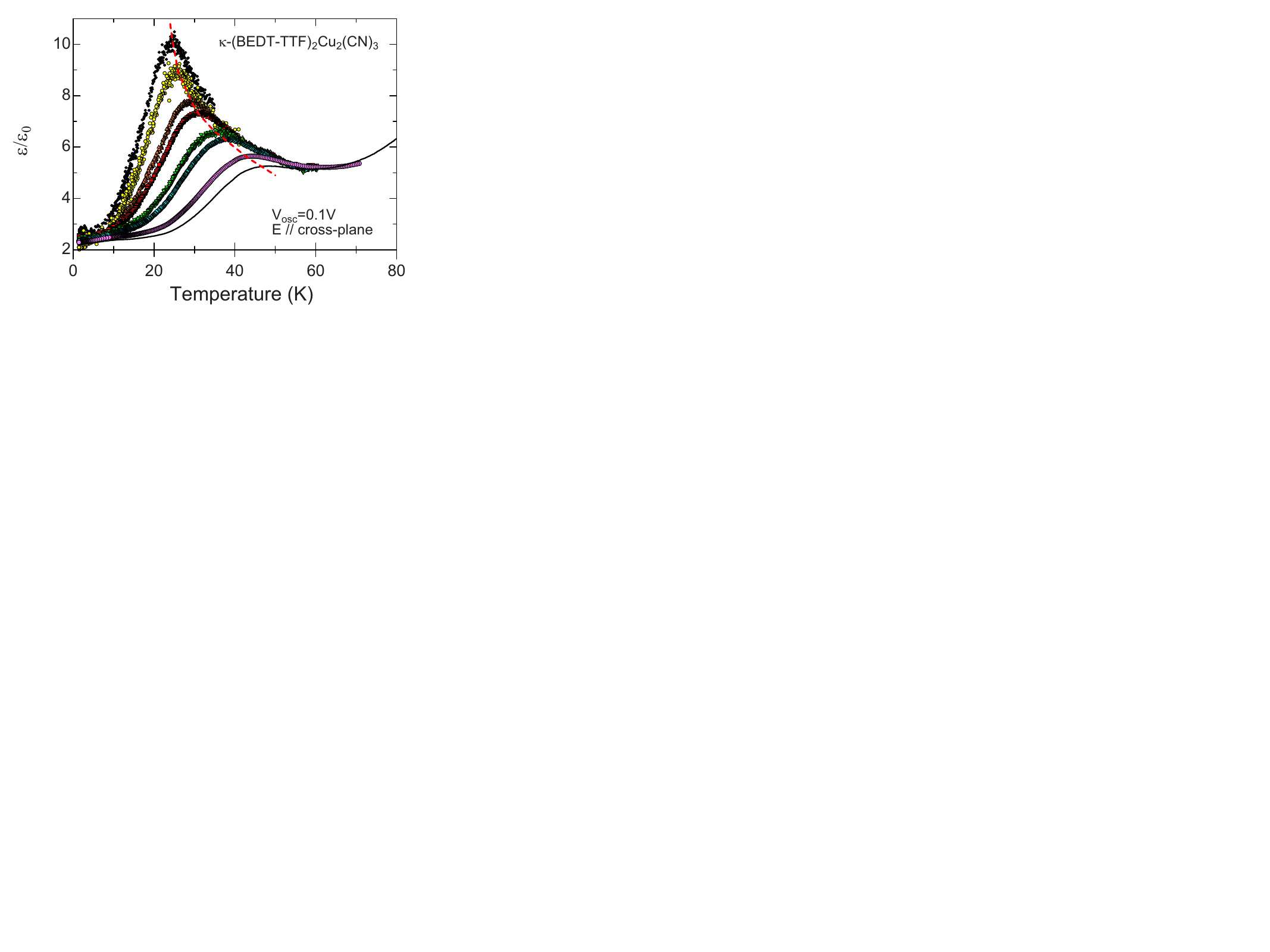}
\caption{Normalized dielectric constant of $\kappa$-(ET)$_2$Cu$_2$(CN)$_3$ measured with electric field aligned along the $a$ axis, i.e., perpendicular to the ET planes for varying frequencies. The broken line connects the maximum positions. Reproduced from Ref.\ \cite{Abdel-Jawad2010}. 
 }\label{kappa-Cu2CN3-epsilon}
\end{figure}

First indications for a ferroelectric response in dimerized quasi-2D organic charge-transfer salts were observed in $\kappa$-(ET)$_2$Cu$_2$(CN)$_3$ by Abdel-Jawad \textit{et al.} \cite{Abdel-Jawad2010} in measurements of the dielectric constant, see Fig.\,\ref{kappa-Cu2CN3-epsilon}. This salt had been considered for a long time as a candidate system for the realization of a quantum spin liquid \cite{Shimizu2003}. This notion was initially supported by this salt being a weak dimer-Mott insulator, i.e., electronic states which are close to the metal-insulator transition, featuring a triangular lattice with a high degree of frustration $t$’/$t$ = 0.84 which increases to 0.86 at 5\,K \cite{Jeschke2012} (see Table \ref{model parameters}). However, the ultimate zero-temperature ground state of this salt has been challenged recently on the basis of new experimental and theoretical insights \cite{Riedl2019,Miksch2021}, indicative of the formation of spin singlets in an ordered or glassy manner at low temperatures, see \cite{Pustogow2022} for a recent comprehensive review. 

Measurements of the dielectric constant along the out-of plane $a$ axis (Fig.\,\ref{kappa-Cu2CN3-epsilon}) \cite{Abdel-Jawad2010} revealed a peak anomaly below about 60\,K, the position and height of which showed a pronounced frequency dependence typical for a relaxor-type ferroelectric; see also Ref.\,\cite{Pinteric2014} for fields aligned parallel to the ET planes. The peak anomaly was found to follow to a good approximation a VFT law $\nu = \nu_0 \exp [-B/(T_\mathrm{max} - T_\mathrm{VFT})]$, corresponding to eq. (\ref{Vogel-Fulcher}) for $\nu=1/(2\pi\tau)$, again typical for relaxor ferroelectrics as discussed in chapter \ref{subsubsecDielFerroel} \cite{Cross1987,Samara2003,Bokov2006}. Here $\nu$ is the frequency, $\nu_0$ a pre-exponential factor, and $T_\mathrm{max}$ is the position of the maximum in $\varepsilon'$. With $T_\mathrm{VFT} \approx 6$\,K a good description of the data was achieved.

\begin{figure}[t]
\centering
\includegraphics[width=0.475\textwidth]{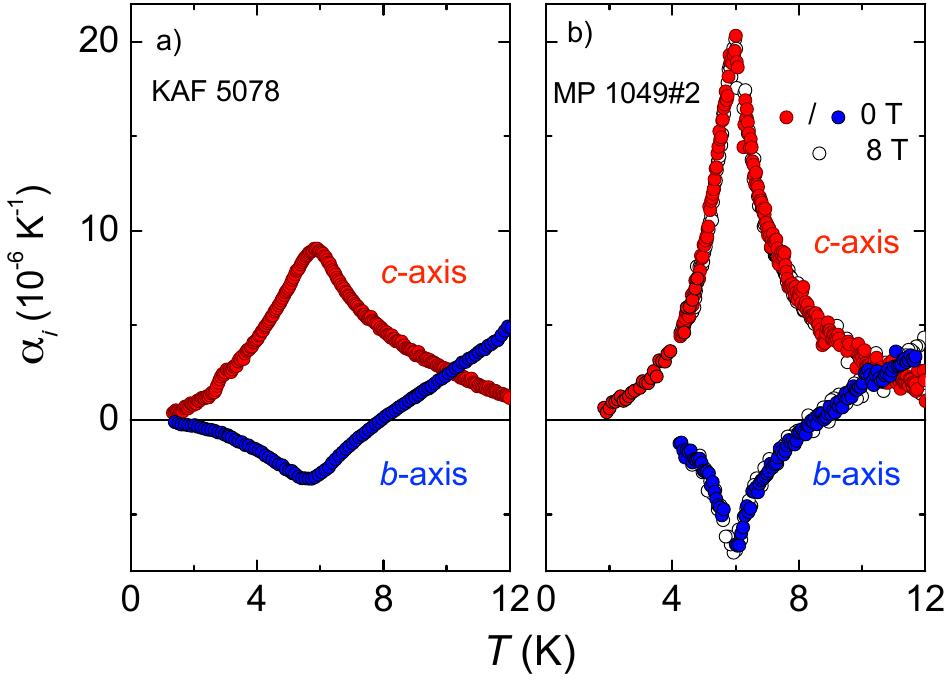}
\caption{Thermal expansion coefficients for two different crystals of $\kappa$-(ET)$_2$Cu$_2$(CN)$_3$ measured along the in-plane $b$ and $c$ axis around the 6\,K phase transition anomaly. The data in (a) were reported in Ref.\,\cite{Manna2010}, the data in (b), where the anomalies are most strongly pronounced, were reported in Ref.\,\cite{Manna2018}.
 }\label{kappa-Cu2CN3-thermal expansion}
\end{figure}

Moreover, as shown in Ref.\ \cite{Abdel-Jawad2010}, the high-temperature envelope curve of the $\varepsilon'$ peaks for the various frequencies, representing the static dielectric constant, was found to follow a Curie-Weiss law with a Curie-Weiss temperature of $T_\mathrm{CW}$ = 6\,K providing a rough estimate of the freezing temperature. It is interesting to note that at this temperature anomalous behavior was observed in a variety of properties --- commonly referred to as the ``\textit{6\,K anomaly}'', see Ref.\ \cite{Pustogow2022}. This anomaly is particularly strongly pronounced in the thermal expansion \cite{Manna2010} indicating a strong coupling to the lattice degrees of freedom. While the thermal expansion results provided clear indications for a phase transition at 6\,K, the size of the phase transition anomaly, rather than its position, revealed significant sample-to-sample variations \cite{Manna2018}, cf. Fig.\,\ref{kappa-Cu2CN3-thermal expansion}, highlighting the role of disorder.  

\begin{figure}[t]
\centering
\includegraphics[width=0.475\textwidth]{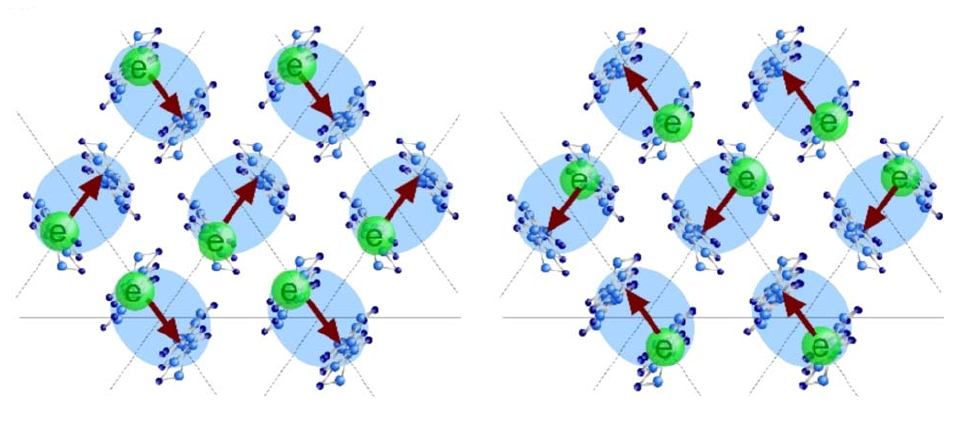}
\caption{Possible short-range domains of collectively fluctuating quantum electric dipoles (arrows) resulting from intra-dimer charge disproportionation. Taken from Ref.\,\cite{Abdel-Jawad2010}.
 }\label{kappa-Cu2CN3-fluctuating dipols}
\end{figure}

The scenario put forward in Ref.\ \cite{Abdel-Jawad2010}, and also discussed by others \cite{Naka2010,Hotta2010,Clay2010,Gomi2013}, involved fluctuating charges (holes) on the dimer site (see Fig.\,\ref{kappa-Cu2CN3-fluctuating dipols}), corresponding to fluctuating dipoles which give rise to the observed relaxational response. The relaxor ferroelectricity revealed at low temperatures thus can be assigned to a finite intra-dimer charge disproportionation, driven by inter-molecular Coulomb interactions, combined with the influence of disorder. 

Spectroscopic studies aiming to identify the origin of the ferroelectric signatures came to conflicting conclusions. Whereas infrared vibrational-spectroscopy studies failed to detect clear indications for a line splitting and therefore suggest the absence of charge disproportionation \cite{Sedlmeier2012}, Raman scattering revealed a clear line broadening consistent with intra-dimer charge fluctuations \cite{Yakushi2015}. 
Strong evidence for fluctuating intra-dimer charge and spin fluctuations were found in recent inelastic neutron scattering experiments by taking advantage of the strong coupling of these fluctuations to certain intra-dimer vibrational modes \cite{Matsuura2022}. In this study a compilation of 47 co-aligned crystals of deuterated  $\kappa$-(D$_8$-ET)$_2$Cu$_2$(CN)$_3$ were investigated with particular focus lying on low-energy breathing modes of the ET dimers. A drastic change was observed in the phonon damping for a low-lying optical mode at an energy $E$ = 4.7\,meV, assigned to an intra-dimer breathing mode \cite{Dressel2016}, upon cooling below 6\,K.  It was argued that the drastic increase in the lifetime of this mode below 6\,K can be attributed to a phase transition involving the lattice and its coupling to the charge and spin degrees of freedom \cite{Matsuura2022}.    

\begin{figure}[t]
\centering
\includegraphics[width=0.425\textwidth]{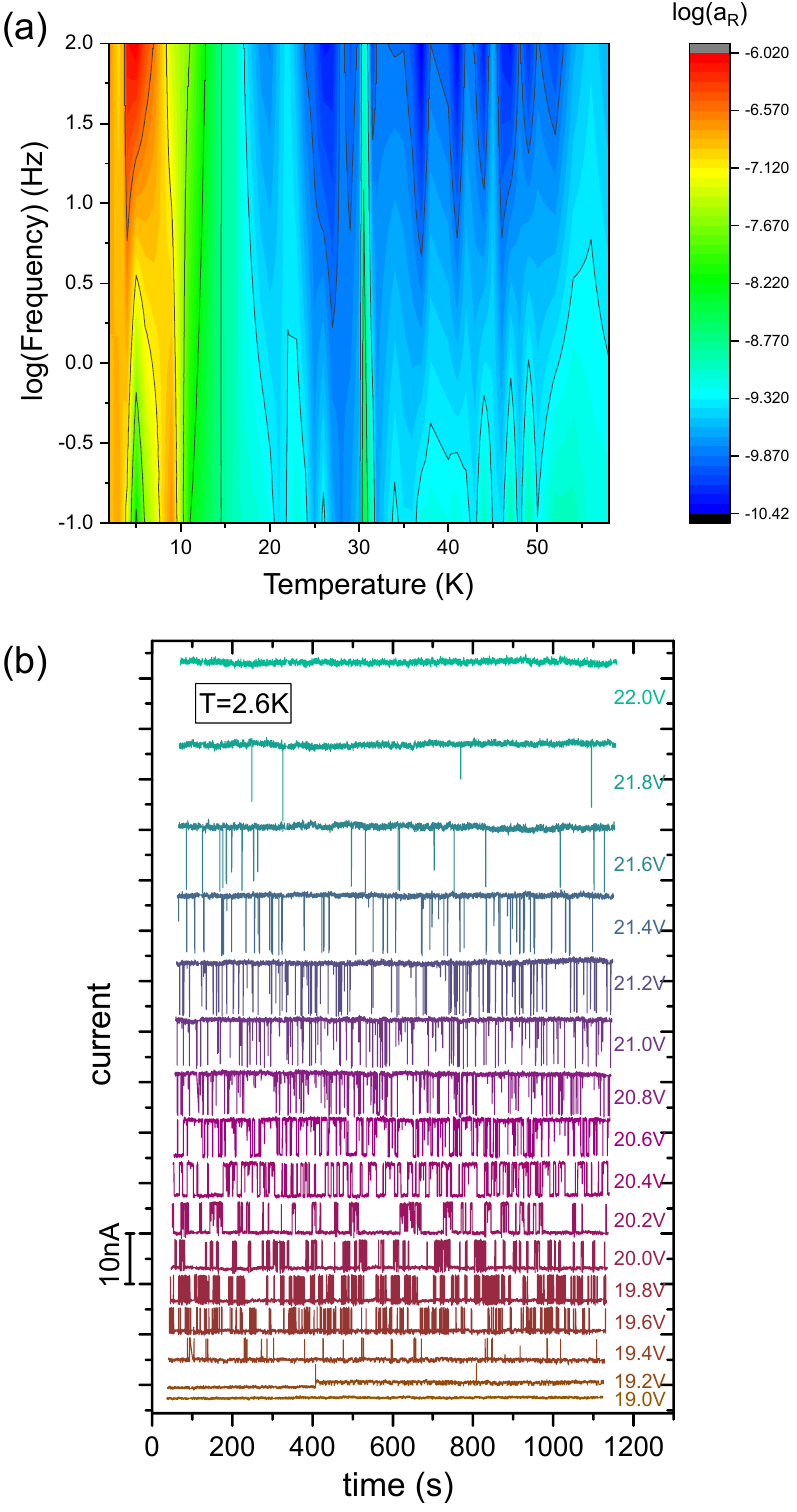}
\caption{(a) Relative conductance noise level $\log(a_\mathrm{R} = f \times S_I/I^2)$ vs.\ $T$ vs.\ $\log(f)$ for $\kappa$-(ET)$_2$Cu$_2$(CN)$_3$. (b) Two-terminal current fluctuations at $T = 2.6$\,K for different applied voltages.}
\label{kappa-Cu2CN3-noise}
\end{figure}
Also for $\kappa$-(ET)$_2$Cu$_2$(CN)$_3$, strongly enhanced low-frequency fluctuations around 6\,K and TLS coupled to the electric field have been observed.
Figure \ref{kappa-Cu2CN3-noise}(a) displays the dimensionless normalized relative conductance noise level $a_\mathrm{R}(T,f) = f \times S_I/I^2(T,f)$ showing a strong increase of the $1/f$-type fluctuations upon cooling to temperatures below about 25\,K which can be roughly described by a model of variable range hopping \cite{Hartmann2017}. Note that there is a pronounced peak in the noise, most prominent at the higher end ($f > 10$\,Hz) of the measured frequency range at 6\,K, i.e.\ yet another anomaly at this temperature which is compatible with a phase transition at this temperature. A nearly frequency-independent sharp peak in the noise above 30\,K coincides with a change in the activation energy of the conductivity in a Mott variable range hopping model.
Related to the dielectric properties is the time train of the measured current at different voltages for $T = 2.6$\,K, i.e.\ deep in the relaxor ferroelectric state, shown in Fig.\,\ref{kappa-Cu2CN3-noise}(b). Interestingly, in a certain interval of the applied electric field $\mathcal{E}$, we observe distinct two-level fluctuations, where the relative lifetime of the states are found to be inverted with varying $\mathcal{E}$. Such a behavior is in agreement with a thermally-activated switching or polar clusters, described for \betaICl, $\kappa$-(ET)$_2$Cu[N(CN)$_2$]Cl and $\kappa$-(BETS)$_2$Mn[N(CN)$_2$]$_3$, which is described for the latter salts in detail in sections \ref{subsec_betaICl_noise}, \ref{subsec_kCL-noise} and \ref{subsec-quasi-2D-kappa-X}, respectively

\subsection{$\kappa$-(BEDT-TTF)$_2$Cu[N(CN)$_2$]Cl}\label{subsec-quasi-2D-kappa-Cl}
\subsubsection{Results from dielectric spectroscopy}
This salt represents a particularly interesting case where strong indications for ferroelectricity were observed which occur simultaneously with the transition to antiferromagnetic order \cite{Lunkenheimer2012}. This finding was remarkable for two reasons: first, it establishes this material as another multiferroic charge-transfer salt beside the above-mentioned (TMTTF)$_2$SbF$_6$  and $\beta$’-(ET)$_2$ICl$_2$. Second, and in contrast to the latter two cases, where the ferroelectric and magnetic order set in at distinctly different temperatures, here the simultaneous occurrence of both orders suggests an intimate coupling between ferroelectricity and magnetism, as reviewed in section \ref{sec_phase_diagrams}. The
$\kappa$-(ET)$_2$Cu[N(CN)$_2$]Cl salt was investigated intensively in the past due to its special character as a weak dimer-Mott insulator, featuring an anisotropic triangular lattice, in close proximity to unconventional superconductivity \cite{Kanoda1997,Toyota2007,Powell2011,Riedl2019}. The system shows long-range antiferromagnetic order below $T_\mathrm{N}$ = 27\,K \cite{Miyagawa1995} consistent with the material’s moderate degree of frustration $t$’/$t$ $\approx$ 0.46 \cite{Kandpal2009,Koretsune2014} (see Table \ref{model parameters}). The pressure-temperature phase diagram (Fig.\,\ref{kappa-Cl-phase-diagram}) has been studied in great detail by numerous groups \cite{Lefebvre2000,Limelette2003,Fournier2003,Kagawa2005,Gati2016}, especially with regard to the Mott transition and its critical behavior, justifying the notion of this material representing a good realization of the dimer-Mott scenario. In fact, the degree of dimerization here amounts to $t_1$/$t$’ = 6.2 \cite{Koretsune2014}  exceeding the values of $t_1$/$t$’ =  4.3 for $\kappa$-(ET)$_2$Cu$_2$(CN)$_3$  and $t_1$/$t$’ = 3.1 for $\kappa$-(ET)$_2$Hg(SCN)$_2$Cl (cf. Table \ref{model parameters}).

\begin{figure}[t]
\centering
\includegraphics[width=0.475\textwidth]{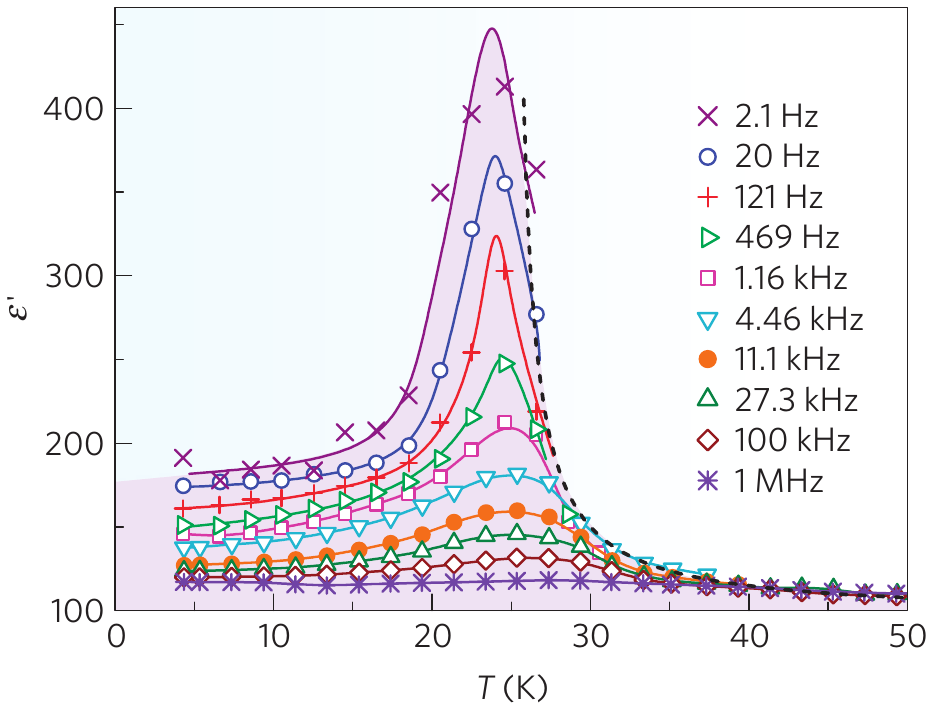}
\caption{Temperature dependence of the dielectric constant for electric fields aligned perpendicular to the ET planes of $\kappa$-(ET)$_2$Cu[N(CN)$_2$]Cl. Data were taken at varying frequencies. The dashed line indicates Curie-Weiss behavior with a Curie-Weiss temperature $T_\mathrm{CW}$ = 25\,K. Taken from Ref.\,\cite{Lunkenheimer2012}.
 }\label{kappa-Cl-epsilon}
\end{figure}

\begin{figure}[t]
\centering
\includegraphics[width=0.45\textwidth]{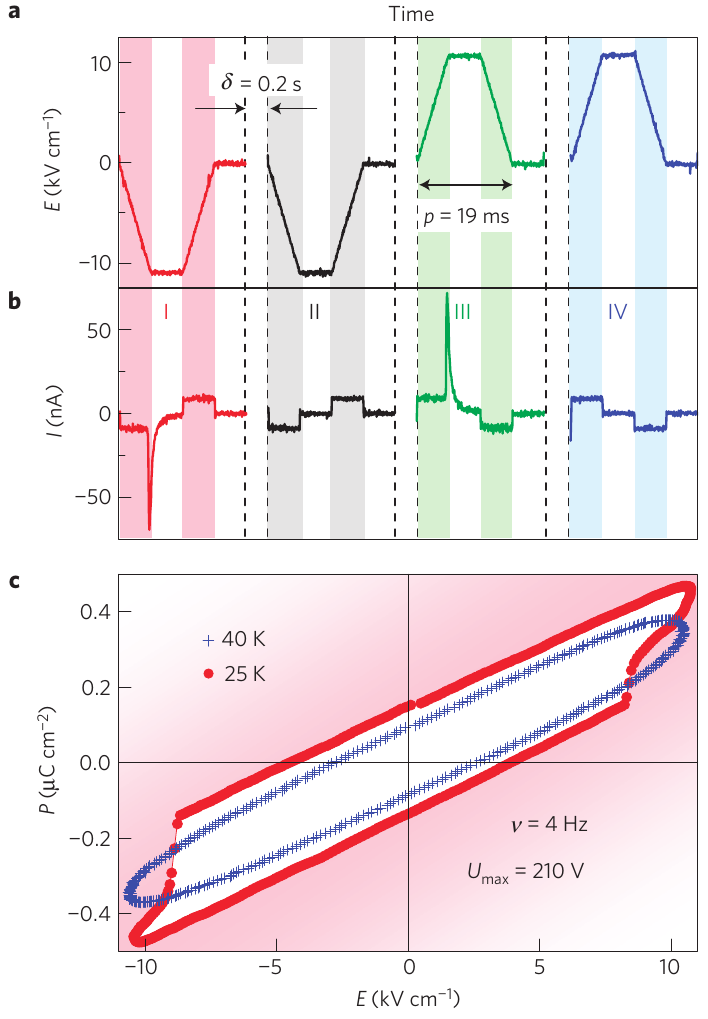}
\caption{(a) Time-dependent excitation signal with electric field perpendicular to the ET planes of the PUND measurements for $\kappa$-(ET)$_2$Cu[N(CN)$_2$]Cl. The data were taken at 25\,K. (b) Resulting time-dependent current with spikes in response to the pulses I and III. (c) Polarization-field hysteresis curves taken at temperatures above (crosses) and just below (red spheres) the occurrence of ferroelectric order. Taken from Ref.\,\cite{Lunkenheimer2012}. 
 }\label{kappa-Cl-PUND}
\end{figure}

It therefore came as a surprise that measurements of the dielectric constant on this salt, performed for electric fields perpendicular to the ET planes, revealed strong indications for ferroelectricity \cite{Lunkenheimer2012}. 
As seen in Fig.\,\ref{kappa-Cl-epsilon}, the $\varepsilon'(T)$ data exhibit a peak around 25\,K which is most strongly pronounced at the lowest frequency. It becomes reduced in size with increasing frequency whereas its position remains practically unaffected. The signatures revealed in these investigations bear the characteristics of an order-disorder-type ferroelectric [cf. Fig.\,\ref{paper-ferroele-eps}(b)]. In Ref.\,\cite{Lunkenheimer2012}, the polar state realized here has been assigned to charge order within the ET dimers \cite{Lunkenheimer2012}, although alternative interpretations have been proposed \cite{Pinteric1999,Tomic2013,Pinteric2015}, see Ref.\,\cite{Lunkenheimer2015a} for a critical discussion. It should be mentioned that so far, attempts to experimentally resolve the intra-dimer charge disproportionation have remained unsuccessful \cite{Sedlmeier2012}, indicating that charge order, if present, is rather weak.   
Further strong evidence for ferroelectricity was provided by PUND measurements (see chapter \ref{subsecPolarization}) where a peak-like current response was observed when the electric field $|E|$ exceeded a threshold field of about 10\,kV/cm for the first and third pulse, indicating the switching of ferroelectric domains, see Fig.\,\ref{kappa-Cl-PUND}(b) \cite{Lunkenheimer2012}. These peaks were absent for the second and fourth pulse as the domains were already switched by the proceeding pulse. Moreover, measurements of the field-dependent polarization $P$($E$) [Fig.\,\ref{kappa-Cl-PUND}(c)] \cite{Lunkenheimer2012} revealed an elliptical curve at 40\,K, i.e., above $T_\mathrm{FE}$, consistent with a linear polarization response (paraelectricity) with some loss contributions from charge transport. In contrast, at lower temperatures, just below the temperature at which ferroelectric order sets in, non-linear behavior was observed above about 8.5\,kV/cm with the tendency to saturation at highest fields, indicative of ferroelectric polarization switching, cf. Fig.\,\ref{pap-PE_PUND}(a).

\begin{figure}[t]
\centering
\includegraphics[width=0.425\textwidth]{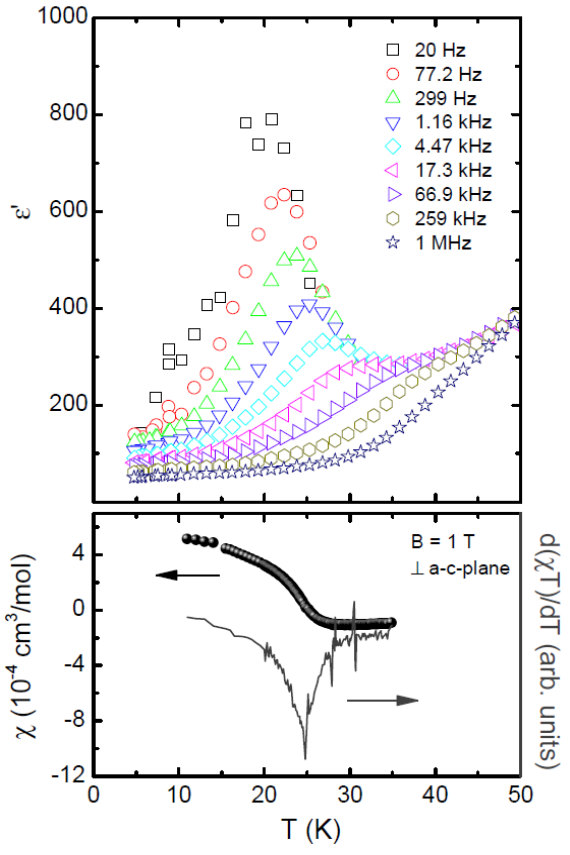}
\caption{Temperature dependence of the out-of-plane dielectric constant of single crystalline $\kappa$-(ET)$_2$Cu[N(CN)$_2$]Cl (different sample than in Fig.\,\ref{kappa-Cl-epsilon}) measured at various frequencies. The lower panel shows the magnetic susceptibility (left scale) together with the derivative d($\chi \cdot T$)/d$T$ (right scale) on the same crystal. The susceptibility was taken at a field of 1\,T applied perpendicular to the ET planes. Taken from Ref.\,\cite{Lang2014}. } \label{kappa-Cl-sample dependence}
\end{figure}

In an attempt to look for sample-to-sample variations with regard to the dielectric response, overall eight $\kappa$-(ET)$_2$Cu[N(CN)$_2$]Cl single crystals from different sources were investigated \cite{Lang2014}. The dielectric response was measured for electric fields perpendicular to the ET planes. The crystals were also subject to a magnetic characterization using a SQUID magnetometer for identifying the transition into antiferromagnetic order. Two different types of dielectric response were revealed. One group of crystals (five out of eight) showed the order-disorder-type dielectric response [cf. Fig.\,\ref{paper-ferroele-eps}(b)] as, e.g., seen in Fig.\,\ref{kappa-Cl-epsilon} discussed above, characterized by a pronounced peak in $\varepsilon'$($T$), the position of which is practically independent of the frequency. The second group of crystals (three out of eight) revealed a significant shift of the peak position to higher temperatures with increasing frequency, typical for relaxor-type ferroelectricity [Fig.\,\ref{paper-ferroele-eps}(c)], as examplarily shown in the upper panel of Fig.\,\ref{kappa-Cl-sample dependence}. Despite their different dielectric response, very similar magnetic behavior was found, yielding a steplike anomaly in the magnetic susceptibility, corresponding to a spike in d($\chi \cdot T$)/d$T$, around 25\,K, cf. lower panel of Fig.\,\ref{kappa-Cl-sample dependence}, consistent with literature results on the magnetic order in this salt \cite{Miyagawa1995}. The study showed that in all cases clear ferroelectric signatures occur around the magnetic transition while sample-specific effects, likely originating from different crystal growth conditions, affect the dielectric properties more than the magnetic ones \cite{Lang2014} changing the order-disorder type ferroelectric response into a relaxor-type one.

\subsubsection{Results from inelastic neutron scattering}\label{subsec_kCL-neutron}
\begin{figure}[!ht]
\centering
\includegraphics[width=0.45\textwidth]{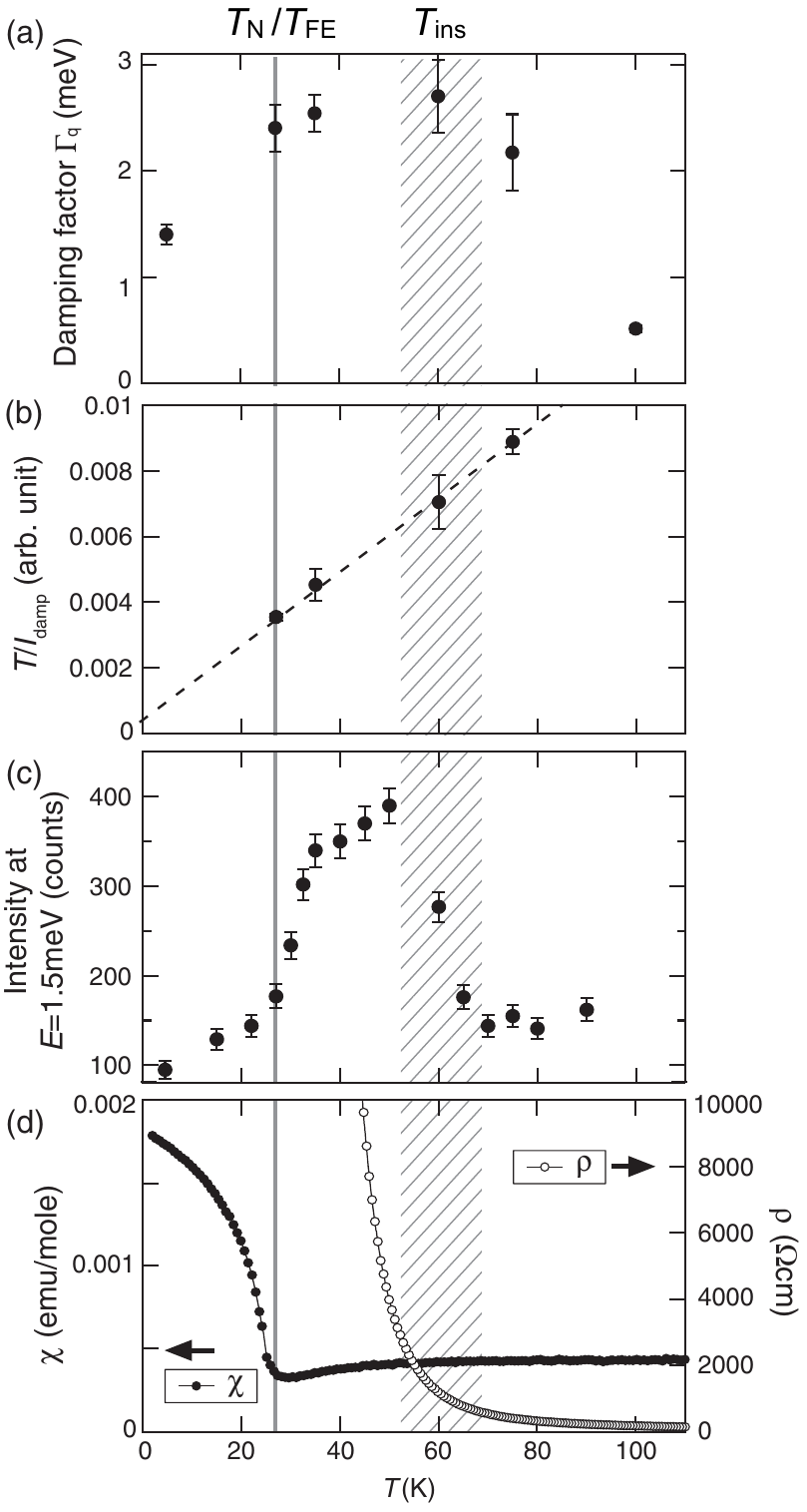}
\caption{Temperature dependence of the damping factor $\Gamma_\mathrm{q}$ (a) for a low-lying intra-dimer breathing mode at
$E$ = 2.6\,meV of $\kappa$-(D$_8$-ET)$_2$Cu[N(CN)$_2$]Cl measured at a momentum transfer \textbf{Q} = 
$(603)$, (b) $T$/$I_\mathrm{damp}$ with $I_\mathrm{damp}$ the integrated intensity of the overdamped mode. The finite
$T$/$I_\mathrm{damp}$ at $T_\mathrm{N}$ = $T_\mathrm{FE}$ indicates that a structural change is not the primary order 
parameter for the phase transition at this temperature. (c) Scattering intensity measured at an energy
$E$ = 1.5\,meV at $(603)$, (d) the out-of-plane electrical resistivity $\rho$($T$) (right scale) 
and magnetic susceptibility $\chi$($T$) (left scale). The hatched area marks the crossover temperature 
$T_\mathrm{ins}$, corresponding to the opening of the charge gap, and $T_\mathrm{N}$ = $T_\mathrm{FE}$ marks the temperature 
of orderings in the spin and charge channel. 
Taken from Ref.\,\cite{Matsuura2019}.} \label{kappa-Cl-neutrons}

\end{figure}

The scenario that has been put forward based on the above phenomenology \cite{Lunkenheimer2012,Lang2014} involves fluctuating charges on the ET dimers above $T_\mathrm{FE}$, corresponding to fluctuating dipoles, and the polar ordering (long range or short range, depending on sample) of these fluctuations below $T_\mathrm{FE}$. In light of significant electron-lattice coupling characterizing these salts, the different dynamics of the charge should also be reflected in the dynamics of certain lattice modes, in particular the intra-dimer breathing modes, which can be sensitively probed by inelastic neutron scattering. 
By using a collection of co-aligned deuterated single crystals $\kappa$-(D$_8$-ET)$_2$Cu[N(CN)$_2$]Cl with a total mass of 9\,mg, the dynamics of low-lying vibrational modes were investigated \cite{Matsuura2019}. In fact, strong renormalization effects could be observed for the low-lying mode at an energy $E$ = 2.6\,meV, which was assigned to an intra-dimer breathing mode, cf. Fig.\,\ref{neutron-dimer}.
Figure \ref{kappa-Cl-neutrons} shows characteristic phonon parameters, obtained from fitting the phonon peak at $E$ = 2.6\,meV by a damped-harmonic oscillator function [cf.\,eq.\,(\ref{DHO})], and their variation with temperature. The data reveal a strong increase of the damping factor $\Gamma_\mathrm{q}$, the inverse of the phonon lifetime, when the charge carriers become localized on the dimer site below $T_\mathrm{ins}$, reflected by the rapid increase in the resistivity [cf. Fig.\,\ref{kappa-Cl-neutrons}(d)]. The damping becomes reduced, once static order develops below $T_\mathrm{N}$ = $T_\mathrm{FE}$ = 25\,K. The phonon anomaly and its peculiar temperature dependence becomes clearly visible in the scattering intensity taken at an energy $E$ = 1.5\,meV [Fig.\,\ref{kappa-Cl-neutrons}(c)], which strongly increases around $T_\mathrm{ins}$ and decrease near $T_\mathrm{N}$ = $T_\mathrm{FE}$. It was argued in Ref. \cite{Matsuura2019} that such a phonon renormalization effect reflects the strong coupling to some relaxational mode. This is the case when the characteristic frequency and wave vector of this relaxational mode matches with the corresponding figures of those phonon modes. The coincidence of the phonon anomaly and $T_\mathrm{ins}$, below which the charge gap opens \cite{Kornelsen1992,Sasaki2004a}, strongly suggests a coupling between the breathing mode and the intra-dimer charge degrees of freedom. This is consistent with the drop in the intensity [Fig.\,\ref{kappa-Cl-neutrons}(c)] and the decrease in the damping [Fig.\,\ref{kappa-Cl-neutrons}(a)] below $T_\mathrm{N}$ = $T_\mathrm{FE}$ = 25\,K, as a decoupling of the modes is expected due to the critical slowing down of the charge and spin fluctuations.

\subsubsection{Results from fluctuation spectroscopy}\label{subsec_kCL-noise}

\begin{figure}[!ht]
\centering
\includegraphics[width=0.425\textwidth]{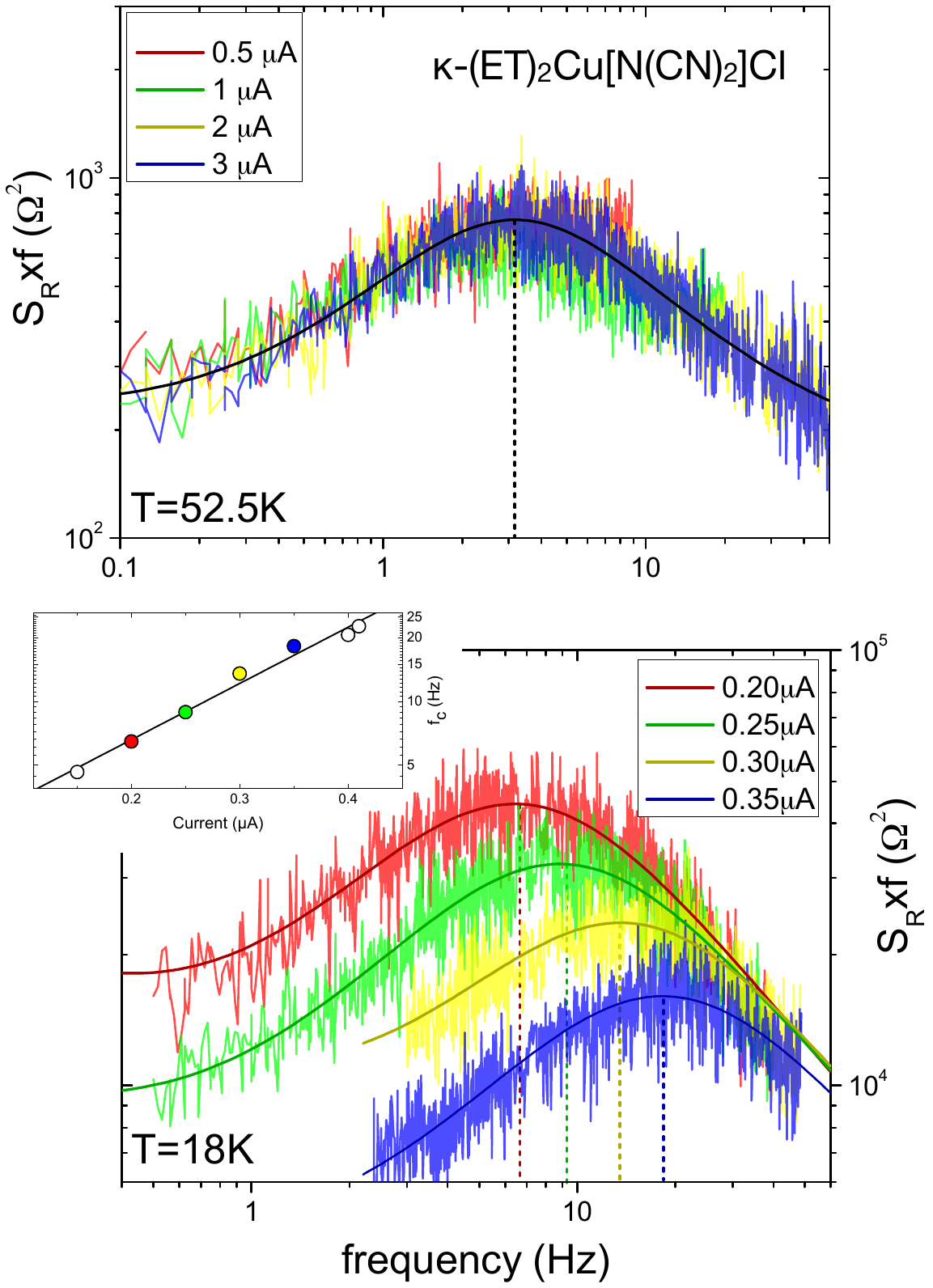}
\caption{Current dependence of the resistance noise PSD of $\kappa$-(ET)$_2$Cu[N(CN)$_2$]Cl, showing a characteristic Lorentzian contribution due to changes in the charge carrier dynamics, above (upper panel) and below (lower panel) the Néel temperature $T_\mathrm{N}$, which --- for samples showing long-range ferroelectric order --- coincides with the ferroelectric transition at $T_\mathrm{FE}$. The sample shown here is the same as the one discussed in \cite{Thomas2019}, exhibiting relaxor-type ferroelectricity similar to $\beta^\prime$-(ET)$_2$ICl$_2$.}\label{kappa-Cl-noise}
\end{figure}

As outlined above, both long-range ferroelectric order (suggested to be caused by charge ordering) of order-disorder type \cite{Lunkenheimer2012} and relaxor-type ferroelectricity are observed for different samples \cite{Lang2014,Thomas2019}. Very similar to the behavior observed for $\beta^\prime$-(ET)$_2$ICl$_2$ [Fig.\ \ref{beta-prime-conductance-Lorentzians} in section \ref{subsec_betaICl_noise}], Fig. \ref{kappa-Cl-noise} displays two-level fluctuations in $\kappa$-(ET)$_2$Cu[N(CN)$_2$]Cl exhibiting a clear shift of the characteristic corner frequency $f_c$ of a Lorentzian contribution superimposed on the underlying $1/f$-type noise which is only seen for temperatures below $T_\mathrm{FE}$. Similar to $\beta^\prime$-(ET)$_2$ICl$_2$, from the shift of $f_\mathrm{c}$ with the electric field measured across the sample, a nano-scale size of the switching entity has been deduced, where at this temperature the threshold electric field, cf.\ Eq.\ \eqref{mod_Arrhenius}, seems to be very small. Thus, the concept of PNR as precursors of relaxor-type electronic ferroelectricity, that can be probed by resistance or conductance noise spectroscopy, may be common to the organic charge-transfer salts prone to electronic ferroelectricity.

\subsection{On the nature of relaxor-type ferroelectricity – the case of $\kappa$-(BETS)$_2X$}\label{subsec-quasi-2D-kappa-X}
A good candidate for electronic ferroelectricity
is the system \BETS\ (in short \kBETS).
Although the compound hosts magnetic Mn$^{2+}$ ions in the acceptor molecules, 
from different experiments \cite{Vyaselev2017,Kartsovnik2017,Riedl2021} it has been inferred that the coupling between the BETS $\pi$- and the Mn $d$-electron spins is rather negligible. 
In theoretical \textit{ab initio} calculations and modeling of magnetic torque and NMR measurements \cite{Vyaselev2012,Vyaselev2017} a spin-vortex-crystal order highlighting the importance of magnetic ring exchange \cite{Riedl2021} has been discussed. 
The system undergoes a transition from a metallic high-temperature phase to an insulating ground state at $T_\mathrm{MI} \sim 20 - 25\,$K caused by strong electron-electron interactions, consistent with band structure calculations \cite{Zverev2010} revealing a narrow bandwidth and a relatively strong dimerization of the BETS molecules. Quantum oscillation measurements under pressure \cite{Kartsovnik2017,Zverev2019} indicate strong electronic correlations suggesting a Mott instability as the origin of the MI transition.
As for $\kappa$-(ET)$_2$Cu[N(CN)$_2$]Cl, see Fig.\,\ref{kappa-Cl-phase-diagram} above, hydrostatic pressure induces a superconducting state with $T_\mathrm{c} = 5.7\,$K at $p = 0.6\,$kbar \cite{Zverev2010}.
The metal-insulator transition in single crystals of \BETS\ and the dynamics related to possible electronic ferroelectricity recently has been characterized in detail by studies of thermal expansion, resistance fluctuation (noise) spectroscopy and dielectric spectroscopy \cite{Thomas2024}. The combination of these spectroscopic methods is particularly useful in this compound exhibiting metallic behavior down to about 25\,K such that the relatively high conductivity leads to pronounced non-intrinsic effects in the dielectric spectra (so-called Maxwell-Wagner relaxations) \cite{Lunkenheimer2009,Bobnar2002}. Therefore, measurements of the dielectric constant are restricted to temperatures $T \lesssim T_{\mathrm{MI}}$ and are complemented by noise measurements performed at temperatures below and above $T_{\rm MI}$.
Due to the coupling of dielectric fluctuations to the sample's resistance, fluctuation spectroscopy is capable of detecting fluctuating polar entities on the nano-/mesoscopic scale \cite{Raquet2000,JMueller2009a,JMueller2020}. The two complementary spectroscopies then reveal a distinct low-frequency dynamics on different length scales, namely (i) an intrinsic relaxation that is typical for relaxor ferroelectrics which classifies the system as a possible new multiferroic, and (ii) two-level processes which we have identified as fluctuating PNR, i.e., clusters of quantum electric dipoles that fluctuate collectively. 

\begin{figure}[!t]
\centering
\includegraphics[width=0.36\textwidth]{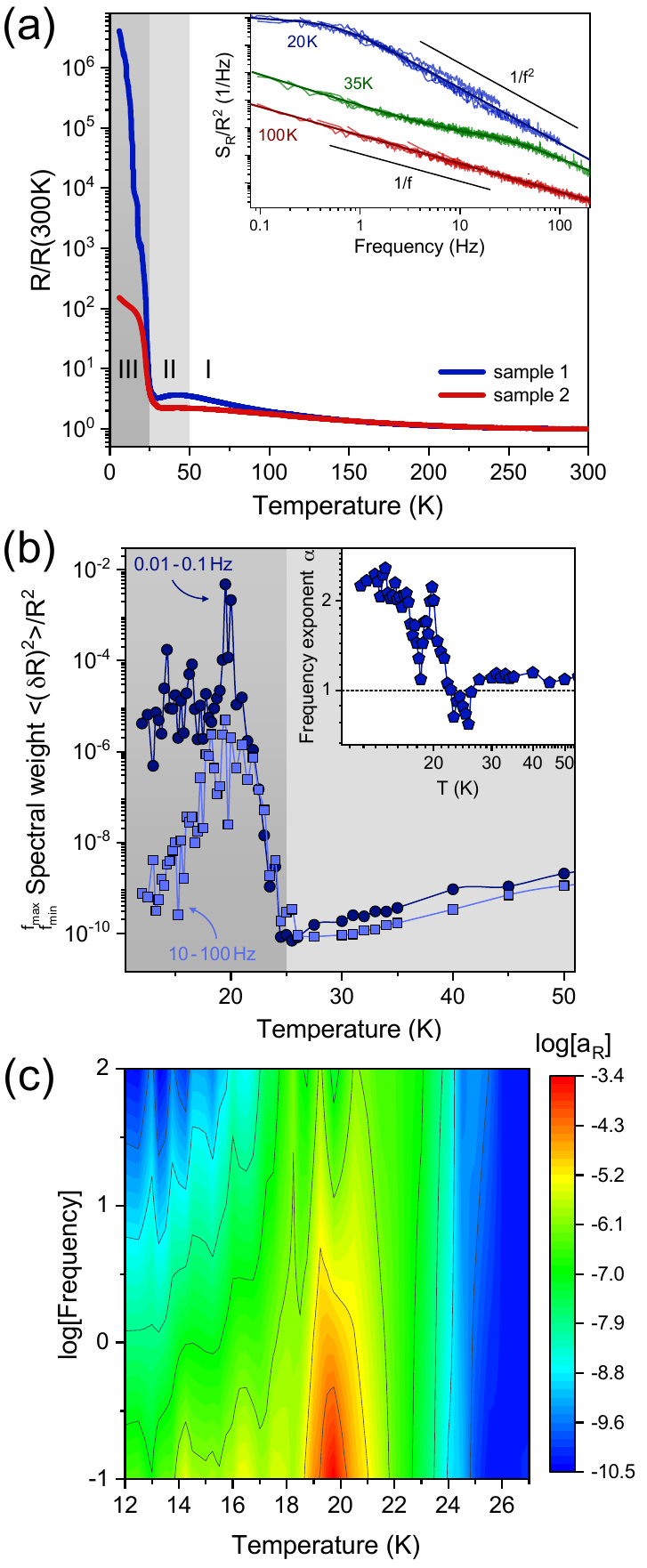}
\caption{(a) Normalized $R(T)$ of two \kBETS\ samples. The gray shaded regions correspond to the temperature regimes, where pure $1/f$-type spectra (region I), $1/f$-type and superimposed Lorentzian spectra (region II) and spectra of $1/f^2$-type with a strong time dependence (region III) were observed. Inset: Typical fluctuation spectra (sample \#1), shown as normalized resistance noise PSD, $S_R/R^2(f)$, in a double-logarithmic plot for selected temperatures representing the regimes I, II and III. Spectra are shifted for clarity. Different lines shown for each temperature are due to subsequent measurements of different frequency spans. Slopes $S_R/R^2 \propto 1/f$ and $\propto 1/f^2$ are indicated. Solid lines are fits to Eq.\ \eqref{noise_spectra}. 
(b) Spectral weight for different frequency ranges
of sample \#1 vs.\ $T$ around the MI transition (light grey: regime II and dark grey: III). Inset: Frequency exponent $\alpha$ vs.\ temperature. (c) Contour plot of the relative noise level $a_{\rm{R}}= f \times S_R/R^2$ in log scale vs.\ $T$ vs.\ $\log{f}$. Reproduced from \cite{Thomas2024}.}
\label{k-BETS-Mn_resistivity-and-noise}
\end{figure}

Figure \ref{k-BETS-Mn_resistivity-and-noise} shows the normalized resistance $R(T)/R(300\,\rm{K})$ of two \kBETS\ samples, where sample 1 seems of better quality in terms of the magnitude of resistance change upon entering the insulating phase. Below a small local hump at about $T\sim42\,$K the resistivity decreases upon cooling before it displays an abrupt increase by six orders of magnitude, marking the MI transition at $T_{\rm MI} \approx 22.5$\,K defined by a peak in ${\rm d}\ln{R}/{\rm d}T$ \cite{Kushch2008} (with the onset of the resistance increase at about 25\,K). This coincides with the occurrence of AFM order at $21 - 23$\,K in NMR  and specific heat measurements \cite{Vyaselev2012,Vyaselev2012b,Riedl2022}.

The inset of Fig.\ \ref{k-BETS-Mn_resistivity-and-noise}(a) shows three resistance noise spectra characterizing three distinctly different temperature regimes of charge carrier dynamics. At high temperatures (region I), the resistance noise PSD shows pure $1/f$-type behavior (red spectrum) the temperature dependence of which is related to enhanced structural dynamics, which we will not discuss here. A more complex behavior of the charge carrier dynamics emerges for temperatures below and just above the MI transition. For $T > T_{\rm MI}$ (region II, light gray shaded area), Lorentzian spectra superimposed on the underlying $1/f$-type noise (green) caused by dominating two-level fluctuators are observed to strongly couple to the applied electric field and are therefore interpreted as fluctuating PNR.
For $T < T_{\rm MI}$ (region III, dark gray shaded area), strongly enhanced Lorentzians with $S_R/R^2 \propto 1/f^2$ for $f > 1$\,Hz (blue) are observed, which are both current dependent and time dependent
suggesting, respectively, (i) a non-linear coupling to the electric field, and (ii) non-equilibrium dynamics and spatial correlations emphasizing the metastable character of the charge carrier dynamics. The latter is confirmed by measurements of the so-called 'second spectrum' $S^{(2)}(f_1,f_2)$ demonstrating ergodicity breaking, see discussion below.

Considering firstly only the $1/f$-type contribution to the fluctuation spectra, i.e.\ without the superimposed Lorentzian contribution in regime III, we show in Fig.\ \ref{k-BETS-Mn_resistivity-and-noise}(b) the spectral weight $\int_{f_{\rm min}}^{f_{\rm max}}S_R(f)/R^2{\rm d}f$, which corresponds to the variance of the signal for the given bandwidth, vs.\ temperature
for two frequency bandwidths $[0.01,0.1\,$Hz] and $[10,100\,$Hz].
The MI transition at these low frequencies is accompanied by a drastic increase of the noise PSD's spectral weight by many orders of magnitude, culminating in a maximum at $T\sim20\,$K, which is more pronounced and sharp for the lower frequency bandwidth [$0.01,0.1\,$Hz]. A striking observation is that upon further cooling, after a drastic increase by more than seven orders of magnitude, the noise level at these low frequencies saturates at a high level below the peak (about five orders of magnitude higher than at the onset of the transition), whereas the spectral weight for the higher frequency window [$10,100\,$Hz] decreases upon further cooling to a value comparable to that above the transition. Thus, the charge carrier dynamics below $T_{\rm{MI}}$ is dominated by rather slow fluctuations. The spectral weight of fluctuations with $S_R \propto 1/f^\alpha$ is reflected by the frequency exponent $\alpha(T) = - \partial\ln S_R(T)/\partial\ln f$, shown in the inset of Fig.~\ref{k-BETS-Mn_resistivity-and-noise}(b), where $\alpha = 1$ corresponds to a homogeneous distribution of the energies of fluctuators contributing to the $1/f$-type noise, and $\alpha > 1$ and $\alpha < 1$ correspond to slower and faster fluctuations in comparison, respectively \cite{JMueller2018}. 
$\alpha$ strongly increases upon cooling through the MI transition with a peak value $\alpha \sim 2$ at 20\,K, indicating a strong shift of spectral weight to lower frequencies and a drastic slowing down of the dynamics. A spectrum with $\alpha = 2$ is a signature of non-equilibrium dynamics and often results from a high-frequency tail of a single Lorentzian [cf.\ blue curve in the inset of Fig.~\ref{k-BETS-Mn_resistivity-and-noise}(a)], which dominates the fluctuations over the entire frequency range of our measurements and implies a switching of the system mainly between two states.
The overall picture of an enhanced noise level and drastic slowing down of charge carrier dynamics is highlighted in a contour plot of the so-called relative noise level $a_{\rm{R}}=S_R/R^2 \times f$ [see Fig.~\ref{k-BETS-Mn_resistivity-and-noise}(c)] vs.\ temperature and frequency. $a_{\rm{R}}$, a dimensionless quantity characterizing the strength of the fluctuations, upon cooling clearly starts to increase below the onset of the MI transition $T = 25$\,K at all frequencies and peaks at about $T = 19 - 20$\,K. The noise maximum is more pronounced at lower frequencies (note that $a_{\rm{R}}$ and $f$ are shown on a logarithmic scale). In particular, at low temperatures, only the slow fluctuations prevail, whereas at higher frequencies, the noise level of $T > T_{\rm MI}$ is almost recovered. 

\begin{figure}[!t]
\centering
\includegraphics[width=0.475\textwidth]{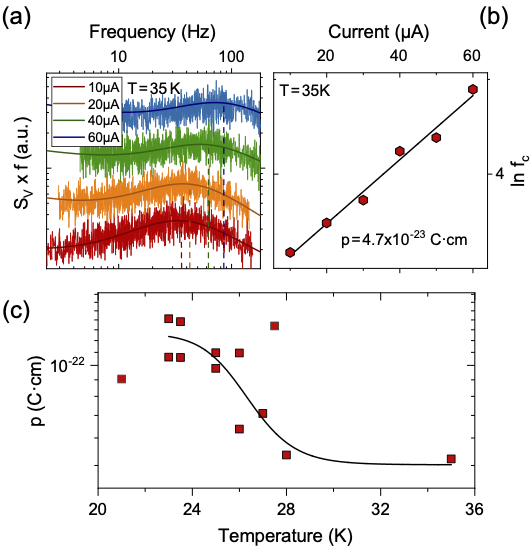}
\caption{
(a) Lorentzian contribution in $\kappa$-(BETS)$_2$Mn[N(CN)$_2$]$_3$ shown as $S_\mathrm{V} \times f$ vs.\ $f$ for $T=35\,$K and different currents, revealing a shift of the corner frequency as illustrated in (b). The dipole moment, which is extracted from the slope of the linear fits according to Eq.\,(\ref{eq:dipole-energy}) shown in (b), is displayed in (c) vs.\ temperature. The black line is a guide to the eye. Reproduced after \cite{Thomas2024}.}
\label{k-BETS-Mn_Lorentzians}
\end{figure}
We now discuss the observed systematic non-linear current dependence of the fluctuations for temperature regime II. 
Analyzing the observed Lorentzian spectra superimposed on the $1/f$-type noise, we describe the fluctuations by Eq.\,\eqref{noise_spectra} introduced above,
\begin{equation*}
\frac{S_R(f,T)}{R^2} = \frac{a}{f^{\alpha}} + \frac{b}{f^2 + f_\mathrm{c}^2},
\end{equation*}
at temperatures from 50\,K down the temperature of the noise peak below the onset of the MIT (regime II), 
see exemplarily the spectra at $T = 35$\,K shown in Fig.\ \ref{k-BETS-Mn_Lorentzians}.
In Fig.\ \ref{k-BETS-Mn_Lorentzians}(a) we show $S_R/R^2 \times f$,
thereby emphasizing the Lorentzian contribution with its maximum at the corner frequency $f_{\rm{c}}$, vs.\ the applied current representing the electric field $\mathcal{E}$ for a fixed temperature. For fixed current/electric field, the corner frequency vs.\ temperature (not shown) exhibits an Arrhenius behavior with temperature, i.e.\ a thermally activated switching behavior with characteristic energies of order $20 - 90$\,meV at temperatures $25 - 50$\,K \cite{Thomas2024}. We assign these two-level processes to clusters of quantum electric dipoles fluctuating collectively, i.e.\ PNR with a distribution of cluster sizes causing the range of activation energies observed at different temperatures. 

The two-level (Lorentzian) fluctuations with characteristic frequency $f_c$ shown in Fig.~\ref{k-BETS-Mn_resistivity-and-noise}(a,b) at fixed temperature, however, strongly shift with the applied current to higher values, whereas their magnitude $b(T)$ gets suppressed, very similar to previous observations in the square- and triangular-lattice Mott insulators and relaxor ferroelectrics $\beta^\prime$-(ET)$_2$ICl$_2$ \cite{JMueller2020} (Fig.\ \ref{beta-prime-conductance-Lorentzians}), and $\kappa$-(ET)$_2$Cu[N(CN)$_2$]Cl (Fig.\ \ref{kappa-Cl-noise}), respectively.
The linear increase observed in a plot $\ln{f_c}$ vs.\ current [Fig.~\ref{k-BETS-Mn_resistivity-and-noise}(b)] suggests that the thermally activated two-level processes depend on the dipole energy $E_{\rm dipole} = p\mathcal{E}$ according to \cite{Raquet2000,JMueller2020}
\begin{equation}
f_{\rm{c}}=f_0\exp\left(\frac{p\mathcal{E}-E_{\rm{a}}}{k_{\rm{B}}T}\right).
\label{eq:dipole-energy}
\end{equation}
As for the compounds discussed above, the dipole moment $p$ at a fixed temperature can be determined from the slope yielding $p=4.7\cdot10^{-23}\,{\rm C \cdot cm}$ at $T=35\,$K, where the electric field is calculated from the applied current, the temperature-dependent resistance and the sample thickness. Very recent vibrational infrared  spectroscopy measurements revealed a small but significant charge disproportionation on the dimer which tentatively has been estimated to $\delta \lesssim 0.02 - 0.05\,e$ \cite{Dressel2024}. This corresponds to a fluctuating nanoscale polar region of radius $3 - 5$\,nm for spherical PNR or $5-8$\,nm for cylindrical PNR with the height of the unit cell (one molecular layer), which is comparable though somewhat smaller than the PNR size estimated for $\beta^\prime$-(ET)$_2$ICl$_2$ with $\delta \leq 0.1\,e$ \cite{Iguchi2013,JMueller2020}, see section \ref{subsec_betaICl}.
The temperature evolution of $p$ is displayed in Fig.~\ref{k-BETS-Mn_Lorentzians}(c) revealing increasing values for decreasing temperatures, which appear to saturate below  $T_{\rm MI}$.

Dielectric measurements have been performed in a broad frequency range $0.1\, \mathrm{Hz < \nu < 1.8}$\,GHz and at temperatures $T = 5 - 300$\,K. However, above the MI transition the rather high conductivity of \kBETS\ gives rise to pronounced non-intrinsic contributions (so-called Maxwell-Wagner relaxations) \cite{Lunkenheimer2009,Bobnar2002}, likely due to the generation of Schottky diodes at the electrode-sample interfaces (see \cite{Thomas2024} for a detailed discussion). These measurements provide complementary information to the resistance noise spectroscopy in several ways: Dielectric spectroscopy covers a broader frequency range and detects the dynamics of individual electric dipoles fluctuating in an ac electric field. Fluctuation spectroscopy is sensitive to the dynamics of larger-scale polar objects like PNR or domains, which couple to the resistivity. Moreover, dielectric spectroscopy requires a relatively insulating behavior of the measured material, whereas resistance noise spectroscopy works for sufficiently conducting samples. 

\begin{figure}[!h]
\centering
\includegraphics[width=0.475\textwidth]{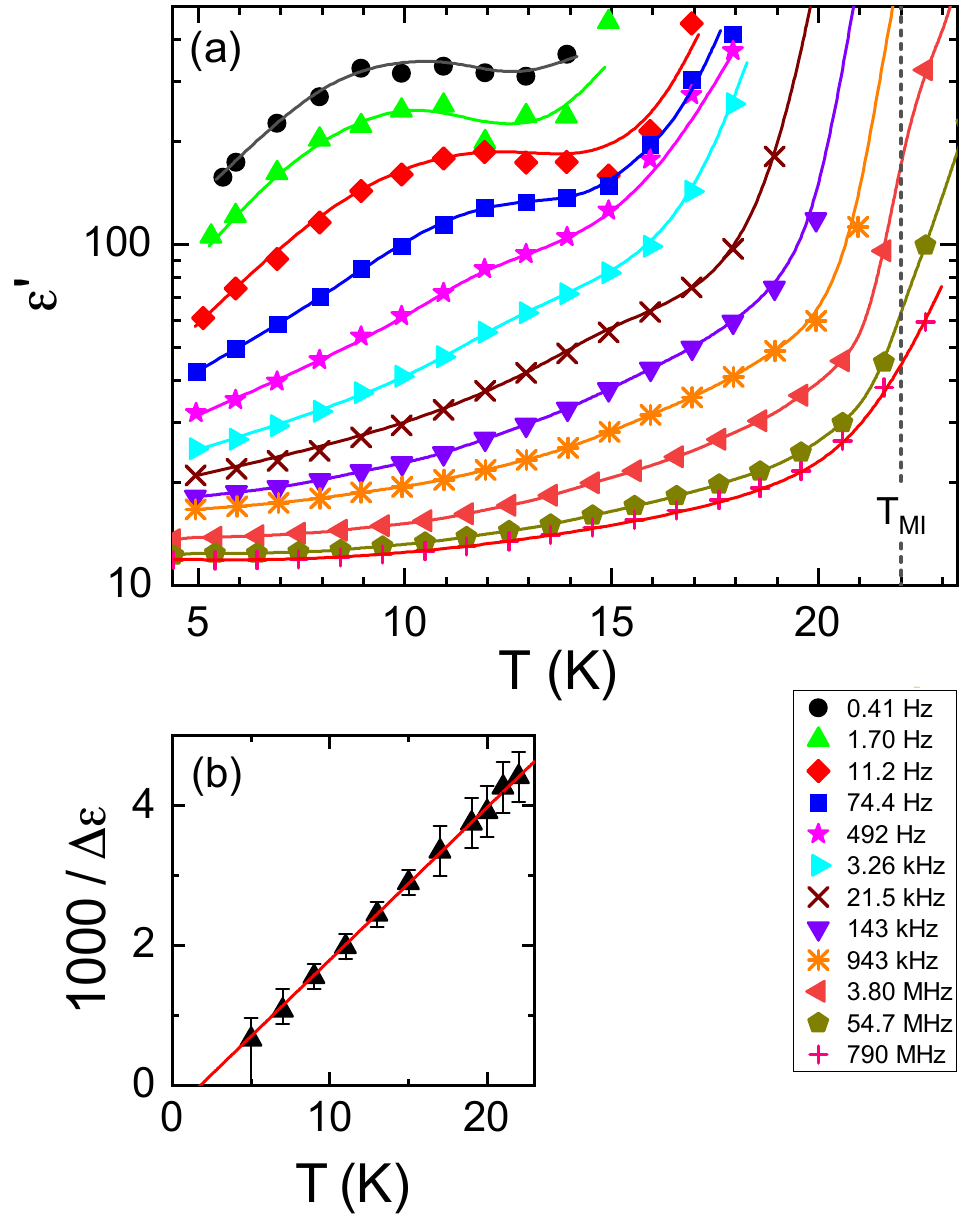}
\caption{(a) Temperature dependence of \eps\ of $\kappa$-(BETS)$_2$Mn[N(CN)$_2$]$_3$ measured at various frequencies. The onset temperature of the metal-insulator transition (22\,K for this sample) is indicated by the vertical dashed line. The lines are guides for the eye. (b) Inverse relaxation strength of the intrinsic process at $T<T_{\mathrm{MI}}$. The line indicates a Curie-Weiss law with $T_{\mathrm{CW}} = 1.8$\,K. Reproduced from \cite{Thomas2024}.}

\label{k-BETS-Mn_dielecric}
\end{figure}

Figure~\ref{k-BETS-Mn_dielecric}(a) shows the temperature dependence of the dielectric constant \eps\ at various frequencies for temperatures $T \lesssim T_{\mathrm{MI}}$ (the pronounced increase of \eps\ at the highest temperatures arises from the onset of the Maxwell-Wagner relaxations mentioned  above). The sigmoidal curve shape can be ascribed to an intrinsic relaxation process, shifting to lower temperatures with decreasing frequency. At the lowest frequencies, \eps($T$) exhibits a peak and reaches rather high values of several hundred. Aside of the mentioned non-intrinsic contributions, the overall behavior in Fig.~\ref{k-BETS-Mn_dielecric}(a) reveals the typical signatures of relaxor ferroelectricity [cf. Fig.\,\ref{paper-ferroele-eps}(c)] \cite{Cross1987,Samara2003}. Therefore, \BETS\ can be regarded as another example of a possible organic multiferroic.

As discussed above, relaxor ferroelectricity, believed to arise from short-range, cluster-like ferroelectric order \cite{Viehland1990,Bokov2006,Fu2009}, was also reported for various other charge-transfer salts \cite{Abdel-Jawad2010,Abdel-Jawad2013,Lang2014,Lunkenheimer2015b,Fischer2021,Canossa2021,Lunkenheimer2015a}. For \kBETS\, the intrinsic nature of the detected relaxor behavior was checked by measuring several samples using two different experimental setups and different contact materials \cite{Thomas2024}. For a further confirmation of relaxor ferroelectricity in \kBETS\, polarization measurements as reported, e.g., in \cite{Lunkenheimer2015b,Fischer2021,Thurn2021} and as shown in Fig.\ \ref{beta-prime-polarization} for \betaICl\ \cite{Iguchi2013}, would be desirable. However, due to the relatively high conductivity of this material \cite{Kushch2008,Zverev2010} such measurements are impracticable.

In Ref. \cite{Thomas2024}, the dielectric spectra of \kBETS\ were fitted using an equivalent-circuit approach accounting for the mentioned Maxwell-Wagner contributions. This enabled, e.g., the deduction of the relaxation strength $\Delta \varepsilon$ of the intrinsic, relaxor-like relaxation process. In Fig. \ref{k-BETS-Mn_dielecric}(b), the inverse of the obtained $\Delta \varepsilon (T)$ is shown. Its linear increase evidences Curie-Weiss behavior, $\Delta \varepsilon \propto 1/(T-T_{\mathrm{CW}})$ with $T_\mathrm{CW}=1.8$\,K. The latter provides an estimate of the quasistatic dipolar freezing temperature. It is even lower than the relatively low $T_\mathrm{CW}=6$\,K reported for $\kappa$-(ET)$_2$Cu$_2$(CN)$_3$ \cite{Abdel-Jawad2010} and it is significantly lower than the $T_\mathrm{CW}$ values between 35 and 206\,K found for other charge-transfer salts revealing relaxor ferroelectricity  \cite{Iguchi2013,Lunkenheimer2015b,Fischer2021,Canossa2021}.

We want to point out again that the relaxation processes detected by dielectric spectroscopy and resistance fluctuation mirror different microscopic processes, although their frequency and temperature ranges overlap. The 
latter can be ascribed to PNR switching, involving rather large effective energy barriers, which leads to relatively slow dynamics. The 
former reflect
much faster, reorientational processes, occurring on smaller length scales than the dynamics of the whole PNR. Interestingly, the theoretical treatment in Ref.\ \cite{Vugmeister2006} assumes local polarization dynamics {\it inside} the PNR of relaxor ferroelectrics. Alternatively, faster dynamics in relaxor ferroelectrics may also be explained by motions of PNR boundaries as discussed by the breathing model \cite{Glazounov1999}.
Overall, the application of these complementary experimental methods to electronic ferroelectrics leads to a more comprehensive picture of the relaxation dynamics in these systems, which reveal emergent electronic phase separation due to disorder and/or competing interactions. Notably, the PNR are fluctuating units that an electric field can stabilize already above the MI transition. Thus, they can be regarded as precursors of the relaxor ferroelectricity observed below $T_{\rm MI}$.
 
\begin{figure}[!h]
\centering
\includegraphics[width=0.475\textwidth]{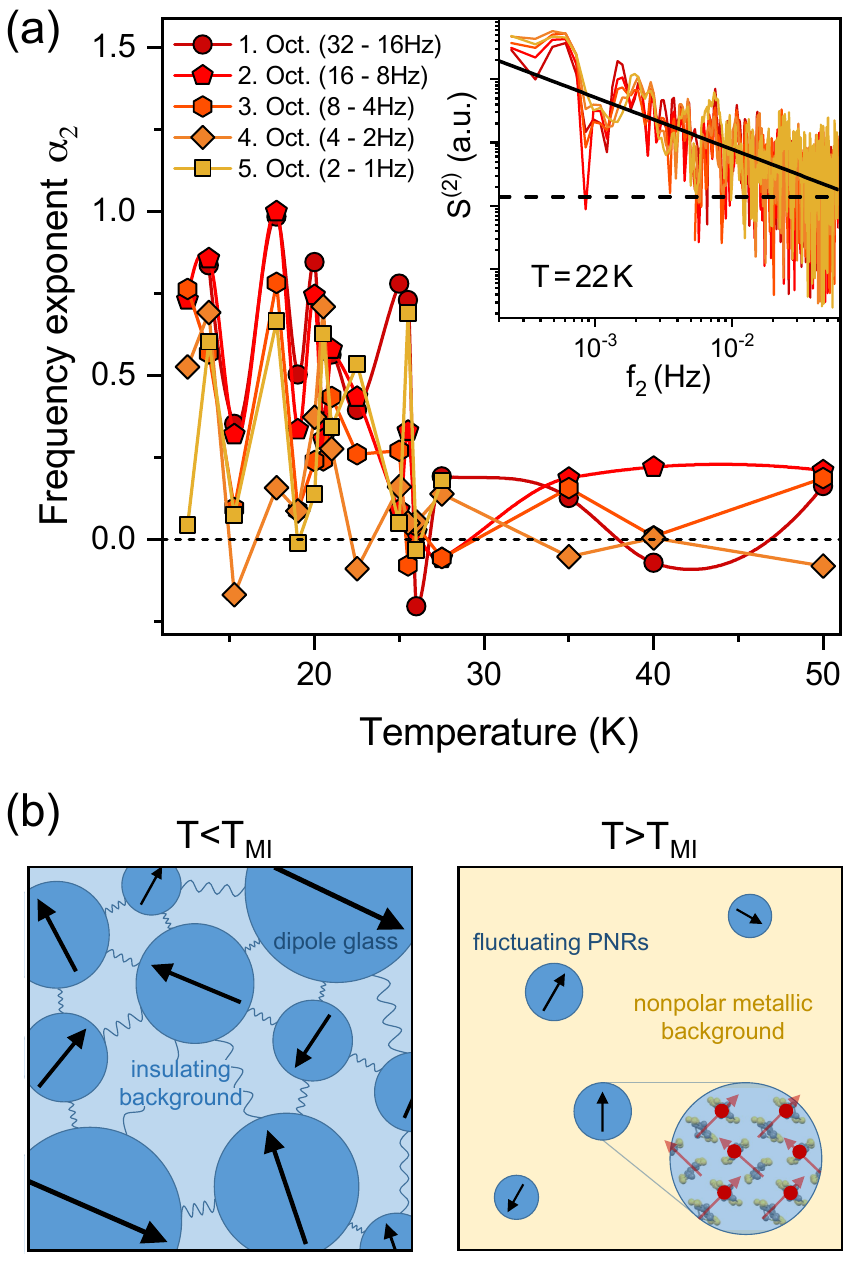}
\caption{(a) Frequency exponent $\alpha_2$ of the second spectrum $S^{(2)} \propto 1/f_2^{\alpha_2}$ of $\kappa$-(BETS)$_2$Mn[N(CN)$_2$]$_3$ against temperature, which is extracted from a linear fit of the PSD in a double-logarithmic plot, as shown in the inset. Different red to orange colors mark different octaves. (b) Schematic illustration of the dipolar dynamics that leads to the observed behavior in the resistance noise and dielectric properties. Left: Frozen dipole-glass state below $T_{\rm{MI}}$ exhibiting relaxor-type ferroelectricity and strong non-equilibrium dynamics. Right: Polar nanoregions (PNR) preformed in the metallic phase above $T_{\rm{MI}}$. Whereas above $T_{\rm{MI}}$, the PNR are independently fluctuating, they exhibit spatial correlations in the insulating/ferroelectric regime (indicated by wavy lines). Taken from  \cite{Thomas2024}.
}
\label{k-BETS-Mn_non-equilibrium}
\end{figure}

Finally, we point out the unusual non-equilibrium dynamics and ergodicity breaking in the relaxor-ferroelectric phase. A sudden onset of a strong time dependence of the resistance/conductance noise PSD in the insulating state occurs below $T_{\rm MI}$, which results in fluctuations that are not statistically stationary anymore, i.e.\ variations of the spectral weight for repeated measurements at the same temperature. This so-called spectral wandering is reflected by a non-Gaussian probability distribution of the time signal, often caused by spatially correlated fluctuators \cite{Bogdanovich2002,Jaroszynski2002,Kar2003,Jaroszynski2004,Hartmann2015}.

Below $T_{\rm{MI}}$, we observe deviations from a Gaussian distribution (not shown) very similar to the dynamics of the first-order electronic phase transition in complex transition metal oxides \cite{Ward2009}, which was ascribed to electronic phase separation.  
In order to identify interacting/spatially-correlated fluctuators, we investigated the higher-order correlation function 
by measurements of the second spectrum $S^{(2)}(f_2,f_1,T)$, which corresponds to the PSD of the fluctuating first spectrum (see, e.g., \cite{Restle1983,Weissman1992,Weissman1993,Seidler1996,Yu2004a,Yu2004b} for more detailed information). Here, $f_1 \equiv f$ and $f_2$ correspond to frequencies of the first and second spectrum, respectively, where $f_2$ results from the time dependence of $S(f,T) \equiv S^{(1)}(f_1,t)$ at fixed frequency $f_1$. In the case of correlated fluctuators, the second spectrum often shows a frequency dependence according to $S^{(2)} \propto 1/ f_2^{\alpha_2}$ with $\alpha_2 > 0$, whereas $\alpha_2 = 0$ for statistically stationary, Gaussian fluctuations \cite{Weissman1988}.\\ 
A typical spectrum of $S^{(2)}(f_2)$ is displayed in the inset of Fig.\ \ref{k-BETS-Mn_non-equilibrium}(a) at $T=22\,$K and indeed reveals a $S^{(2)} \propto 1/f_2$ behavior (black line). 

The second spectrum, analyzed for different octaves which are indicated by red to orange colors in Fig.\ \ref{k-BETS-Mn_non-equilibrium}(a), correspond to varying frequency ranges of the first spectrum, see \cite{Thomas2024}. 
Strikingly, $S^{(2)}$ is roughly frequency independent ($\alpha_2 \sim 0$) above the MI transition, and a sudden increase of the frequency exponent up to $\alpha_2\sim1$ occurs upon cooling through $T_{\rm{MI}}$, coinciding with the strong increase in the magnitude of slow fluctuations $S^{(1)}(f_1)$ shown in Fig.\,\ref{k-BETS-Mn_resistivity-and-noise} above. Clearly, the MI transition is accompanied by the onset of strong non-equilibrium charge dynamics indicative of spatially-correlated fluctuators and a dipolar glass.

The following tentative scenario is therefore suggested for the dynamics of the electric dipoles residing on the (BETS)$_2$ dimers. Upon approaching $T_{\mathrm{MI}}$ from above, the $\pi$-holes become localized on the dimers due to strong electron-electron correlations driving the system into the Mott insulating state. This results in a strong increase in both resistance and resistance noise. However, the holes remain delocalized on their respective dimers and the hopping of the holes between the two molecules corresponds to the reorientation of an associated dipolar moment, detected by dielectric spectroscopy.\\
With decreasing temperature, this local motion starts to slow down and, simultaneously local ferroelectric correlations (which have developed in part already above $T_{\mathrm{MI}}$) lead to the cluster-like ferroelectric order, typical for relaxor ferroelectrics. Upon cooling through the MI transition, those PNR which dominate the resistance noise and preexist as fluctuating entities above $T_{\mathrm{MI}}$ undergo a
transition of interacting TLS.
The abrupt and drastic change of the statistical properties of the fluctuations and onset of non-equilibrium dynamics --- a signature of spatial correlations \cite{Bogdanovich2002,Jaroszynski2002,Kar2003,Jaroszynski2004,Hartmann2015} --- right at the onset of the metal-insulator transition can be explained by a change of the background matrix, which is metallic above $T_{\rm{MI}}$ providing screening of the dipole moments. Below $T_{\rm{MI}}$, the insulating background allows the PNR to interact, see Fig.~\ref{k-BETS-Mn_non-equilibrium}(b) for a schematic illustration. 
 
The strong enhancement of the noise magnitude $S_R/R^2$ at the MI transition at low frequencies below $\sim 1$\,kHz shown in Fig.\ \ref{k-BETS-Mn_resistivity-and-noise}(b,c) can be considered a signature of the first-order phase transition accompanied by emergent electronic phase separation \cite{Chen2007,Ward2009,Daptary2019}. The enhancement of the frequency exponent $\alpha$ corresponding to a drastic shift of spectral weight to low frequencies and slowing down of charge carrier dynamics which persists at low temperatures, is consistent with the localization of charge carriers and the onset of the freezing of PNR switching.  
Moreover, whereas in conventional relaxors the correlated dipolar dynamics arises from ionic motions and exhibits glass-like freezing upon further cooling \cite{Cross1987,Samara2003,Viehland1990}, in \kBETS\ tunneling starts to dominate at low temperatures \cite{Thomas2024}, preventing the further slowing down and final 
arrest of the essentially electronic dynamics. This may explain the saturation of the noise level/persistence of slow dynamics at low frequencies and low temperatures.

\section{Epilogue}
\label{sec-summary}
In this review, we have discussed the ferroelectric and multiferroic properties of organic charge-transfer salts. The main focus has been placed on the more recent developments in the field, covering the quasi-2D systems of the (BEDT-TTF)$_2X$ and (BETS)$_2X$ varieties. These studies have followed the seminal work on the quasi-1D (TMTTF)$_2X$ systems where ferroelectricity was well established. The charge-order-driven polar state observed in the latter systems represents a prime example for electronic ferroelectricity which is controlled by electronic degrees of freedom, instead of the shifts of atomic positions as in conventional displacive ferroelectrics.

It has turned out that due to their molecular degrees of freedom, involving electronic states extending over arrays of molecules, correlated 2D organic charge-transfer salts close to a Mott- or charge-order metal-insulator transition, are prone to a particular type of electronic ferroelectricity, where the polar state is caused by charge disproportionation in a modulated or bond-alternated dimerized lattice of organic donor molecules. Here the centrosymmetric dimer-Mott state features a single $\pi$-hole with spin 1/2, whose center of gravity is localized at the midpoint of the molecular dimer unit (ET)$_2$, representing a charge-centered or charge-averaged phase. However, if the charge shifts toward one of the molecules within the dimer, resulting in a charge-ordered phase, this imbalance gives rise to a quantum electric dipole. Unlike canonical ferroelectrics, where ions are involved, the much lighter mass of electrons in charge-driven ferroelectrics can enable significantly faster switching processes. \\

The proof of ferroelectricity in the various classes of organic charge-transfer salts has been based on a combination of different experimental probes often initiated and accompanied by theoretical considerations.\\ 

On the experimental side, dielectric spectroscopy, although often experimentally challenging to apply to organic charge-transfer salts, has become a most successful tool to study the response to ac electric fields and to explore the ferroelectric properties.
Clear evidence for order-disorder-type electronic ferroelectricity has been revealed for $\kappa$-(ET)$_2$Cu[N(CN)$_2$]Cl and $\kappa$-(ET)$_2$Hg(SCN)$_2$Cl. Whereas the polar order in both cases is driven by charge order, in the former compound the CO is weaker in magnitude and emerges within the Mott insulating state, remarkably occurring simultaneously with long-range antiferromagnetic order. In the latter compound, CO and ferroelectricity coincide with the metal-insulator transition.
A similar strong charge-order-based scenario was also found for $\theta$-(BEDT-TTF)$_2$RbZn(SCN)$_4$.

A relaxor-type ferroelectric response, indicative of interacting electric dipoles in the presence of a random potential or competing interactions, is more prevalent and was first observed in $\kappa$-(ET)$_2$Cu$_2$(CN)$_3$. Similarly, relaxor-type ferroelectricity occurs in other $\kappa$-(ET)$_2X$ compounds, such as $X$ = Ag$_2$(CN)$_3$, as well as in systems with different packing motifs or donor molecules, like $\beta^\prime$-(ET)$_2$ICl$_2$, 
$\alpha$-(ET)$_2$l$_3$, and various Pd(dmit)$_2$ systems.

Structural aspects like lattice deformation accompanying the ferroelectric transition as well as the lattice dynamics coupled to the charge and/or spin degrees of freedom have been investigated by thermal expansion measurements and inelastic neutron scattering, respectively. Infrared spectroscopy and Raman scattering have played an important role in determining possible charge disproportionations on the molecules.

From theoretical considerations, it is expected that the intra-dimer electron mobility leads to large temporal and spatial dielectric fluctuations, which may play a critical role in the dielectric and optical properties, as well as in the ferroelectric phase transition.
In this context, resistance or conductance fluctuation (noise) spectroscopy are an ideal complement to dielectric spectroscopy since the former measurements reveal additional information on fluctuating polar entities of different length and time scales and thus the physics underlying electronic ferroelectricity, and at temperatures where samples are too conductive for dielectric measurements.
Comparing the results of different organic charge-transfer salts reveals striking similarities regarding the formation of PNR or fluctuating mesoscopic domains and their interaction with electric fields. These results suggest that fluctuating PNR, sometimes stabilized in electric fields at temperatures above the coherent localization/ordering of charges,
may be a common characteristic of electronic ferroelectricity of relaxor type in molecular materials. Likewise, strong non-equilibrium dynamics and ergodicity breaking at low temperatures consistent with droplet models in spin glasses \cite{Thomas2024} has been observed in different systems. 

Multiferroicity, the simultaneous ordering of both electric and magnetic dipoles, is observed in some few organic salts with weak magnitudes of charge order, including $\kappa$-(ET)$_2$Cu[N(CN)$_2$]Cl. Microscopically, the primary mechanism for the interplay between charge and spin orders is the modification of the effective magnetic couplings by charge disproportionation between different molecules. This mechanism 
differs from other multiferroics, often based on displacive ferroelectricity where distortions of the ions themselves lead to electric polarization and modification of the magnetic couplings.
As such, mechanisms such as the inverse Dzyaloshinskii-Moriya are not relevant in organic charge-transfer salts. Instead, it is the modification of the magnitudes of isotropic couplings that serves as the primary mechanism of magnetoelectric coupling. For strong charge order, the resulting magnetic couplings typically become increasingly one-dimensional, likely leading to a competition between charge and magnetic orders. For weak charge order, the reduction of symmetry may lead to a relief of magnetic frustration, promoting magnetic order. The coupling of the charge and spin degrees of freedom leads to a variety of effects, such as inhomogeneous magnetic responses in charge glass states. \\

Clearly, the observation of ferroelectric order driven by electronic degrees of freedom, has added another dimension to the research on molecular systems in general and organic charge-transfer salts in particular. Studying the static and dynamical properties of correlated electrons in reduced dimensions under the influence of  geometric frustration, coupling to the lattice and random potential, are of great importance. Research in this area shows great promise not only for fundamental research. It also bears the potential for future applications in electronic devices with ultra-fast response time.  

\begin{acknowledgements}
We acknowledge support from the Deutsche Forschungsgemeinschaft (DFG, German Research Foundation) through the Transregional Collaborative Research Center TRR 288 - 422213477 (projects A06(terminated) and B02). Work in Augsburg was supported by the DFG through TRR 80.

\end{acknowledgements}

The authors declare that they have no conflict of interest. 

\bibliography{review-article}

\end{document}